\documentclass[twocolumn, secnumarabic, amssymb, nobibnotes, aps, prd, longbibliography, nofootinbib]{revtex4-2}

\usepackage[hidelinks]{hyperref}
\usepackage{amsmath}
\usepackage{mathtools}
\usepackage{bbm}
\usepackage{xcolor}
\usepackage{xfrac}

\usepackage{graphicx}
\usepackage{lipsum}

\begin{document}

\title{
Intuitive dissection of the Gaussian information bottleneck method
with an application to optimal prediction
}

\author{Vahe Galstyan}
\email{v.galstyan@amolf.nl}
\affiliation{AMOLF, Science Park 104, 1098 XG Amsterdam, The Netherlands}
\author{Age Tjalma}
\affiliation{AMOLF, Science Park 104, 1098 XG Amsterdam, The Netherlands}
\author{Pieter Rein ten Wolde}
\email{tenwolde@amolf.nl}
\affiliation{AMOLF, Science Park 104, 1098 XG Amsterdam, The Netherlands}
\date{\today}

\begin{abstract}

Efficient signal representation is essential for the functioning of living and artificial systems operating 
under resource constraints.
A widely recognized framework for deriving such representations is the information bottleneck method, 
which yields the optimal strategy for encoding a random variable, such as the signal,
in a way that preserves maximal 
information about a functionally relevant variable, subject to an explicit constraint on 
the amount of information encoded.
While in its general formulation the information bottleneck method is a numerical scheme, 
it admits an analytical solution in an important special case where 
the variables involved are jointly Gaussian.
In this setting, the solution predicts discrete transitions in the dimensionality of the optimal representation 
as the encoding capacity is increased.
Although these signature transitions, along with other features of the optimal strategy,
can be derived from a constrained optimization problem, 
a clear and intuitive understanding of their emergence is still lacking.
In our work, we advance our understanding of the Gaussian information bottleneck method 
through multiple mutually enriching perspectives, including geometric and information-theoretic ones.
These perspectives offer novel intuition about the set of optimal encoding directions, 
the nature of the critical points where the optimal number 
of encoding components changes, and about the way the optimal strategy navigates 
between these critical points.
We then apply our treatment of the information bottleneck to a previously studied signal prediction problem,
obtaining new insights on how different features of the signal are encoded across multiple components 
to enable optimal prediction of future signals.
Altogether, our work deepens the foundational understanding of the information bottleneck method 
in the Gaussian setting, motivating the exploration of analogous perspectives in broader,
non-Gaussian contexts.

\end{abstract}

\maketitle

\section{Introduction}

\begin{figure}[!ht]
\centering
\includegraphics{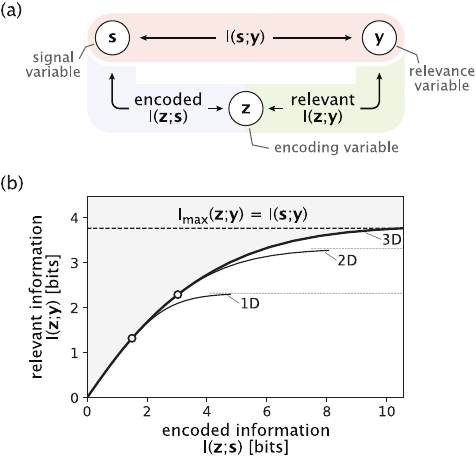}
\caption{Elements and operation of the information bottleneck method.
(a) Three key variables of the problem, together with the information metrics defined for each pair.
$I({\bf z}; {\bf s})$ is the information encoded in the variable ${\bf z}$ about the signal ${\bf s}$, and 
$I({\bf z}; {\bf y})$ is the information that ${\bf z}$ retains about the relevance variable ${\bf y}$.
Independently of the encoding procedure, 
intrinsic correlations between ${\bf s}$ and ${\bf y}$
set the mutual information $I({\bf s}; {\bf y})$, 
which serves as an upper bound on the relevant information $I({\bf z}; {\bf y})$.
(b) Information curve arising in the Gaussian information bottleneck method for an example setting 
with three-dimensional ${\bf s}$ and ${\bf y}$ variables.
The white dots mark the transitions from lower- to higher-dimensional optimal signal representations
as the encoding capacity is increased.
Thinner curves that saturate at lower levels of relevant information 
correspond to suboptimal strategies where the dimension of ${\bf z}$ 
is less than three and does not vary with increasing encoding capacity.
The Lagrange multiplier $\gamma$ in the objective functional 
$\mathcal{L} = I({\bf z}; {\bf y}) - \gamma I({\bf z}; {\bf s})$
decreases as the curves rise from the origin.
In the figure, the three curves terminate at the same value of $\gamma$.
The gray region above the information curve is inaccessible.}
\label{fig:intro}
\end{figure}

Many engineered and natural systems process stimuli received from their environment
and subsequently adjust their behavior in response 
\cite{BowsherSwain2014, VictorJNeuro2019, Palmer2015, NelsonAutoRobot2018, GoyalICLR2019}.
A key step during processing is the creation of an internal representation of the environment 
via stimulus encoding.
Intuitively, more accurate representations are harder to obtain due to stricter requirements on 
available resources and, potentially, on the architecture of the signal processing mechanism.
Since the representation only partially captures the full signal
due to limited encoding capacity, the system may choose which of the signal features to store.
What features are chosen ultimately depends on their relevance to the tasks of the system.

The information bottleneck method is a general information-theoretic approach for finding optimal 
representations \cite{Tishby1999}.
In this framework, 
a stochastic encoding procedure compresses the signal ${\bf s}$ into the encoding variable ${\bf z}$,
both of which are, in general, multi-dimensional random variables.
A key element of the problem that dictates which signal features must be encoded 
is the so-called relevance variable, denoted here by ${\bf y}$.
Correlated with the signal, ${\bf y}$ is itself a random variable known to have functional
importance for the system.
In the biological context of cell signaling, 
for example, the signal ${\bf s}$ may represent the time-dependent concentration of nutrients, 
${\bf z}$ can correspond to the levels of intracellular readout molecules, 
with the cellular signaling network defining the ${\bf s} \rightarrow {\bf z}$ mapping rule.
If the system needs to anticipate the changes in its environment, 
then ${\bf y}$ may represent the future value or future derivative of the stochastic signal,
the knowledge of which would enable the system to mount a response in advance
\cite{Sachdeva2021, Tjalma2023, Palmer2015}.
In the broader context of machine learning, if the goal is to perform signal classification, 
then ${\bf y}$ may represent a low-dimensional signal category label \cite{Goldfeld2020}.
In all these cases, the optimal encoding strategy should selectively preserve those features of the signal 
that are maximally informative about the relevance variable.

In the bottleneck method, the cost of encoding the signal is captured via the mutual information $I({\bf z}; {\bf s})$,
while the quality of encoding is captured through $I({\bf z}; {\bf y})$ 
-- a measure of how informative the signal representation ${\bf z}$ is about the relevance variable ${\bf y}$
(Fig.~\ref{fig:intro}a).
The method yields the best encoding strategy via a stochastic ${\bf s} \rightarrow {\bf z}$ mapping rule 
that maximizes the relevant information $I({\bf z}; {\bf y})$ for the given amount of encoded information $I({\bf z}; {\bf s})$.
This problem is typically formulated in the literature as one of constrained optimization where the functional 
$\mathcal{L}[p({\bf z} | {\bf s})] = I({\bf z}; {\bf y}) - \gamma I({\bf z}; {\bf s})$ is maximized over encoding strategies 
$p({\bf z} | {\bf s})$, with $\gamma$ being a Lagrange multiplier.
Intuitively, the more the constraint on the encoding capacity is relaxed (smaller $\gamma$),
the larger will the relevant information become under the optimal strategy.
In the $\gamma \rightarrow 0$ limit where the encoding capacity is no longer constrained,
${\bf z}$ will perfectly capture all signal features correlated with the relevance variable,
making the relevant information $I({\bf z}; {\bf y})$ reach its limit $I({\bf s}; {\bf y})$ set 
by the intrinsic correlations between ${\bf s}$ and ${\bf y}$ variables.

In its general formulation, the information bottleneck method provides the optimal encoding strategy through
a numerical iterative algorithm \cite{Tishby1999}.
An analytical procedure, however, becomes available in an important special case, namely when
all random variables of the problem are Gaussian together with their joint distributions \cite{Chechik2005}.
Signals with such statistical properties typically emerge from linear models, 
which are widely applied in studies of natural systems as well as in engineering 
\cite{deRondePRE2010, ChalkPNAS2018, Kalman1960, ScaglioneIEEE2002}.

In the Gaussian information bottleneck method, 
the optimal signal representation is achieved via noisy linear encoding.
A key feature of the framework is the progressively increasing 
dimensionality of this representation with the growing capacity to encode the full signal \cite{Chechik2005}.
For example, if the signal and relevance variables are three-dimensional vectors, 
then the optimal signal representation will be a scalar if the amount of information 
that can be encoded is less than a certain threshold;
once this threshold is crossed, a second encoding component will emerge 
that will capture an additional linearly independent feature of the signal.
As the encoding capacity is increased even further, past another threshold 
it becomes optimal to incorporate a third encoding component.
This suggests that the ``information curve'' 
capturing the dependence of the relevant information $I({\bf z}; {\bf y})$ on the encoded information 
$I({\bf z}; {\bf s})$
will be composed of distinct
analytic segments, each corresponding to a different dimension 
of the optimal signal representation (Fig.~\ref{fig:intro}b).
While this feature of discrete transitions 
together with other properties of optimal \mbox{encoding}
can be demonstrated mathematically,
our understanding of their emergence is still incomplete.
For instance, why is it that, instead of using multiple encoding components from the outset, the optimal strategy introduces 
them one by one at special transition points? And what are the defining properties of these points?

The main aim of the current work is to provide an intuitive understanding 
of the Gaussian information bottleneck method through a combination of analytical 
and geometric arguments with clear graphical interpretations.
As we will discuss in more detail later, a distinguishing element of our approach that 
makes many of these intuitive arguments possible is the initial standardization of 
the marginal distributions $P({\bf s})$ and $P({\bf y})$ which makes the covariance 
matrices of ${\bf s}$ and ${\bf y}$ variables diagonal, containing equal entries.
Due to this procedure, the structural features of the problem, originally contained 
in the full joint probability distribution $P({\bf s}, {\bf y})$, get concentrated in the stochastic 
${\bf s} \rightarrow {\bf y}$ and ${\bf y} \rightarrow {\bf s}$ mapping rules set 
by the conditional distributions 
$P({\bf y} | {\bf s}) = \frac{P({\bf s}, {\bf y})}{P({\bf s})}$ and 
$P({\bf s} | {\bf y}) = \frac{P({\bf s}, {\bf y})}{P({\bf y})}$, respectively.
While not compromising the generality of treatment,
this allows disentangling aspects of the problem that would otherwise be 
impossible to separate and interpret geometrically.

Additionally, in our treatment we leverage the symmetry in the definition of mutual information 
to gain deeper insights on optimality from distinct and mutually enriching perspectives. 
This complements the original approach by Chechik \textit{et al.} \cite{Chechik2005} 
where the encoded and relevant informations $I({\bf z}; {\bf s})$ and $I({\bf z}; {\bf y})$ 
were computed as reductions in the entropy of the encoding variable ${\bf z}$ when 
the signal ${\bf s}$ and the relevance variable ${\bf y}$, respectively, were given, 
i.e. $I({\bf z}; {\bf s}) = H({\bf z}) - H({\bf z} | {\bf s})$ and analogously
 for $I({\bf z}; {\bf y})$.
Here, we will often consider the alternative ``decoding'' perspective where these
information metrics are interpreted as reductions in the entropy of the respective variable
(${\bf s}$ or ${\bf y}$) when the encoding value ${\bf z}$ is provided, 
namely $I({\bf z}; {\bf s}) = H({\bf s}) - H({\bf s} | {\bf z})$ and similarly for $I({\bf z}; {\bf y})$.

In our work, we first study the Gaussian information bottleneck method 
for two-dimensional signal and relevance variables.
The complexity of this setting is sufficient for illustrating the key features of optimal 
encoding related to its directionality and dimensionality.
Later we show how the derived results generalize to higher-dimensional cases.

The remaining of the paper is organized as follows.
In Sec. II, we introduce the variables, the linear encoding rule, and the problem of obtaining 
the optimal signal encoding strategy under constrained information cost.
Then in Sec. III, we analyze the one-dimensional encoding scenario and provide interpretations of 
optimality from two distinct perspectives (encoding and decoding).
We extend our analysis to the case of two-dimensional signal encoding in Sec. IV where we offer 
novel intuition on how the information from multiple components gets combined and 
at what point having two encoding components instead of one becomes the preferred strategy.
We end our general discussion of Gaussian information bottleneck in Sec. V by showing how 
our interpretation applies to a three-dimensional setting and, by extension, 
to an arbitrary multi-dimensional case.

To illustrate our new outlook on a concrete example, in Sec. VI we apply it to the problem of signal prediction.
Specifically, we consider signals generated via a stochastically driven mass-spring model -- a canonical setup used 
in prior studies on prediction \cite{Creutzig2009, Becker2015, Palmer2015, Sachdeva2021, Tjalma2023}.
There, the signal is represented by the fluctuating position of the mass.
The dynamics is such that the current position ($x_0$) and its time derivative ($v_0$) fully specify 
the statistics of the future signal.
The goal of the information bottleneck method is to encode these two signal features, 
namely ${\bf s} = [x_0, v_0]^{\rm T}$, in such a way that the representation ${\bf z}$ is maximally 
informative about the future pair of features, ${\bf y} = [x_\tau, v_\tau]^{\rm T}$.
We obtain insightful analytical results on the optimal encoding strategy 
and illustrate how it varies with the forecast interval 
and the dynamical regime of the signal.

We end by summarizing our results and discussing their broader implications in Sec. VII.

\section{Problem formulation}

\begin{figure}[!ht]
\centering
\includegraphics{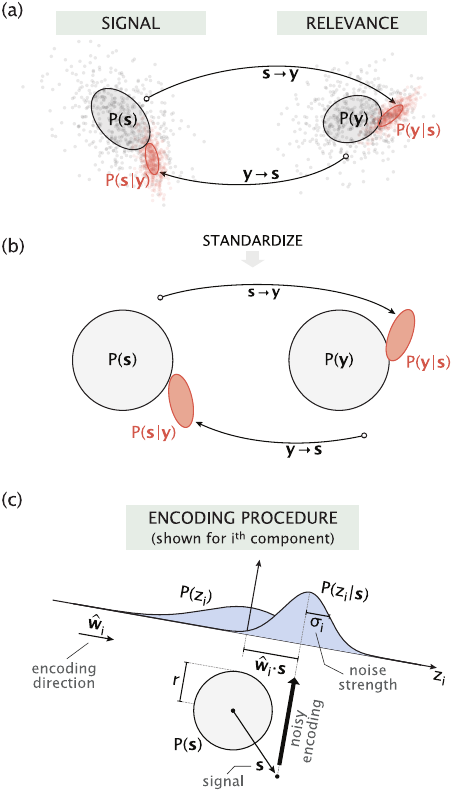}
\caption{
Visualization of the key probability distributions and procedures in the Gaussian information bottleneck problem.
(a) Distributions of the signal and relevance variables, and of the stochastic
mappings between them (${\bf s} \rightarrow {\bf y}$ and ${\bf y} \rightarrow {\bf s}$).
The dots are samples from these distributions, 
while the ellipses represent their $1\sigma$-level contours.
(b) Standardization procedure for the distributions $P({\bf s})$ and $P({\bf y})$ 
that turns their constant-density elliptical contours into circles, thus
concentrating key statistical features of the problem in the conditional
probabilities $P({\bf s} | {\bf y})$ and $P({\bf y} | {\bf s})$.
(c) Illustration of the encoding procedure for a single component 
$z_i \sim \mathcal{N}(\hat{\bf w}_i \cdot {\bf s}, \, \sigma^2_i)$.
It can be viewed as a projection of the signal ${\bf s}$ onto the encoding direction $\hat{\bf w}_i$,
followed by the addition of the encoding noise $\xi_i$.
Centered at the origin, the marginal distribution $P(z_i)$ has a larger variance
$\sigma^2_{z_i} = r^2 + \sigma^2_i$ that is independent of the encoding direction $\hat{\bf w}_i$.
The cross-correlation matrix used for specifying the statistics of standardized variables in panel (b) is 
$\boldsymbol{\Sigma}_{{\bf s} {\bf y}} = 
\begin{pmatrix}
0.83 & -0.50 \\
0.48 & \hspace{7.2pt} 0.70
\end{pmatrix}$.
}
\label{fig:problem_intro}
\end{figure}

Consider a two-component signal ${\bf s} = [s_1, s_2]^{\rm T}$ that has a Gaussian distribution centered
at the origin ($\langle {\bf s} \rangle = {\bf 0}$).
Each signal value maps stochastically to a different Gaussian variable ${\bf y} = [y_1, y_2]^{\rm T}$ 
called the relevance variable.
Fig.~\ref{fig:problem_intro}a shows example distributions of these two variables and of the stochastic mappings between them.
There, the different multivariate Gaussian distributions are depicted visually via their 
$1\sigma$-level elliptical contours (set of all points where the probability equals $e^{-1/2}$
times its maximum value). 
We will be using this geometric way of depicting
distributions throughout the paper.

Since the storage of the full signal may be impossible or impractical, we are interested in its partial 
representation, specifically one in the form of a noisy linear transformation:
\begin{align}
\label{eqn:encoding_general}
{\bf z} = {\bf W}^{\rm T} {\bf s} + \boldsymbol{\xi}.
\end{align}
Here, ${\bf W}$ is a linear transformation matrix, while $\boldsymbol{\xi}$ is 
a Gaussian encoding noise vector with zero mean ($\langle \boldsymbol{\xi}\rangle = {\bf 0}$)
and covariance matrix $\boldsymbol{\Sigma}_{\boldsymbol{\xi}}$.
We note that Eq.~\ref{eqn:encoding_general} is, in fact, the optimal form of signal encoding
when
the signal and relevance variables have jointly Gaussian statistics \cite{Chechik2005}.

As already mentioned in the Introduction,
the cost of signal encoding is measured via 
the mutual information $I({\bf z}; {\bf s})$. Noisier and hence, less accurate representations will 
have a lower corresponding $I({\bf z}; {\bf s})$. Conversely, a perfect signal representation can only be
achieved in the limit of infinite encoded information.
Multiple
encoding strategies defined via
the pair
$({\bf W}, \boldsymbol{\Sigma}_{\boldsymbol{\xi}})$ can have the same information cost
$I({\bf z}; {\bf s})$. 
The goal of Gaussian information bottleneck is to identify the subset of these encoding
strategies that remain maximally informative about the relevance variable, i.e. they maximize 
the relevant information $I({\bf z}; {\bf y})$ for a given encoding capacity $I({\bf z}; {\bf s})$ \cite{Chechik2005}.

To help interpret these and other information metrics intuitively, we initially standardize 
the distributions of signal and relevance variables to make their covariance matrices diagonal,
containing equal entries. 
This procedure does not affect the generality of treatment 
since the information metrics are invariant under standardization,
and the encoding rules for 
the original variables can be worked out easily through a back transformation (Appendix~A1).
Standardization results in the forms $\boldsymbol{\Sigma}_{\bf s} = r^2 \, {\bf I}_{\bf s}$ and 
$\boldsymbol{\Sigma}_{\bf y} = r^2 \, {\bf I}_{\bf y}$, where $r^2$ is the single-component variance that sets 
the scale of these variables (same scale chosen for convenience), while 
${\bf I}_{\bf s}$ and ${\bf I}_{\bf y}$ are identity matrices with their dimensions given by ${\rm dim}({\bf s})$ and
${\rm dim}({\bf y})$, respectively.
\mbox{After} the standardization procedure, 
the stochastic mappings ${\bf s} \rightarrow {\bf y}$ and ${\bf y} \rightarrow {\bf s}$
specified via the conditional probability distributions $P({\bf y} | {\bf s})$ and $P({\bf s} | {\bf y})$, respectively,
fully capture the statistical structure of the problem (Fig.~\ref{fig:problem_intro}b).
The standardization procedure also makes 
the ellipses corresponding to these two distributions geometrically identical,
although their orientations do not match in general (Appendix~A2).
Notably, because the joint statistics of ${\bf s}$ and ${\bf y}$ variables is Gaussian, 
the shapes and orientations of $P({\bf s} | {\bf y})$ and $P({\bf y} | {\bf s})$ ellipses
are independent of what specific ${\bf y}$ and ${\bf s}$ values are used in conditioning.

Of particular interest is the mutual information $I({\bf s}; {\bf y})$ between the signal and relevance variables.
It sets an upper bound on the relevant information $I({\bf z}; {\bf y})$ since the encoding ${\bf z}$ is only 
a partial representation of the signal. This mutual information can be written in two alternative ways as
\begin{subequations}
\label{eqn:Isy_alt_ways}
\begin{align}
I({\bf s}; {\bf y}) &= H({\bf s}) - H({\bf s} | {\bf y}) \\
&\qquad \text{or} \nonumber\\
I({\bf s}; {\bf y}) &= H({\bf y}) - H({\bf y} | {\bf s}).
\end{align}
\end{subequations}
Since all variables of the problem are Gaussian, and the entropy of a $d$-dimensional Gaussian random variable
is of the form $H = \frac{1}{2} \log \big( (2\pi e)^d | \boldsymbol{\Sigma}| \big)$, we can write
\begin{align}
I({\bf s}; {\bf y}) = \frac{1}{2} \log \bigg( 
\frac{
\left| \boldsymbol{\Sigma}_{\bf s} \right|
}{
\left| \boldsymbol{\Sigma}_{ {\bf s} | {\bf y}} \right|
}
\bigg) = \frac{1}{2} \log \bigg( 
\frac{
\left| \boldsymbol{\Sigma}_{\bf y} \right|
}{
\left| \boldsymbol{\Sigma}_{ {\bf y} | {\bf s}} \right|
}
\bigg).
\end{align}
Now, it is known that the constant-probability elliptical contour of a multivariate Gaussian distribution 
has an area that scales with the determinant of the covariance matrix as $\propto |\boldsymbol{\Sigma}|^{1/2}$.
This area sets the entropy of the corresponding distribution.
The mutual information $I({\bf s}; {\bf y})$, viewed as a difference of entropies (Eq.~\ref{eqn:Isy_alt_ways}b),
can thus be interpreted as a measure of how much 
the localization space of the relevance variable shrinks upon knowing the signal.
For instance, $I({\bf s}; {\bf y}) = 3\, {\rm bits}$ would mean that the area of the ellipse corresponding to $P({\bf y} | {\bf s})$
is eight times smaller than that of the circle corresponding to the standardized marginal distribution $P({\bf y})$. 
Due to the symmetric definition of mutual information, an analogous
interpretation can be made with the variables ${\bf s}$ and ${\bf y}$ interchanged.
In the rest of our work, we will often use this geometric picture for interpreting information measures.

Next, to help illustrate the encoding procedure more clearly, we consider a particular form for the pair of 
encoding parameters (${\bf W}, \boldsymbol{\Sigma}_{\boldsymbol{\xi}}$). 
Specifically, we represent the linear transformation matrix as a collection of unit encoding vectors, i.e.
${\bf W} = [\hat{\bf w}_1, \hat{\bf w}_2, \dots]$, 
and the Gaussian noise vector $\boldsymbol{\xi}$ as one with independent components, implying a diagonal 
form for its covariance matrix, namely $\boldsymbol{\Sigma}_{\boldsymbol{\xi}}^{ij} = \sigma^2_i \delta_{ij}$.
We justify this set of considerations in Appendix~A3, 
showing why they do not reduce the generality of the problem treatment.
We note that in their original work on the Gaussian information bottleneck, 
Chechik \textit{et al.} \cite{Chechik2005} also diagonalized the noise covariance matrix 
for mathematical convenience.
The main difference, however, is that they set $\boldsymbol{\Sigma}_{\boldsymbol{\xi}}$ equal to an identity matrix
and considered tunable amplitudes for the encoding vectors,
whereas in our approach, we normalize the encoding vectors and treat the noise strengths $\{ \sigma^2_i \,\}$
as distinct tunable parameters. 
We find that the latter approach, in which the mean encoding variable 
$\langle {\bf z} | {\bf s} \rangle = {\bf W}^{\rm T} {\bf s}$ stays the same when tuning the noise strengths, 
yields a more informative geometric picture of the encoding procedure.

Given the above considerations for the pair $({\bf W}, \boldsymbol{\Sigma}_{\boldsymbol{\xi}})$,
individual components of the encoding variable ${\bf z}$ can be written as
\begin{align}
z_i =  \hat{\bf w}_i \cdot {\bf s} + \xi_i.
\end{align}
Formation of the $i^{\rm th}$ 
encoding component can thus be interpreted as the projection of the signal vector ${\bf s}$
onto the encoding direction $\hat{\bf w}_i$, followed by the addition of encoding noise $\xi_i$ (Fig.~\ref{fig:problem_intro}c). 
The conditional distribution of the encoding component $z_i$
is therefore Gaussian with mean $\hat{\bf w}_i \cdot {\bf s}$ and variance $\sigma^2_i$, i.e.
$P(z_i | {\bf s}) \sim \mathcal{N}(\hat{\bf w}_i \cdot {\bf s}, \sigma^2_i)$.
Importantly, because we have standardized the signal distribution
($\boldsymbol{\Sigma}_{\bf s} = r^2 \,{\bf I}_{\bf s}$), the variance of $z_i$ is independent of the corresponding
encoding direction $\hat{\bf w}_i$ and is conveniently given by $\sigma^2_{z_i} = r^2 + \sigma^2_i$.
While this implies that information $I(z_i; {\bf s})$ encoded in each component $z_i$ about the signal is also independent 
of the direction $\hat{\bf w}_i$, information $I(z_i; {\bf y})$ retained in $z_i$ about the relevance variable ${\bf y}$ 
will, in general, depend on $\hat{\bf w}_i$ and dictate its optimal choice.

The problem of finding the optimal $({\bf W}, \boldsymbol{\Sigma}_{\boldsymbol{\xi}})$ pair is 
therefore reduced 
to obtaining the optimal sets of encoding directions $\{ \hat{\bf w}_i \}$ 
and corresponding encoding noise strengths $\{ \sigma^2_i \}$
that maximize $I({\bf z}; {\bf y})$ for a given $I({\bf z}; {\bf s})$.
To build intuition on how the optimal strategies emerge, we will first thoroughly study the case of one-dimensional encoding 
and afterwards consider the more general scenarios.

\section{One-dimensional encoding}

\begin{figure*}
\centering
\includegraphics{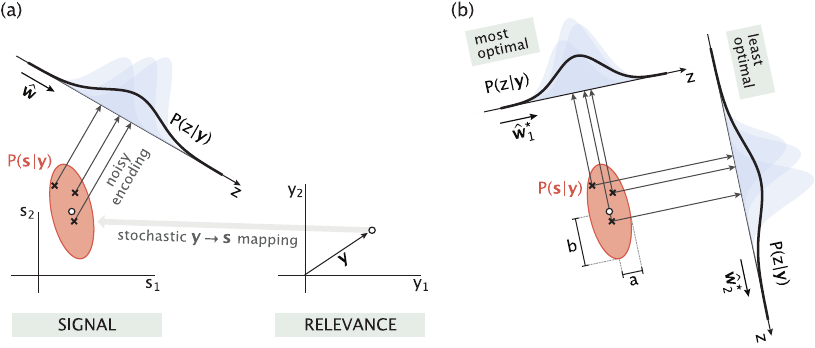}
\caption{
Graphical perspective on optimality in the one-dimensional encoding setting.
(a) Formation of the conditional distribution $P(z | {\bf y})$ presented as the projection of 
${\bf y} \rightarrow {\bf s}$ mapping statistics $P({\bf s} | {\bf y})$ onto the encoding direction $\hat{\bf w}$,
followed by the addition of noise $\xi \sim \mathcal{N} (0, \sigma^2)$. 
The three example Gaussian distributions shown in blue represent the noisy encoding of different 
signals ${\bf s}$, with the heights of the Gaussians proportional to the corresponding probabilities 
$P({\bf s} | {\bf y})$.
(b) Most and least optimal encoding strategies leading 
to the narrowest and widest distributions $P(z | {\bf y})$, respectively.
}
\label{fig:1d_perspective1}
\end{figure*}

Suppose the two-component signal ${\bf s}$ is encoded in the scalar representation $z$ via
\begin{align}
\label{eqn:encoding_rule_1d}
z = \hat{\bf w} \cdot {\bf s} + \xi.
\end{align}
The mutual information between $z$ and ${\bf s}$, which serves as a measure of representation cost,
can be computed from the definition $I(z; {\bf s}) = H(z) - H(z | {\bf s})$ as
\begin{align}
\label{eqn:I_enc_1d}
I(z; {\bf s}) = \frac{1}{2} \log  \left( \frac{\sigma^2_z}{\sigma^2_{z | {\bf s}}}  \right)
= \frac{1}{2} \log \left( 1 + \frac{r^2}{\sigma^2}\right).
\end{align}
Here, we substituted the variances of distributions $P(z)$ and $P(z | {\bf s})$ plotted in Fig.~\ref{fig:problem_intro}c.
As can be seen, $I(z; {\bf s})$ 
depends on the encoding noise but not on the encoding direction.
This means that fixing $I(z; {\bf s})$ will fix $\sigma^2$, and hence, maximization of the relevant information $I(z; {\bf y})$
under fixed $I(z; {\bf s})$ must be done by optimally choosing the encoding vector $\hat{\bf w}$.

Using the definition $I(z; {\bf y}) = H(z) - H(z | {\bf y})$, we write the relevant information as
\begin{align}
\label{eqn:Irel_1d_general}
I(z; {\bf y}) = \frac{1}{2} \log \left( \frac{\sigma^2_z}{\sigma^2_{z | {\bf y}}} \right).
\end{align}
We already know that $\sigma^2_z = r^2 + \sigma^2$ which is constant for a given $I(z; {\bf s})$.
To understand what factors contribute to the conditional variance $\sigma^2_{z | {\bf y}}$,
we represent the corresponding distribution $P(z | {\bf y})$ as
\begin{align}
P(z | {\bf y}) = \int {\rm d}{\bf s}
\begingroup
      \color{gray}
\underbracket[0.140ex]{\color{black} P(z | {\bf s}) }_{\substack{ {\rm noisy}  \\ {\rm encoding}}}
\endgroup \,
\begingroup
      \color{gray}
\underbracket[0.140ex]{\color{black} P ({\bf s} | {\bf y})}_{\substack{\text{stochastic} \\ {\bf y} \, \rightarrow \, {\bf s} \\ \text{mapping}}}.
\endgroup
\end{align}
The two conceptually distinct probabilities inside the integral were already discussed in the previous section;
specifically, $P(z | {\bf s})$ stands for the noisy encoding procedure for a given signal ${\bf s}$
(Fig.~\ref{fig:problem_intro}c), while 
$P({\bf s} | {\bf y})$ captures the stochastic ${\bf y} \rightarrow {\bf s}$ mapping (Fig.~\ref{fig:problem_intro}b)
that does not depend on $\hat{\bf w}$.
The formation of $P(z | {\bf y})$ can thus be interpreted mechanistically as the projection of 
the distribution $P({\bf s} | {\bf y})$ onto the encoding direction $\hat{\bf w}$, followed by 
the addition of an independent encoding noise (Fig.~\ref{fig:1d_perspective1}a).
This translates into the following expression for the conditional variance:
\begin{align}
\label{eqn:sigma_z_cond_y}
\sigma^2_{z | {\bf y}} = \hat{\bf w}^{\rm T} \boldsymbol{\Sigma}_{{\bf s} | {\bf y}} \hat{\bf w} + \sigma^2.
\end{align}
The first term on the right-hand side depends on the encoding direction $\hat{\bf w}$ and therefore, can be
tuned, while the second term ($\sigma^2$) is fixed for given $I(z; {\bf s})$.

Now, we know from Eq.~\ref{eqn:Irel_1d_general} that the relevant information is the largest 
when $\sigma^2_{z | {\bf y}}$ is minimal 
or, equivalently, the distribution $P(z | {\bf y})$ is the narrowest. 
In view of Fig.~\ref{fig:1d_perspective1}a, 
the condition for minimizing the width of the $P(z | {\bf y})$ distribution becomes straightforward:
the encoding vector $\hat{\bf w}$ must be parallel to the minor axis of the ${\bf y} \rightarrow {\bf s}$ mapping ellipse.
This means that the optimal $\hat{\bf w}$ is an eigenvector of the covariance matrix $\boldsymbol{\Sigma}_{{\bf s} | {\bf y}}$ 
with the smaller corresponding eigenvalue, i.e. 
$\boldsymbol{\Sigma}_{{\bf s} | {\bf y}} \hat{\bf w}_1^* = a^2 \hat{\bf w}_1^*$,
where $a$ is the length of the semi-minor axis (see Appendix~B1 for details).
Similarly, the least favorable encoding direction $\hat{\bf w}_2^*$ is parallel to the major axis of the ellipse, 
satisfying the criterion $\boldsymbol{\Sigma}_{{\bf s} | {\bf y}} \hat{\bf w}_2^* = b^2 \hat{\bf w}_2^*$,
where $b$ is the semi-major axis length ($b \ge a$). 
These two options are shown in Fig.~\ref{fig:1d_perspective1}b.

Using the optimal encoding direction $\hat{\bf w} = \hat{\bf w}_1^*$ in Eq.~\ref{eqn:sigma_z_cond_y} 
and substituting the resulting expression for $\sigma^2_{z | {\bf y}}$ into Eq.~\ref{eqn:Irel_1d_general}, 
we find the relevant information under optimal encoding to be
\begin{align}
\label{eqn:I_rel_opt_scalar}
I^{\rm opt} (z; {\bf y}) = \frac{1}{2} \log \left( \frac{r^2 + \sigma^2}{a^2 + \sigma^2 }\right).
\end{align}
As expected, $I^{\rm opt}(z; {\bf y})$ is close to zero when the encoding noise dominates the signal ($\sigma \gg r$).
Conversely, it is the largest in the limit of noiseless encoding ($\sigma \rightarrow 0$)
and is given by $I^{\rm opt}_{\rm max} (z; {\bf y}) = - \log \tilde a$,
where $\tilde{a} = a/r$ is the normalized semi-minor axis length ($0 < \tilde{a} < 1$).

\begin{figure}[!t]
\centering
\includegraphics{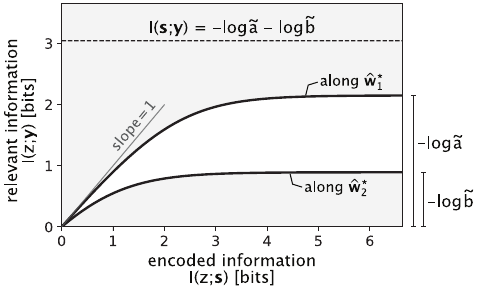}
\caption{
Relevant vs. encoded information in the scalar encoding setting.
The two curves correspond to the most and least optimal strategies
with encoding directions $\hat{\bf w} = \hat{\bf w}_1^*$ and 
$\hat{\bf w} = \hat{\bf w}_2^*$, respectively.
Inaccessible regions of the information plane are colored in gray.
}
\label{fig:1d_information_curves}
\end{figure}

Fig.~\ref{fig:1d_information_curves} shows the plot of the relevant information $I^{\rm opt}(z; {\bf y})$ 
under the optimal strategy as a function of the encoded information $I(z; {\bf s})$. 
The initial slope of the curve, given by $1 - \tilde{a}^2$, is necessarily less than 1 
(see Appendix~B2, as well as Chechik \textit{et al.} \cite{Chechik2005}).
Below the optimal curve, we also plot $I(z; {\bf y})$ vs. $I(z; {\bf s})$ 
for the least favorable encoding strategy with $\hat{\bf w} = \hat{\bf w}_2^*$. 
There, the curve has an initial slope of $1 - \tilde{b}^2$ and saturates at the value $-\log \tilde{b}$.

Since both ${\bf s}$ and ${\bf y}$ are multidimensional, the scalar encoding variable $z$
cannot capture all the information that the full signal ${\bf s}$ contains about relevance variable ${\bf y}$,
even in the noiseless encoding limit.
This information, expressed in terms of 
the parameters $\tilde{a}$ and $\tilde{b}$, 
can be written as $I({\bf s}; {\bf y}) = -\log \tilde{a} - \log \tilde{b}$.
It is shown in Fig.~\ref{fig:1d_information_curves} as an unattainable bound.
As the figure hints, this bound 
can be reached when a second encoding component with maximum information contribution 
$-\log \tilde{b}$ is combined with the first component with contribution $-\log \tilde{a}$.
In Sec.~IV, we will study this scenario in detail.

\subsection*{Alternative ``decoding'' perspective}

\begin{figure*}
\centering
\includegraphics{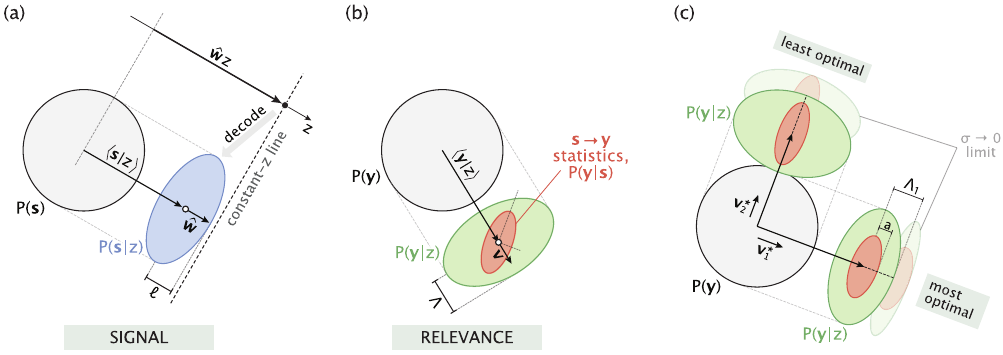}
\caption{Alternative perspective on optimality in the scalar encoding scenario.
(a) Decoding of the signal ${\bf s}$ from the encoding variable $z$. 
The decoding distribution $P({\bf s} | z)$ collapses onto the constant-$z$ line in the $\sigma \rightarrow 0$ limit.
(b) Formation of the distribution $P({\bf y} | z)$ viewed as the stochastic mapping of $P({\bf s} | z)$
onto the relevance plane. 
The label ``${\bf s} \rightarrow {\bf y}$ statistics'' represents the distribution
$P({\bf y} \, |\,  {\bf s} \hspace*{-0.4mm}  = \hspace*{-1.0mm} \langle {\bf s} | z \rangle)$.
(c) The distribution $P({\bf y} | z)$ arising under the most and least optimal strategies (shown as green ellipses), 
overlaid by the ellipses capturing the ${\bf s} \rightarrow {\bf y}$ mapping statistics (shown in red).
Transparent ellipses represent the noiseless encoding limit ($\sigma \rightarrow 0$).
${\bf v}_1^*$ and ${\bf v}_2^*$ are the directions on the ${\bf y}$-plane along 
which information about the relevance variable is preserved under the two respective strategies. 
No information on ${\bf y}$ is preserved in directions perpendicular to the ${\bf v}$-vectors,
which is the reason why the semi-major axes of $P({\bf y} | z)$ ellipses are not compressed 
and have a length equal to $r$ -- the radius of the $P({\bf y})$ circle.
In making the plots, the encoding value $z$ was kept the same.
}
\label{fig:1d_perspective2}
\end{figure*}

Before considering the two-dimensional encoding scenario, 
we present here an alternative perspective on optimality that will complement 
our understanding in the scalar encoding case and later
serve as a key framework for studying the higher-dimensional cases.
We start off by revisiting the encoded information, this time based on the definition
$I(z; {\bf s}) = H({\bf s}) - H({\bf s}| z)$, 
rather than $I(z; {\bf s}) = H(z) - H(z | {\bf s})$ used in the previous subsection. 
It represents the reduction in the uncertainty about the signal 
achieved when knowing the encoding value.
Using the encoding rule in Eq.~\ref{eqn:encoding_rule_1d}, we can derive the statistical properties of the 
``decoding'' distribution $P({\bf s} | z)$ (see Appendix~B3). Specifically, its mean is given by
\begin{align}
\label{eqn:signal_decoding_mean_1d}
\langle {\bf s} | z \rangle = \frac{r^2}{r^2 + \sigma^2} \hat{\bf w} z,
\end{align}
while the conditional covariance matrix is
\begin{align}
\label{eqn:cov_signal_decoding_1d}
\boldsymbol{\Sigma}_{{\bf s} | z} = r^2 \left( {\bf I}_{\bf s} - \frac{r^2}{r^2 + \sigma^2} 
\hat{\bf w} \hat{\bf w}^{\rm T} \right).
\end{align}

As can be seen from Eq.~\ref{eqn:signal_decoding_mean_1d} 
and illustrated in Fig.~\ref{fig:1d_perspective2}a, the mean vector
$\langle {\bf s} | z \rangle$ is parallel to the encoding direction $\hat{\bf w}$,
with its magnitude dependent both on the encoding value $z$
and on the encoding noise strength $\sigma$.
In the noiseless limit ($\sigma \rightarrow 0$), it has the largest magnitude (equal to $\hat{\bf w} z$) and ends
on the constant-$z$ line (the line perpendicular to $\hat{\bf w}$ that passes through ${\bf s} = \hat{\bf w}z$),
while in the infinite noise limit ($\sigma \rightarrow \infty$), it reaches the origin,
retaining no information about the signal.

Although the position of the ellipse corresponding to $P({\bf s} | z)$ depends on
the encoding value $z$, 
the shape of the ellipse does not (a property of Gaussian statistics);
the latter only depends on the signal magnitude
$r$ and the encoding noise level $\sigma$.
The ellipse is compressed along the encoding direction $\hat{\bf w}$, with its semi-minor axis
given by $\ell = r  \sigma/\sqrt{r^2 + \sigma^2}$.
In the noiseless limit ($\sigma \rightarrow 0$), the ellipse gets localized on the constant-$z$ line. 
Although the corresponding encoded information $I(z; {\bf s})$ is infinite in this limit
(due to the $P({\bf s} | {\bf z})$ ellipse having zero area),
recovering the full vector signal ${\bf s}$ remains impossible since scalar encoding only 
informs on a single projection of ${\bf s}$, providing no information in the perpendicular direction.
In the opposite limit ($\sigma \rightarrow \infty$), information about the signal is obscured completely,
and the ellipse turns into a circle centered at the origin.

We next use a similar approach to interpret the relevant information, writing it as 
$I(z; {\bf y}) = H({\bf y}) - H({\bf y} | z)$. 
The marginal distribution $P({\bf y})$, as discussed earlier, is standardized and is
presented graphically as a circle of radius $r$ (Fig.~\ref{fig:1d_perspective2}b).
To better understand the conditional distribution $P({\bf y} | z)$ corresponding to the second entropy
term $H({\bf y} | z)$, we write
\begin{align}
P({\bf y} |z) = \int {\rm d}{\bf s}
\begingroup
      \color{gray}
\underbracket[0.140ex]{\color{black} P({\bf s} |z) }_{\substack{ {\rm noisy}  \\ {\rm decoding}}}
\endgroup \,
\begingroup
      \color{gray}
\underbracket[0.140ex]{\color{black} P ({\bf y} | {\bf s})}_{\substack{\text{stochastic} \\ {\bf s} \, \rightarrow \, {\bf y} \\ \text{mapping}}}.
\endgroup
\end{align}
In view of the above integral expression, the formation of $P({\bf y} | z)$ can be understood as 
the stochastic mapping of the decoded signal distribution $P({\bf s} | z)$ onto the relevance plane.
Derived in Appendix~B3, the mean of this distribution is
\begin{align}
\langle {\bf y} | z \rangle = \frac{r^2}{r^2 + \sigma^2}
{\bf v} z,
\end{align}
and the covariance matrix is given by
\begin{align}
\boldsymbol{\Sigma}_{{\bf y} | z} = r^2 \left( {\bf I}_{\bf y}  - \frac{r^2}{r^2 + \sigma^2} 
{\bf v} {\bf v}^{\rm T} \right).
\end{align}
Here, we have introduced the vector
\begin{align}
\label{eqn:v_definition}
{\bf v} = \tilde{\boldsymbol{\Sigma}}_{{\bf y} {\bf s}} \hat{\bf w},
\end{align}
with $\tilde{\boldsymbol{\Sigma}}_{{\bf y} {\bf s}} = \boldsymbol{\Sigma}_{{\bf y} {\bf s}}/r^2$
and $||{\bf v}|| \le 1$.
It depends both on the encoding direction $\hat{\bf w}$ and on the ${\bf s} \leftrightarrow {\bf y}$ statistics
characterized by $\tilde{\boldsymbol{\Sigma}}_{{\bf y} {\bf s}}$. Functionally,
the vector ${\bf v}$ sets the direction on the relevance plane along which information about the ${\bf y}$-variable
is preserved in the encoding value $z$.
This property is reflected in the shape of the ellipse corresponding to the distribution $P({\bf y} | z)$,
which is compressed along the direction of ${\bf v}$ (Fig.~\ref{fig:1d_perspective2}b).
When decreasing the encoding noise $\sigma$, the $P({\bf y} | z)$ ellipse will get more compressed along ${\bf v}$.
However, in the direction perpendicular to ${\bf v}$, the size of the $P({\bf y} | z)$ ellipse 
will remain unchanged and equal to that of the $P({\bf y})$ circle
because $z$ stores no information about ${\bf y}$ in that direction.

Now, independent of the encoding strategy,
there \mbox{exists} a direction on the ${\bf y}$-plane along which 
the uncertainty about the relevance variable gets reduced the most upon knowing the signal value ${\bf s}$.
This direction is set by the eigenvector of the covariance matrix $\boldsymbol{\Sigma}_{{\bf y} | {\bf s}}$
with the smaller corresponding eigenvalue, which is along the minor axis of the ellipse representing 
the ${\bf s} \rightarrow {\bf y}$ mapping statistics (Fig.~\ref{fig:1d_perspective2}b).

For an arbitrary choice of $\hat{\bf w}$,
the direction along which relevant information is preserved (${\bf v}$) does not match the direction of 
least uncertainty in the ${\bf s} \rightarrow {\bf y}$ mapping (Fig.~\ref{fig:1d_perspective2}b).
Under optimal encoding (with $\hat{\bf w} = \hat{\bf w}_1^*$), however, 
these directions have to match (see Appendix~B4 for the proof).
This optimality feature is illustrated in Fig.~\ref{fig:1d_perspective2}c where the ellipses corresponding to 
the decoded relevance distribution $P({\bf y} | z)$ and ${\bf s} \rightarrow {\bf y}$ statistics 
have their minor axes aligned. 
The semi-minor axis of the optimal $P({\bf y} | z)$ ellipse has a length equal to
$\Lambda_1 = r \sqrt{(a^2 + \sigma^2)/(r^2 + \sigma^2)}$, which approaches $r$ as the encoding noise $\sigma$
becomes very large
and converges to $a$ in the limit of noiseless encoding.
The result derived earlier for the maximum relevant information under fixed encoded information 
(Eq.~\ref{eqn:I_rel_opt_scalar}) 
follows directly from this semi-minor axis expression via $I^{\rm opt}(z; {\bf y}) = \log (r/\Lambda_1)$, where 
$r/\Lambda_1$ is the area ratio of ellipses corresponding to the marginal distribution $P({\bf y})$ 
and the optimal $P({\bf y} | z)$, respectively.

In an analogous way, one can show that the least optimal strategy preserves information about the relevance 
variable in the most uncertain direction of ${\bf s} \rightarrow {\bf y}$ mapping, which is along the major axis 
of the mapping ellipse (Fig.~\ref{fig:1d_perspective2}c).
Notably, just as the most and least optimal encoding directions, namely $\hat{\bf w}_1^*$ and $\hat{\bf w}_2^*$
(see Fig.~\ref{fig:1d_perspective1}b), are perpendicular to one another, so are the corresponding vectors
${\bf v}_1^*$ and ${\bf v}_2^*$ (Fig.~\ref{fig:1d_perspective2}c).
This set of properties will be very useful for extending the perspective developed here 
to the two-dimensional encoding scenario, which we will study next.

\section{Two-dimensional encoding}

Earlier in Fig.~\ref{fig:1d_information_curves}, we saw that a scalar representation $z$ is unable to 
fully encode the mutual information $I({\bf s}; {\bf y})$ between two-dimensional signal and 
relevance variables. 
In this section, we will consider the option of encoding the signal into a two-dimensional vector,
will understand when it becomes preferred over the scalar encoding option 
and how it allows reaching the bound on relevant information given by $I({\bf s}; {\bf y})$.

Vector encoding means that we now have two encoding components defined as
$z_1 = \hat{\bf w}_1 \cdot {\bf s} + \xi_1$ and $z_2 = \hat{\bf w}_2 \cdot {\bf s} + \xi_2$.
These components are specified via encoding directions $\{ \hat{\bf w}_1, \hat{\bf w}_2 \}$
and encoding noise strengths $\{ \sigma^2_1, \sigma^2_2 \}$.
The optimal encoding problem is about finding optimal choices of these four parameters 
that maximize the relevant information $I({\bf z}; {\bf y})$ under fixed encoded information $I({\bf z}; {\bf s})$.

We start off by considering a natural choice for the optimal directions $\hat{\bf w}_1$ and $\hat{\bf w}_2$ 
consisting of the optimal $1$D encoding direction $\hat{\bf w}^*_1$ and the direction perpendicular 
to it, $\hat{\bf w}^*_2$, shown in Fig.~\ref{fig:1d_perspective1}b.
Under this assignment, 
$z_1$ and $z_2$ encode linearly independent projections of the signal
and, furthermore, retain information about the relevance variable along perpendicular directions
(${\bf v}_1^*$ and ${\bf v}_2^*$, see Fig. \ref{fig:1d_perspective2}c).
Notably, the directions $\{ \hat{\bf w}^*_i, \hat{\bf v}^*_i \}$ correspond to the basis vectors used in canonical 
correlation analysis -- a statistical method which identifies relationships between multivariate 
sets of variables \cite{Hotelling1936}. This parallel was drawn also in the original work by 
Chechik \textit{et al.} \cite{Chechik2005}.
We discuss the degenerate space of optimal solutions briefly at the end of the next section
and in greater detail in Appendix~D2, 
where we show how correlated $z_i$-components which encode along non-perpendicular directions 
can also yield equally optimal solutions.
In the rest of this section, we will focus on the problem of optimally assigning the encoding noise 
strengths $\{ \sigma_1^2, \sigma_2^2 \}$ 
corresponding to the principal directions $\hat{\bf w}_1 = \hat{\bf w}_1^*$ and $\hat{\bf w}_2 = \hat{\bf w}_2^*$, 
which is at the heart of the Gaussian information bottleneck method.

Following the approach developed in the previous section, we examine the formation 
of the decoding distribution $P({\bf s} | {\bf z})$. The mean of this distribution, given by
\begin{align}
\label{eqn:s_mean_given_vector_z}
\langle {\bf s} | {\bf z} \rangle = \sum_{i\in \{1, 2\}} \frac{r^2}{r^2 + \sigma^2_i} \, \hat{\bf w}_i  
\hspace{0.08em} z_i,
\end{align}
represents the sum of mean signals decoded from the two separate ${\bf z}$-components, i.e.
$\langle {\bf s} | {\bf z} \rangle = \sum_i \langle {\bf s} | z_i \rangle$. 
Note that the scalar encoding case is recovered in the $\sigma_2 \rightarrow \infty$ limit,
since $\langle {\bf s} | z_2 \rangle = {\bf 0}$ in that limit. 
The covariance matrix of $P({\bf s} | {\bf z})$, which is of the form
\begin{align}
\label{eqn:cov_s_given_vector_z}
\boldsymbol{\Sigma}_{{\bf s} | {\bf z}} = r^2 \bigg( 
{\bf I}_{\bf s} - 
\sum_{i\in \{1, 2\}} \frac{r^2}{r^2 + \sigma^2_i} \, \hat{\bf w}_i^{\,}  \hat{\bf w}_i^{\rm T}
\bigg),
\end{align}
also reduces to the scalar encoding result (Eq.~\ref{eqn:cov_signal_decoding_1d}) 
in the limit $\sigma_2 \rightarrow \infty$ (see Appendix~C1 for details).

\begin{figure}
\centering
\includegraphics{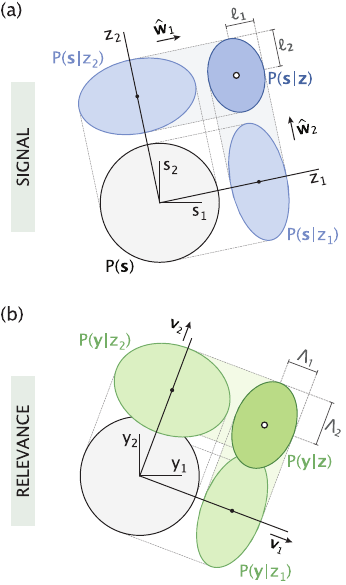}
\caption{
Formation of the distributions $P({\bf s} | {\bf z})$ and $P({\bf y} | {\bf z})$ (panels (a) and (b), respectively)
in the two-dimensional encoding scenario with optimal choices of perpendicular encoding directions, namely
$\hat{\bf w}_1 = \hat{\bf w}^*_1$ and $\hat{\bf w}_2 = \hat{\bf w}_2^*$. The corresponding ${\bf v}$-vectors
(with ${\bf v}_i = \tilde{\boldsymbol{\Sigma}}_{{\bf y} {\bf s}} \hat{\bf w}_i$) are also perpendicular to each other.
}
\label{fig:2d_components}
\end{figure}

Fig.~\ref{fig:2d_components}a illustrates the formation of the decoding distribution $P({\bf s} | {\bf z})$.
As can be seen, the ellipse corresponding to $P({\bf s} | {\bf z})$ is compressed
along both of its axes, which have half-lengths 
$\ell_i = r \sigma_i/\sqrt{r^2 + \sigma^2_i}$.
These lengths are the semi-minor axis lengths of the scalar encoding ellipses corresponding to
the distributions $P({\bf s} | z_1)$ and $P({\bf s} | z_2)$.
Note that in the limit $\sigma_2 \rightarrow \infty$ we have $\ell_2 \rightarrow r$, which results
in the $P({\bf s} | z_2)$ ellipse becoming a circle centered at the origin 
and the $P({\bf s} | {\bf z})$ ellipse becoming identical to the $P({\bf s} | z_1)$ ellipse.

Formation of the distribution $P({\bf y} | {\bf z})$ on the relevance plane 
takes place in an analogous way (Fig.~\ref{fig:2d_components}b). 
The ellipse corresponding to $P({\bf y} | {\bf z})$ has its axes oriented along the vectors
${\bf v}_1$ and ${\bf v}_2$ (with ${\bf v}_i = \tilde{\boldsymbol{\Sigma}}_{{\bf y} {\bf s}} \hat{\bf w}_i$).
Akin to Eq.~\ref{eqn:s_mean_given_vector_z} and Eq.~\ref{eqn:cov_s_given_vector_z}, 
the mean of $P({\bf y} | {\bf z})$ is given by
\begin{align}
\langle {\bf y} | {\bf z} \rangle = \sum_{i\in \{1, 2\}} \frac{r^2}{r^2 + \sigma^2_i} \, {\bf v}_i  
\hspace{0.08em} z_i
\end{align}
and its covariance matrix is
\begin{align}
\boldsymbol{\Sigma}_{{\bf y} | {\bf z}} = r^2 \bigg( 
{\bf I}_{\bf y} - 
\sum_{i\in \{1, 2\}} \frac{r^2}{r^2 + \sigma^2_i} \, {\bf v}_i^{\,}  {\bf v}_i^{\rm T}
\bigg).
\end{align}
The eigenvalues of $\boldsymbol{\Sigma}_{{\bf y} | {\bf z}}$ corresponding to 
the eigenvectors ${\bf v}_1$ and ${\bf v}_2$ are $\Lambda_1^2$ and $\Lambda_2^2$, respectively,
where $\Lambda_1$ and $\Lambda_2$ represent the half-lengths of the $P({\bf y} | {\bf z})$ ellipse.
They satisfy the relation 
$\Lambda_i^2 = r^2 \left(1 - \frac{r^2}{r^2 + \sigma_i^2} |{\bf v}_i|^2 \right)$.
Substituting the identities $|{\bf v}_1|^2 = 1 - \tilde{a}^2$ and $|{\bf v}_2|^2 = 1 - \tilde{b}^2$ 
(see Appendix~C1), one obtains
$\Lambda_1 = r \sqrt{(a^2 + \sigma_1^2)/ (r^2 + \sigma_1^2 )}$ and 
$\Lambda_2 = r \sqrt{(b^2 + \sigma_2^2)/ (r^2 + \sigma_2^2 )}$.
Notably, in the limit of noiseless encoding ($\sigma_1, \sigma_2 \rightarrow 0$),
$\Lambda_1$ and $\Lambda_2$ converge to the semi-axis lengths
$a$ and $b$ of the $P({\bf y} | {\bf s})$ ellipse, respectively.

\subsection*{Optimal noise allocation and dimensionality of representation}

In view of Fig.~\ref{fig:2d_components}a, fixing the encoded information $I({\bf s}; {\bf z})$ 
in the information bottleneck problem means fixing the area of the $P({\bf s} | {\bf z})$ ellipse, 
which is equivalent to fixing $\ell_1 \times \ell_2$ -- the product of the ellipse's major and minor semi-axis 
lengths.
Crucially, there are many ways of choosing $\ell_1$ and $\ell_2$ (via $\sigma_1$ and $\sigma_2$ assignments) 
that keep the product $\ell_1 \times \ell_2$ and hence $I({\bf s}; {\bf z})$ the same.

\begin{figure}
\centering
\includegraphics{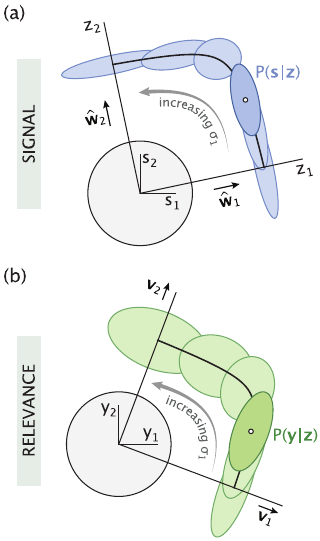}
\caption{
Distributions $P({\bf s} | {\bf z})$ and $P({\bf y} | {\bf z})$ (panels (a) and (b), respectively) 
for different $\{\sigma_1, \sigma_2\}$ assignment options with fixed encoded information $I({\bf z}; {\bf s}) = 2.5$ bits
(increase in $\sigma_1$ is accompanied by a simultaneous decrease in $\sigma_2$
to keep the area of $P({\bf s} | {\bf z})$ ellipses unchanged).
Ellipses in each panel corresponding to the optimal noise assignment 
(maximizing the relevant information $I({\bf z}; {\bf y})$ by minimizing the $P({\bf y} | {\bf z})$ ellipse area)
are drawn in darker colors.
Curved lines in both panels represent the mean vectors 
$\langle {\bf s} | {\bf z} \rangle$ and $\langle {\bf y} | {\bf z} \rangle$ drawn when the noise levels are continuously tuned.
During noise tuning, the encoding values are kept fixed at
$z_1 = \hat{\bf w}_1 \cdot {\bf s}$ and $z_2 = \hat{\bf w}_2 \cdot {\bf s}$
for a specified signal ${\bf s}$.
}
\label{fig:croissant}
\end{figure}

A range of possible options is shown in Fig.~\ref{fig:croissant}a for an example case with
$I({\bf z}; {\bf s}) = 2.5$ bits of encoded information. 
There, the noise assignments vary from finite $\sigma_1$ and infinite $\sigma_2$ 
(all information contained in the $z_1$ component) to infinite $\sigma_1$ and finite $\sigma_2$
(all information contained in the $z_2$ component), with different intermediate cases in between
(both $\sigma_1$ and $\sigma_2$ finite).
Now, each $\{ \sigma_1, \sigma_2 \}$ assignment option also has its corresponding 
$P({\bf y} | {\bf z})$ distribution on the relevance plane (Fig.~\ref{fig:croissant}b).
In contrast to $P({\bf s} | {\bf z})$ ellipses which all have the same area (due to encoded
information $I({\bf z}; {\bf s})$ being fixed), the $P({\bf y} | {\bf z})$ ellipses generally have different areas,
which reflects the fact that the relevant information $I({\bf y}; {\bf z})$ 
depends on the noise assignment strategy.
The optimal strategy is the one that maximizes $I({\bf z}; {\bf y})$ by minimizing the area
of the corresponding $P({\bf y} | {\bf z})$ ellipse
for a given amount of $I({\bf s}; {\bf z})$.

In Fig.~\ref{fig:croissant}, distributions $P({\bf s} | {\bf z})$ and $P({\bf y} | {\bf z})$ corresponding to 
the optimal noise assignment are shown highlighted. 
As can be seen, the optimal $P({\bf s} | {\bf z})$ ellipse is more compressed along the $\hat{\bf w}_1$ direction
compared to the $\hat{\bf w}_2$ direction, and similarly, the corresponding $P({\bf y} | {\bf z})$ ellipse is 
more compressed along ${\bf v}_1$ compared to ${\bf v}_2$.
This reflects the intuitive expectation that the optimal strategy would prioritize the more 
informative $\hat{\bf w}_1$ direction over the less informative $\hat{\bf w}_2$ direction through
a biased allocation of encoding noises, i.e. $\sigma_1 \le \sigma_2$ (hence, $\ell_1 \le \ell_2$).

\begin{figure*}
\centering
\includegraphics{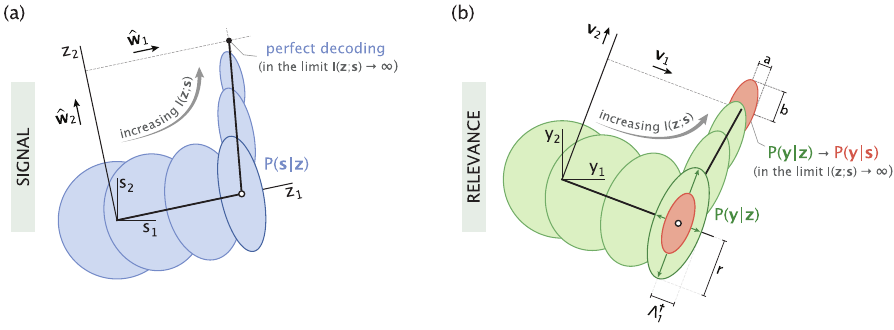}
\caption{
Geometric illustration of the increasing representation complexity under the optimal strategy.
Ellipses corresponding to distributions $P({\bf s} | {\bf z})$ and $P({\bf y} | {\bf z})$ 
under optimal encoding (panels (a) and (b), respectively) are shown for different levels of the encoded
information $I({\bf z}; {\bf s})$, ranging from very low to very high.
Each pair of corresponding ellipses in panels (a) and (b) is the optimal pair among the family of options 
for given $I({\bf z}; {\bf s})$ shown in Fig.~\ref{fig:croissant}, maximizing $I({\bf z}; {\bf y})$ for that given $I({\bf z}; {\bf s})$.
The solid line segments represent the mean vectors $\langle {\bf s} | {\bf z} \rangle$ and $\langle {\bf y} | {\bf z} \rangle$.
The white dots mark the transition points from a scalar to two-dimensional optimal representation.
When tuning $I({\bf z}; {\bf s})$, the encoding values are kept unchanged at $z_1 = \hat{\bf w}_1 \cdot {\bf s}$
and $z_2 = \hat{\bf w}_2 \cdot {\bf s}$, where ${\bf s}$ is the true signal getting encoded.
}
\label{fig:2d_optimality}
\end{figure*}

By what principle is this bias set? 
The answer is illustrated in Figs.~\ref{fig:2d_optimality}a and \ref{fig:2d_optimality}b.
When the encoding capacity represented via $I({\bf z}; {\bf s})$ is low,
the optimal strategy is to encode the signal only 
along the most informative $\hat{\bf w}_1$ direction.
This is reflected in $P({\bf s} | {\bf z})$ ellipses being located near the origin and 
compressed along $\hat{\bf w}_1$ but not $\hat{\bf w}_2$
(Fig.~\ref{fig:2d_optimality}a), and similarly, $P({\bf y} | {\bf z})$ ellipses near the origin 
being compressed along ${\bf v}_1$ but not ${\bf v}_2$ (Fig.~\ref{fig:2d_optimality}b).
Thus, for low values of $I({\bf z}; {\bf s})$, the optimal noise assignment for the second encoding component is
$\sigma_2 \rightarrow \infty$ (implying $I(z_2; {\bf s}) = 0$), while the encoding noise for the first component 
is set by the encoding capacity via
$1/\tilde{\sigma}_1^2 = 2^{2I({\bf z}; {\bf s})} - 1$
(follows from Eq.~\ref{eqn:I_enc_1d} with $\tilde{\sigma}_1 = \sigma_1/r$ and $I(z_1;{\bf s}) = I({\bf z}; {\bf s})$).

In the opposite extreme with very large encoded information $I({\bf z}; {\bf s})$,
the optimal strategy corresponds to low values for both $\sigma_1$ and $\sigma_2$. 
This is intuitive because such an assignment allows for an accurate decoding of the vector signal ${\bf s}$ 
from the two encoding components $z_1$ and $z_2$. 
In the limit $I({\bf z}; {\bf s}) \rightarrow \infty$, encoding noises $\sigma_1$ and $\sigma_2$ approach zero
(hence $\ell_1, \ell_2 \rightarrow 0$), 
the decoding distribution $P({\bf s} | {\bf z})$ becomes localized at the true signal ${\bf s}$ 
(Fig.~\ref{fig:2d_optimality}a), and the ellipse corresponding to $P({\bf y} | {\bf z})$ on the relevance plane 
reduces to the ellipse of $P({\bf y} | {\bf s})$ with semi-axis lengths $a$ and $b$,
as set by the statistics of ${\bf s} \rightarrow {\bf y}$ mapping (Fig.~\ref{fig:2d_optimality}b).
Notably, in this limit 
the relevant information $I({\bf z}; {\bf y})$ reaches its bound set by the correlations between ${\bf s}$ and ${\bf y}$
variables, namely $I_{\rm max}({\bf z}; {\bf y})  = I({\bf s}; {\bf y}) = -\log \tilde{a} - \log{\tilde b}$.

Arguably the most interesting aspect of the Gaussian information bottleneck method 
is the qualitative change in the optimal encoding strategy that 
occurs at a critical encoding capacity $I^\dagger({\bf z}; {\bf s})$.
As illustrated geometrically in Fig.~\ref{fig:2d_optimality}b,
when the encoded information $I({\bf z}; {\bf s})$ reaches its critical value,
the $P({\bf y} | {\bf z})$ decoding ellipse on the relevance plane 
becomes identical in shape to the $P({\bf y} | {\bf s})$ ellipse 
representing the ${\bf s} \rightarrow {\bf y}$ mapping statistics;
at this point, the optimal strategy switches from scalar encoding 
to vector encoding.
Using the $\Lambda_1$ and $\Lambda_2$ notation for
the minor and major semi-axis lengths of the $P({\bf y} | {\bf z})$ ellipse, respectively,
we note that at the transition point $\Lambda_2^\dagger = r$, 
and write the similarity condition of $P({\bf y} | {\bf z})$ and $P({\bf y} | {\bf s})$ ellipses simply as
\begin{align}
\label{eqn:equal_ratio_condition}
\Lambda_1^\dagger \, \big/ \, r = a \, \big/ \, b.
\end{align}
Here, $\Lambda_1^\dagger$ is the minor axis of the $P({\bf y} | {\bf z})$ ellipse at the transition point.
It is achieved when the corresponding encoding noise is 
$\sigma_1^\dagger = \sqrt{\frac{r^2-b^2}{(b/a)^2-1}}$ (see Appendix~C2).

Two special cases are of particular interest.
When $b \rightarrow a$, corresponding to the case where the two encoding directions are equally informative about 
the relevance variable, we obtain $\Lambda_1^\dagger \rightarrow r$ and $\sigma_1^\dagger \rightarrow \infty$.
This means that the optimal strategy is to simultaneously encode along both directions as soon as 
$I({\bf z}; {\bf s})$ becomes nonzero, which is intuitive because 
neither of the equally informative directions is given priority over the other. 
The opposite limit with $b \rightarrow r$ corresponds to the case where the second direction is completely uninformative.
In this case, we find $\Lambda_1^\dagger \rightarrow a$ and $\sigma_1^\dagger \rightarrow 0$ 
(i.e. $I^\dagger({\bf z}; {\bf s}) \rightarrow \infty$), indicating that 
the optimal strategy is to encode only along the most informative direction $\hat{\bf w}_1$
for all values of the encoded information $I({\bf z}; {\bf s})$.

\begin{figure}[!ht]
\centering
\includegraphics{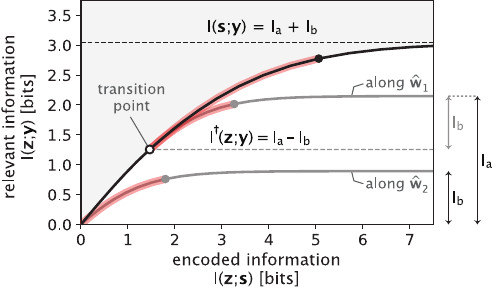}
\caption{
Increase in encoding complexity and behavior past the transition point illustrated 
on the information plane. The second encoding component gets introduced when the amount of 
relevant information that can still be stored in
the first component, $I_a - I^\dagger(z_1; {\bf y})$, becomes equal to the maximum relevant information 
available to the second component, $I_b$ (Eq.~\ref{eqn:Irel_condition_transition_pt}). 
The additional encoding capacity past the transition point 
is allocated equally between the two encoding components. 
This property is captured through the highlighted segments which indicate the correspondence 
between a point $[I({\bf z}; {\bf s}), I({\bf z}; {\bf y})]$
on the information bound past the transition point (black dot) and the locations 
[$I(z_i; {\bf s})$, $I(z_i; {\bf y})$] $(i=1,2)$ on 
the corresponding single-component curves (two gray dots).
The inaccessible region of the information plane beyond the information bound is colored in gray.
}
\label{fig:2d_information_curve}
\end{figure}

In addition to offering a simple geometric interpretation of the transition point, the relation in Eq.~\ref{eqn:equal_ratio_condition} 
also motivates an information-theoretic explanation. 
Recalling that the information contained in the encoding component $z_1$ about the relevance variable 
is $I(z_1; {\bf y}) = \log (r / \Lambda_1)$, we can write the condition of transition
in Eq.~\ref{eqn:equal_ratio_condition} in an alternative form as
\begin{align}
\label{eqn:transition_info_perspective_2d}
I^\dagger(z_1; {\bf y}) = I_a - I_b,
\end{align}
where we introduced $I_a = -\log \tilde{a}$ and $I_b = -\log \tilde{b}$. 
Discussed in the context of Fig.~\ref{fig:1d_information_curves}, 
$I_a = I_{\rm max}(z_1; {\bf y})$ and $I_b = I_{\rm max}(z_2; \bf{y})$ 
are the maximum amounts of relevant information that the encoding components $z_1$ and $z_2$,
respectively, can contain in the limit of noiseless encoding.
Using the $I_{\rm max}$ notation for $I_a$ and $I_b$, 
we rearrange the terms in Eq.~\ref{eqn:transition_info_perspective_2d} and rewrite it as 
\begin{align}
\label{eqn:Irel_condition_transition_pt}
I_{\rm max}(z_1; {\bf y}) - I^\dagger(z_1; {\bf y}) = I_{\rm max}(z_2; {\bf y}).
\end{align}
This illuminating form suggests the following interpretation of the transition point:
the optimal strategy transitions from scalar to vector encoding when 
the amount of relevant information 
that can still be stored in
the most informative encoding component,
namely $I_{\rm max}(z_1; {\bf y}) - I^\dagger(z_1; {\bf y})$,
becomes equal to the maximum relevant information $I_{\rm max}(z_2; {\bf y})$ available to the yet unused, 
less informative encoding component (see Fig.~\ref{fig:2d_information_curve}).

Referring back to the geometric picture in Fig.~\ref{fig:2d_optimality}b,
we now discuss what happens past the transition point as the encoding capacity
$I({\bf z}; {\bf s})$ is increased further.
After being established at the transition point, the property of $P({\bf y} | {\bf z})$
and $P({\bf y} | {\bf s})$ ellipses having an identical aspect ratio persists, i.e.
$\Lambda_1 / \Lambda_2 = a/b$ when $I({\bf z}; {\bf s}) > I^\dagger({\bf z}; {\bf s})$.
A similar property also holds in the ${\bf s}$-plane (Fig.~\ref{fig:2d_optimality}a)
where the aspect ratio of 
the decoding $P({\bf s}| {\bf z})$ ellipse (generally different from that of the $P({\bf y} | {\bf z})$ ellipse) remains unchanged 
as the $P({\bf s} | {\bf z})$ distribution becomes more and more localized with increasing $I({\bf z}; {\bf s})$.
These properties are achieved through a special assignment of encoding noises 
$\sigma_1$ and $\sigma_2$ that obey the relations
$1/\tilde{\sigma}_1^2 = 2^{I({\bf z}; {\bf s}) + I^\dagger({\bf z}; {\bf s})} - 1$
and 
$1/\tilde{\sigma}_2^2 = 2^{I({\bf z}; {\bf s}) - I^\dagger({\bf z}; {\bf s})} - 1$
(for details, see Appendix~C3).

Additional insights about the behavior beyond the transition point can be obtained by looking at information measures.
Specifically, following the approach in Eq.~\ref{eqn:Irel_condition_transition_pt},
the condition $\Lambda_1 / \Lambda_2 = a/b$ can be recast 
in terms of relevant information amounts as
\begin{align}
\label{eqn:equal_Irel_allocation}
I_{\rm max} (z_1; {\bf y}) - I(z_1; {\bf y}) = I_{\rm max}(z_2; {\bf y}) - I(z_2; {\bf y}).
\end{align}
One can also show that the component-wise encoded information amounts are related via
\begin{align}
\label{eqn:equal_Ienc_allocation}
I(z_1; {\bf s}) - I^\dagger(z_1; {\bf s}) = I(z_2; {\bf s}).
\end{align}
These relations capture two important behavioral properties of the optimal strategy 
beyond the transition point. First, the additional encoded 
information, namely $I({\bf z}; {\bf s}) - I^\dagger({\bf z}; {\bf s})$, 
is distributed evenly among the two components (Eq.~\ref{eqn:equal_Ienc_allocation}); 
and second, once equalized at the transition point, 
the amounts of relevant information that can still be stored
in the two encoding components continue to be equal 
(Eq.~\ref{eqn:equal_Irel_allocation}).
These properties are captured on the information plane in Fig.~\ref{fig:2d_information_curve}.
In particular, the segment of the $\hat{\bf w}_1$ curve after the transition point 
is, except for an offset, identical to the entire $\hat{\bf w}_2$ curve for the second component;
furthermore, the two curves are traversed identically as $I({\bf z}; {\bf s})$ 
continues to increase past its critical value $I^\dagger({\bf z}; {\bf s})$,
demonstrating the even allocation of additional encoding capacity 
and the equality of relevant information amounts that can still be stored
in the two components.

\section{Higher-dimensional case}

Having studied the principles of optimality in the case where the signal and relevance 
variables are two-dimensional vectors, we proceed in this section to 
generalizing these principles to the three-dimensional case and, by extension, 
to an arbitrary multi-dimensional scenario.

The principal directions of encoding, like in the 2D case, are specified by 
the eigenvectors of the conditional covariance matrix $\boldsymbol{\Sigma}_{{\bf s} | {\bf y}}$.
When the variables ${\bf s}$ and ${\bf y}$ are three-dimensional, 
the three eigenvectors $\hat{\bf w}_1$, $\hat{\bf w}_2$, and $\hat{\bf w}_3$
are parallel to the three axes of the ${\bf y} \rightarrow {\bf s}$ mapping ellipsoid,
analogous to the 2D case where $\hat{\bf w}_1$ and $\hat{\bf w}_2$ are parallel 
to the axes of the ${\bf y} \rightarrow {\bf s}$ mapping ellipse (Fig.~\ref{fig:1d_perspective1}b).
We denote the eigenvalues corresponding to these three eigenvectors as $a^2$, $b^2$, and $c^2$, 
where $a$, $b$, and $c$ represent the half-lengths of the axes of the $P({\bf s} |  {\bf y})$ ellipsoid 
(with $a \le b \le c$).

\begin{figure*}[!ht]
\centering
\includegraphics[width=0.88\textwidth]{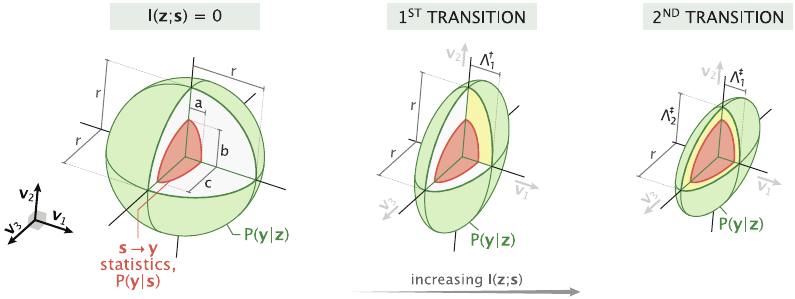}
\caption{
Geometric perspective on the transitions in representation complexity
with growing encoding capacity in a three-dimensional setting.
The red ellipsoid corresponding to the ${\bf s} \rightarrow {\bf y}$ mapping statistics (shown only in the first octant)
has dimensions that are independent of $I({\bf z}; {\bf s})$.
The three axes of this ellipsoid are oriented along the perpendicular vectors ${\bf v}_1$, ${\bf v}_2$, and ${\bf v}_3$.
When no information is encoded, the $P({\bf y} | {\bf z})$ ellipsoid takes the spherical shape of the marginal $P({\bf y})$.
As $I({\bf z}; {\bf s})$ increases from zero, the encoding component $z_1$ 
retains information about the relevance variable ${\bf y}$
along the direction ${\bf v}_1$, and the $P({\bf y} | {\bf z})$ ellipsoid gets compressed along that direction.
At a critical value $I^\dagger({\bf z}; {\bf s})$, the shapes of $P({\bf y} | {\bf z})$ and $P({\bf y} | {\bf s})$ 
ellipsoids projected onto 
the ${\bf v}_1 \text- {\bf v}_2$ plane (highlighted in yellow) become identical. 
At that point, the second encoding component $z_2$ is introduced that retains information about ${\bf y}$
along the direction ${\bf v}_2$.
After shrinking simultaneously along both ${\bf v}_1$ and ${\bf v}_2$, at the second critical value $I^\ddagger({\bf z}; {\bf s})$,
the $P({\bf y} | {\bf z})$ ellipsoid becomes identical in shape to $P({\bf y} | {\bf s})$ 
also on the ${\bf v}_2 \text- {\bf v}_3$ plane.
At that point, the third encoding component $z_3$ gets introduced which informs on ${\bf y}$ along the direction ${\bf v}_3$.
In the limit $I({\bf z}; {\bf s}) \rightarrow \infty$ where the signal ${\bf s}$ can be fully recovered from ${\bf z}$, 
the $P({\bf y} | {\bf z})$ and $P({\bf y} | {\bf s})$ ellipsoids become identical.
}
\label{fig:3d_geometric}
\end{figure*}

Due to the initial standardization performed on ${\bf s}$ and ${\bf y}$ variables, 
the ${\bf s} \rightarrow {\bf y}$ mapping ellipsoid 
in the relevance space has shape and dimensions that are 
identical to that of the ${\bf y} \rightarrow {\bf s}$ mapping ellipsoid
in the signal space (see Fig.~\ref{fig:problem_intro}b for the analogue in two dimensions).
Therefore, the three axes of the $P({\bf y} | {\bf s})$ ellipsoid in relevance space have 
the same half-lengths, namely $a$, $b$, and $c$ (Fig.~\ref{fig:3d_geometric}, left).
These three parameters set the mutual information between signal and relevance variables via
\begin{align}
I({\bf s}; {\bf y}) &= \log \left( \frac{r^3}{a \times b \times c} \right) \nonumber\\
&= -\log \tilde{a} - \log \tilde{b} - \log \tilde{c},
\end{align}
where $\tilde{a} = a/r$ (similarly for $\tilde{b}$ and $\tilde{c}$). From a geometric point of view, 
$I({\bf s}; {\bf y})$ is the logarithm of the volume ratio of 
the marginal $P({\bf y})$ sphere with radius $r$ and the more localized $P({\bf y} | {\bf s})$ ellipsoid 
representing ${\bf s} \rightarrow {\bf y}$ mapping statistics.

Now, the component $z_i = \hat{\bf w}_i \cdot {\bf s} + \xi_i$ that encodes the signal along the direction $\hat{\bf w}_i$ 
preserves information about the ${\bf y}$-variable along the vector ${\bf v}_i = \tilde{\boldsymbol{\Sigma}}_{{\bf y} {\bf s}} \hat{\bf w}_i$
in the relevance space. 
The three vectors $\{ {\bf v}_1, {\bf v}_2, {\bf v}_3 \}$ corresponding 
to the set of principal encoding directions $\{ \hat{\bf w}_1, \hat{\bf w}_2, \hat{\bf w}_3 \}$ described above
are perpendicular to one another and are oriented along the axes of the ${\bf s} \rightarrow {\bf y}$ mapping ellipsoid,
with ${\bf v}_1$ along the shortest axis and ${\bf v}_3$ along the longest (Fig.~\ref{fig:3d_geometric}, left).
How much the encoding values $\{ z_i \}$ inform about the relevance variable ${\bf y}$ 
depends on the respective encoding noise strengths $\{ \sigma^2_i \}$. To express this dependence, 
we write the relevant information $I({\bf z}; {\bf y})$ in the form
\begin{align}
I({\bf z}; {\bf y}) &= 
\log \left( \frac{r^3}{\Lambda_1 \times \Lambda_2 \times \Lambda_3} \right) \nonumber\\
&= 
\begingroup
      \color{gray}
\underbracket[0.140ex]{\color{black} \log \frac{r}{\Lambda_1}}_{I(z_1; {\bf y})}
\endgroup
+
\begingroup
      \color{gray}
\underbracket[0.140ex]{\color{black} \log \frac{r}{\Lambda_2}}_{I(z_2; {\bf y})}
\endgroup
+
\begingroup
      \color{gray}
\underbracket[0.140ex]{\color{black} \log \frac{r}{\Lambda_3}}_{I(z_3; {\bf y})}
\endgroup,
\end{align}
with
\begin{align}
\Lambda_i(\sigma_i) = r \sqrt{\frac{\lambda_i^2 + \sigma_i^2}{r^2 + \sigma_i^2}}.
\end{align}
Here, $\Lambda_i(\sigma_i)$ is the half-length of $P({\bf y} | {\bf z})$ ellipsoid's axis oriented along the vector ${\bf v}_i$,
and $\lambda_i \in \{ a, b, c\}$. 
In the limit where all three $\sigma_i \rightarrow \infty$, no information about the signal is encoded ($I({\bf z}; {\bf s}) = 0$),
and we find $\Lambda_i \rightarrow r$. This is the scenario shown in the left panel of Fig.~\ref{fig:3d_geometric} where the $P({\bf y} | {\bf z})$
ellipsoid turns into a sphere corresponding to the marginal $P({\bf y})$.
In the opposite limit of noiseless encoding where all three $\sigma_i \rightarrow 0$ (hence, $I({\bf z}; {\bf s}) \rightarrow \infty$
and the signal ${\bf s}$ can be fully recovered from ${\bf z}$),
we obtain $\Lambda_i \rightarrow \lambda_i$, which means that in this limit,
the $P({\bf y} | {\bf z})$ ellipsoid will match the ${\bf s} \rightarrow {\bf y}$ mapping ellipsoid of $P({\bf y} | {\bf s})$.

To understand the assignment of encoding noise strengths $\{ \sigma_i^2 \}$ used for optimally navigating 
between the above two limits, namely the assignment that maximizes the relevant information $I({\bf z}; {\bf y})$
for a given finite value of the encoded information
\begin{align}
I({\bf z}; {\bf s}) =  \sum_{i=1}^3 
\begingroup
      \color{gray}
\underbracket[0.140ex]{\color{black} \frac{1}{2} \log \left( 1 + \frac{r^2}{\sigma_i^2} \right) }_{
I(z_i; {\bf s}) },
\endgroup
\end{align}
we again first consider a geometric perspective to the solution of this problem (Fig.~\ref{fig:3d_geometric}).
When the encoded information $I({\bf z}; {\bf s})$ is increased from zero, the $P({\bf y} | {\bf z})$ 
ellipsoid initially gets compressed along one direction only, namely the direction ${\bf v}_1$.
This corresponds to the optimal strategy of the scalar encoding along the first principal direction $\hat{\bf w}_1$.
At a critical point where the shape of the $P({\bf y} | {\bf z})$ ellipsoid projected onto the ${\bf v}_1 \text- {\bf v}_2$ plane (highlighted in yellow)
becomes identical to the shape of the $P({\bf y} | {\bf s})$ ellipsoid on that plane, a transition happens where the second component
is introduced.
The condition of shape identity at the first transition is the same as that in the two dimensional case (Eq.~\ref{eqn:equal_ratio_condition}), 
namely $\Lambda_1^\dagger / r = a/b$.
As the encoded information is increased further, the $P({\bf y} | {\bf z})$ ellipsoid gets compressed 
along two directions, namely 
simultaneously along both ${\bf v}_1$ and ${\bf v_2}$, 
while keeping its shape on the ${\bf v}_1 \text- {\bf v}_2$ plane constant, i.e. $\Lambda_1/\Lambda_2 = a/b$ after the first transition.
When the shape of the $P({\bf y} | {\bf z})$ ellipsoid becomes identical to that of $P({\bf y} | {\bf s})$ not only on the ${\bf v}_1 \text- {\bf v}_2$ plane, 
but also on the ${\bf v}_2 \text- {\bf v}_3$ plane, the second transition takes place where now it becomes optimal to have a third encoding component
(Fig.~\ref{fig:3d_geometric}, right panel).
The similarity condition of $P({\bf y} | {\bf z})$ and $P({\bf y} | {\bf s})$ ellipsoids at the second transition is
\begin{align}
\label{eqn:2nd_transition_geometric_cond}
\frac{\Lambda_2^\ddagger}{r} = \frac{b}{c}.
\end{align}
Past this transition, $P({\bf y} | {\bf z})$ gets compressed along all three directions proportionally, i.e.
\begin{align}
\label{eqn:Lambda_123_ratio}
\Lambda_1 \div \Lambda_2 \div \Lambda_3 = a \div b \div c,
\end{align}
until the distributions $P({\bf y} | {\bf z})$ and $P({\bf y} | {\bf s})$ and their corresponding shapes fully overlap in the limit $I({\bf z}; {\bf s}) \rightarrow \infty$.

\begin{figure*}[!ht]
\centering
\includegraphics[width=1.00\textwidth]{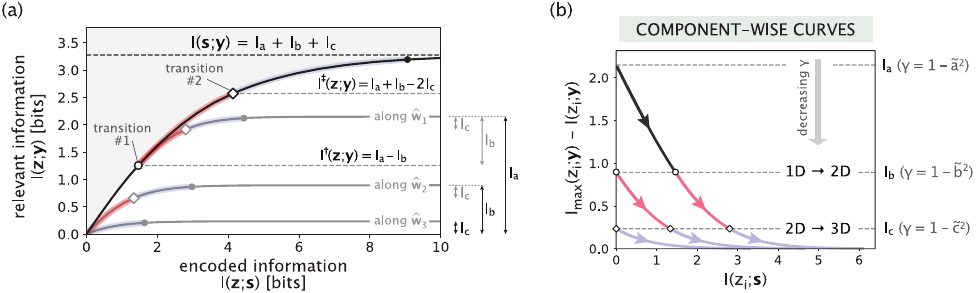}
\caption{
Increasing representation complexity illustrated on the information plane in the case where 
signal and relevance variables are three-dimensional.
(a) Piece-wise formation of the information bound $I({\bf z}; {\bf y})$ vs. $I({\bf z}; {\bf s})$.
Before the first transition, only the most informative first component (encoding along $\hat{\bf w}_1$) is used.
After the first and before the second transition, components encoding along $\hat{\bf w}_1$ and $\hat{\bf w}_2$ 
are used simultaneously (highlighted in red).
The second transition occurs when the relevant information that can still be stored in each of the first two components 
becomes equal to the maximum amount ($I_c$) that can be stored in the third component encoding along $\hat{\bf w}_3$.
Beyond the second transition, the three single-component curves are traversed identically, yielding the blue 
highlighted segment on the information bound.
The three gray dots at the ends of single-component curves with coordinates 
$[I(z_i; {\bf s}), I(z_i; {\bf y})]$ $(i \in \{1, 2, 3\})$
correspond to the black dot on the information bound via
$I({\bf z}; {\bf s}) = \sum_i I(z_i; {\bf s})$ and $I({\bf z}; {\bf y}) = \sum_i I(z_i; {\bf y})$.
(b) Component-wise plots of the relevant information that can still be stored in a component vs. 
the information encoded in that component ($i = 1, 2, 3$ correspond to the three respective curves,
from top to bottom).
Arrows indicate the top-down traversal of the curves, with a given horizontal level corresponding to 
a unique location on the information bound specified by the Lagrange multiplier $\gamma$.
The black, red, and blue colors of curve segments indicate scalar, 2D, and 3D optimal encoding regimes, respectively.
}
\label{fig:3d_information_curve}
\end{figure*}

Next, we discuss how these features of the optimal strategy get reflected on the information plane (Fig.~\ref{fig:3d_information_curve}a).
As in the two-dimensional scenario considered earlier (Fig.~\ref{fig:2d_information_curve}), 
the transition from scalar to 2D optimal encoding happens when the amount of relevant information that can still be stored in the first 
component, namely $I_a - I^\dagger(z_1; {\bf y})$, 
becomes equal to the maximum amount that can be stored in the second component, $I_b$.

To deduce the condition for the second transition (${\rm 2D} \rightarrow {\rm 3D}$ optimal encoding), we recast 
Eq.~\ref{eqn:2nd_transition_geometric_cond} in terms of information metrics as 
\begin{align}
\label{eqn:2nd_transition_info}
I_b - I^\ddagger(z_2; {\bf y}) = I_c.
\end{align}
Here we used the identities $I_b = -\log \tilde{b}$, $I_c = -\log \tilde{c}$, 
and $I^\ddagger(z_2; {\bf y}) = \log (r/ \Lambda_2^\ddagger)$.
The second transition thus 
occurs when the amount of relevant information that $z_2$ can still store, $I_b - I^\ddagger(z_2; {\bf y})$,
which is the same as the amount of relevant information that $z_1$ can still store, $I_a - I^\ddagger(z_1; {\bf y})$,
becomes equal to the maximum amount that can be stored in the third component, namely $I_c$.
The total relevant information at the second transition is then equal to
\begin{align}
I^\ddagger({\bf z}; {\bf y}) &= I^\ddagger(z_1; {\bf y}) + I^\ddagger(z_2; {\bf y}) \nonumber\\
&= I_a + I_b - 2 I_c.
\end{align}
After the second transition, as in the two-dimensional case, any additional encoded information is allocated equally 
among the three components, yielding identical amounts of additional relevant information from each of the components.

The information-theoretic principles behind the increasing representation complexity and 
optimal behavior between distinct transitions can be seen even more directly  
in the component-wise information plots shown in Fig.~\ref{fig:3d_information_curve}b. 
There, the relevant information that can still be stored in component $i$, namely 
$I_{\rm max}(z_i; {\bf y}) - I(z_i; {\bf y})$, is plotted against the amount of encoded 
information allocated to that component, $I(z_i; {\bf s})$. 
As indicated by the arrows, the curves are traversed top-down, 
with a given horizontal level corresponding to a location on 
the information bound in Fig.~\ref{fig:3d_information_curve}a,
which is uniquely specified by the Lagrange multiplier $\gamma$
that sets the derivative via $\gamma = \partial I({\bf z}; {\bf y}) / \partial I({\bf z}; {\bf s})$.
In fact, this derivative is the same as that along the component-wise curves which 
together give rise to the information bound, i.e. $\gamma = \partial I(z_i; {\bf y}) / \partial I(z_i; {\bf s})$
after the $i^{\rm th}$ component is introduced.
As $\gamma$ decreases, new components get introduced at special moments where 
the horizontal level, indicating the amount of relevant information that can still be stored 
in each of the existing components, matches the maximum level of the new component
(Fig.~\ref{fig:3d_information_curve}b).
Once introduced, the curve of the new component is traversed identically to the curves of existing components.

In fact, one can also draw a thermodynamic analogy between the optimal allocation of encoded information 
across different $z$-components and the problem of optimally distributing particles among boxes of varying sizes.
Specifically, the boxes of varying sizes represent the different amounts of maximum relevant information that 
the $z$-components can store. 
Placing a particle in a given box corresponds to allocating additional encoded information to a given $z$-component.
Thermodynamically, the optimal distribution of particles in the boxes should be such that the total free energy 
is minimized.
In general, the optimal strategy is to initially fill the larger box and begin filling the smaller box only when the chemical 
potentials of particles in the two boxes become equal.
One can also show that if the particles are distinguishable, 
the second box starts to fill when the remaining available volume in the first box 
equals that of the smaller second box.
From that point onward,
both boxes should be filled simultaneously, maintaining equal available volumes and chemical potentials.
The same principle extends to the case of more than two boxes (see Appendix~D3 for details).
This is analogous to how in the optimal encoding strategy it is best to use the most informative component first
until the relevant information that can still be stored in that component equals the maximum amount that can be 
stored in the second, yet unused component.

Taken together, the diverse perspectives presented above
offer an intuitive understanding of the various aspects of the optimal encoding strategy
-- from the optimal choice of encoding directions, 
to the allocation of encoding capacity among these directions, to the emergence 
of distinct transitions in the complexity of signal representation.\\

\begin{center}
\bf{General high-dimensional case}
\end{center}

The intuition and principles demonstrated earlier for the two- and three-dimensional cases 
extend naturally to arbitrary dimensions. Reserving the detailed discussion of the general 
solution to the bottleneck problem to Appendix~D1, here we present some of its main results 
which go in parallel to the results of Chechik {\it et al.} \cite{Chechik2005}.
In particular, the noise strength associated with the $i^{\rm th}$ encoding direction can be expressed as
\begin{align}
\sigma^2_i(\gamma) = r^2 \,\frac{\gamma \tilde{\lambda}_i^2}{(1 - \gamma) - \tilde{\lambda}_i^2},
\end{align}
where $\gamma$ is the Lagrange multiplier, 
$r$ is the scale of the standardized signal ($\boldsymbol{\Sigma}_{\bf s} = r^2 {\bf I}_{\bf s}$),
and $\lambda_i$ is the $i^{\rm th}$ semi-axis length 
of the $P({\bf s} | {\bf y})$ ellipsoid (with $\tilde{\lambda}_i = \lambda_i/r$); equivalently, $\lambda_i^2$ is the $i^{\rm th}$ eigenvalue 
of the conditional covariance matrix $\boldsymbol{\Sigma}_{{\bf s} | {\bf y}}$.
The $i^{\rm th}$ encoding component is introduced when the value of the Lagrange multiplier 
gets below the critical value 
\begin{align}
\gamma_i^c = 1 - \tilde{\lambda}_i^2.
\end{align}

With $\sigma^2_i(\gamma)$ at hand, we next compute the $i^{\rm th}$ semi-axis length $\ell_i(\gamma)$
of the decoding $P({\bf s} | {\bf z})$ ellipsoid, namely
\begin{align}
\ell_i(\gamma) = r \sqrt{\frac{\gamma}{1 - \gamma}} \sqrt{\frac{\tilde{\lambda}_i^2}{1 - \tilde{\lambda}_i^2}}.
\end{align}
The fact that the $\gamma$-dependence appears as a multiplicative factor indicates that 
the aspect ratio $\ell_i(\gamma) \div \ell_j(\gamma)$ stays independent of $\gamma$
once the $i^{\rm th}$ and $j^{\rm th}$ components are both introduced.

Similarly, we can compute the $i^{\rm th}$ semi-axis length $\Lambda_i(\gamma)$ of the $P({\bf y} | {\bf z})$
ellipsoid in the relevance space as
\begin{align}
\Lambda_i(\gamma) = \frac{\lambda_i}{\sqrt{1- \gamma}}.
\end{align}
One can easily see how the expression for the aspect ratio follows, namely
$\Lambda_i(\gamma) \div \Lambda_j(\gamma) = \lambda_i \div \lambda_j$.
This shows the generality of the geometric interpretation discussed earlier
in cases of 2D (Fig.~\ref{fig:2d_optimality}) and 3D (Fig.~\ref{fig:3d_geometric}) encoding.

Finally, we note that the encoded information $I({\bf z}; {\bf s})$ and 
the relevant information $I({\bf z}; {\bf y})$ can be obtained directly 
from the semi-axis length expressions via
\begin{align}
I({\bf z}; {\bf s}) &= \log \prod_{i = 1}^{n(\gamma)} \frac{r}{\ell_i(\gamma)} 
= -\sum_{i=1}^{n(\gamma)} \log \tilde{\ell}_i(\gamma), \\
I({\bf z}; {\bf y}) &= \log \prod_{i = 1}^{n(\gamma)} \frac{r}{\Lambda_i(\gamma)}
= - \sum_{i=1}^{n(\gamma)} \log \tilde{\Lambda}_i(\gamma),
\end{align}
where $n(\gamma)$ is the number of encoding components that have already been 
introduced for the given value of the Lagrange multiplier $\gamma$.\\

\begin{center}
\bf{Comment on the degenerate space of optimal solutions}
\end{center}

In our general treatment of the bottleneck problem so far, 
we considered the encoding components $z_i = \hat{\bf w}_i \cdot {\bf s} + \xi_i$
to be statistically independent of one another.
This was achieved by choosing independent encoding noises $( \xi_i )$
and perpendicular encoding directions $( \hat{\bf w}_i )$ corresponding 
to the eigenvectors of $\boldsymbol{\Sigma}_{{\bf s} | {\bf y}}$.
A broader space of optimal solutions, however, becomes accessible
when relaxing the assumption of independent components.
Specifically, the same $I({\bf z}; {\bf s})$ and $I({\bf z}; {\bf y})$
can be achieved by considering correlated noises and/or non-perpendicular 
encoding directions which correlate $z_i$ values through the signal ${\bf s}$.

A particularly interesting class of degenerate optimal solutions is the one 
where the encoding noises assigned to non-perpendicular directions are uncorrelated 
but have the same strength.
This is in contrast to the principal solution where the more informative components 
have a lower associated noise (hence, a higher allocated encoding capacity).
The identical noise strength in such  alternative scheme,
given by 
$\sigma^{-2}_{\rm alt} = \left \langle \sigma_i^{-2} \right \rangle$
(averaging performed over different components $i$),
falls between the highest
and the lowest ones assigned to the principal directions, i.e. 
$\sigma_1 \le \sigma_{\rm alt} \le \sigma_n$, where $n$ is the number of components.
An analogous inequality then holds for individual encoded informations, namely
$I(z_1; {\bf s}) \ge I(z_i^{\rm alt}; {\bf s}) \ge I(z_n; {\bf s})$,
where $I(z_i^{\rm alt}; {\bf s})$ is the same for all $i$ because $\sigma_{\rm alt}$ is the same.
While $I({\bf z}^{\rm alt}; {\bf s}) = I({\bf z}; {\bf s})$ from the optimality condition, 
due to the correlations between $z_i^{\rm alt}$,
the sum of component-wise information amounts in the alternative strategy
will be greater than that for the principal strategy, namely
\begin{align}
\label{eqn:I_enc_alternative_inequality}
\begingroup
      \color{gray}
\underbracket[0.140ex]{\color{black} \sum_{i=1}^n I(z^{\rm alt}_i; {\bf s}) }_{n I(z_i^{\rm alt}; {\bf s}) }
\endgroup
\ge 
\begingroup
      \color{gray}
\underbracket[0.140ex]{\color{black} \sum_{i=1}^n I(z_i; {\bf s}) }_{ I({\bf z}; {\bf s}) }
\endgroup
\end{align}

The relation in Eq.~\ref{eqn:I_enc_alternative_inequality} may have functional implications 
in terms of energy cost.
For example, if the cost of encoding the signal into
different components were to scale linearly with the sum of the individual 
informations, then the principal solution with perpendicular encoding 
directions would be the preferred one.
In the principal solution, however, encoding capacity is distributed unequally across 
components (i.e. $I(z_1; {\bf s}) \ge I(z_2; {\bf s}) \ge ...$).
This means that if the encoding cost were to scale nonlinearly with information \cite{MulderPRR2025},
then the alternative strategy with non-perpendicular directions may become the preferred one.
We refer the reader to Appendix~D2 for a more detailed discussion of the degenerate space of solutions.

\section{Signal prediction problem}

\begin{figure*}[!ht]
\centering
\includegraphics[width=1.00\textwidth]{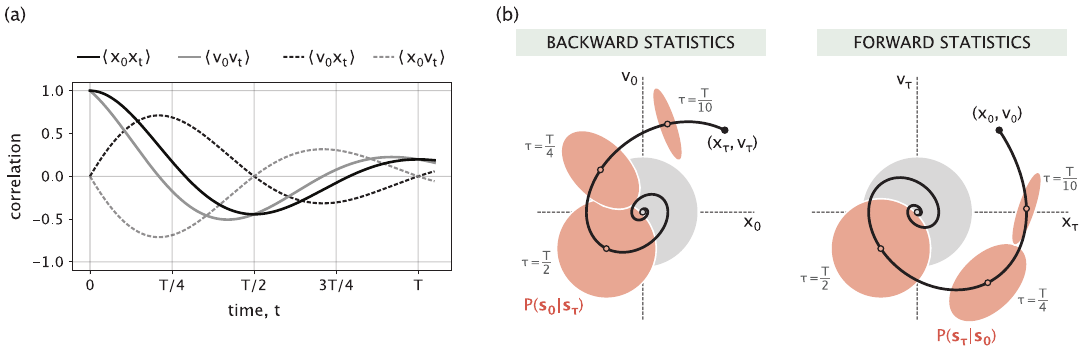}
\caption{
Statistical structure of an example underdamped signal with $\eta = 0.5$.
(a) Auto- and cross-correlation functions of the signal and its derivative.
(b) Backward and forward signal statistics represented by the distribution $P({\bf s}_0 | {\bf s}_\tau)$
and $P({\bf s}_\tau | {\bf s}_0)$, respectively, for difference values of $\tau$.
Backward statistics capture the distribution of present states ${\bf s}_0$ 
that could lead to the given future state ${\bf s}_\tau$,
whereas forward statistics capture the distribution of future states ${\bf s}_\tau$ that may be reached  
from the given present state ${\bf s}_0$.
Gray circles of unit radius correspond to the marginal $P({\bf s})$ reached in the $\tau \rightarrow \infty$ limit.
}
\label{fig:signal_prediction}
\end{figure*}

As an application of the different perspectives on the bottleneck method, 
we now examine the problem of signal prediction for a canonical 
signal-generation model
\cite{Creutzig2009, Becker2015, Palmer2015, Sachdeva2021, Tjalma2023}.
In particular, we consider signals arising from a stochastically driven harmonic oscillator,
the dynamics of which is governed by the following set of dimensionless equations:
\begin{align}
\label{eqn:signal_dynamics_x}
\frac{{\rm d}x}{{\rm d}t} &= v, \\
\label{eqn:signal_dynamics_v}
\frac{{\rm d}v}{{\rm d}t} &= -x - \eta v + \sqrt{2\eta} \, \psi(t).
\end{align}
Here, the position $x$ represents the signal, $v$ is the signal derivative,
$\psi(t)$ denotes unit white noise, and $\eta$ is the damping coefficient.
The value of $\eta$ dictates the qualitative behavior of signal dynamics: 
underdamped for $\eta < 2$, critically damped at $\eta = 2$, and overdamped when $\eta > 2$.

In the context of information bottleneck, the prediction problem is to encode 
the past signal trajectory in such a way that the encoding preserves maximum information 
about the future trajectory, subject to a constraint on encoding capacity.
Importantly, while the dynamics of $x$ is non-Markovian, the joint dynamics of $x$ and $v$ is Markovian,
as per Eqs.~\ref{eqn:signal_dynamics_x} and \ref{eqn:signal_dynamics_v}.
As a result, predicting the future trajectory $x_{[\tau, \infty)}$, where $\tau$ is the forecast interval, 
reduces to predicting the pair $(x_\tau, v_\tau)$.
Similarly, out of the entire past trajectory $x_{(-\infty, 0]}$, only the present value and its derivative,
namely $(x_0, v_0)$, need to be encoded for the purpose of predicting $(x_\tau, v_\tau)$.

The problem of optimal encoding can then be framed in 
the two-dimensional setting studied earlier in Sec. IV.
Specifically, the goal is to optimally encode the current position $x_0$ and derivative $v_0$
into the variables
\begin{subequations}
\label{eqn:z1_z2_prediction}
\begin{align}
z_1 &= \hat{w}_{1,1} \, x_0 + \hat{w}_{1,2} \, v_0 + \xi_1, \\
z_2 &= \hat{w}_{2,1} \, x_0 + \hat{w}_{2,2} \, v_0 + \xi_2,
\end{align}
\end{subequations}
such that ${\bf z} = [z_1, z_2]^{\rm T}$ is maximally informative about 
the future position $x_\tau$ and derivative $v_\tau$, subject to a constraint 
on the encoded information $I({\bf z}; {\bf s}_0)$.
Here we introduced the state vector ${\bf s}_0 = [x_0, v_0]^{\rm T}$ to represent 
the initial pair $(x_0, v_0)$. Similarly, we will use ${\bf s}_\tau = [x_\tau, v_\tau]^{\rm T}$
to denote the pair at time $\tau$,
which, in the information bottleneck terminology, will be the relevance variable ${\bf y}$.

We note that this problem setup was also studied in the prior work of 
Sachdeva {\it et al.} \cite{Sachdeva2021}.
Our approach, however, differs in several key aspects.
Firstly, the marginal distributions $P({\bf s}_0)$ and $P({\bf s}_\tau)$ resulting from the
dimensionless equations of dynamics (Eqs.~\ref{eqn:signal_dynamics_x} and \ref{eqn:signal_dynamics_v})
are bivariate Gaussians with covariance matrices equal to the identity matrix,
i.e. $\langle x^2 \rangle = \langle v^2 \rangle = 1$ and $\langle x v \rangle = 0$
(see Appendices~E1 and E2).
With this standardized form, the orientations of encoding vectors $\hat{\bf w}_1$ and $\hat{\bf w}_2$ 
directly reflect the relative importance assigned to $x_0$ and $v_0$ during encoding.
In the treatment by Sachdeva, {\it et al.} \cite{Sachdeva2021}, 
however, a different nondimensionalization approach leads 
to unequal marginal variances for $x_0$ and $v_0$.
As a result, the weights they derive are not direct indicators of 
the relative importance assigned to $x_0$ and $v_0$,
and the corresponding analytical expressions are more complex
and harder to interpret.
Secondly, Sachdeva {\it et al.} \cite{Sachdeva2021}
focus on the study of the leading encoding component $z_1$, 
addressing the second component $z_2$ only briefly and in general terms.
In this work, we place equal emphasis on the second component
and its predictive capabilities across varying forecast intervals, uncovering novel behaviors.

We start off with the question of finding the optimal encoding directions $\hat{\bf w}_1$ and $\hat{\bf w}_2$.
These directions are determined by the future $({\bf s}_\tau)$ $\rightarrow$ present $({\bf s}_0)$ mapping statistics.
In turn, these statistics -- captured by the conditional covariance matrix 
$\boldsymbol{\Sigma}_{ {\bf s}_0 | {\bf s}_\tau}$ -- are set by the pairwise
auto- and cross-correlation functions of position and velocity.
As derived in Appendix~E3, the pairwise correlations determine
the orientation angles of $\hat{\bf w}_1$ and $\hat{\bf w}_2$ via
\begin{align}
\tan(2\varphi) = \frac{2 \langle v_0 x_\tau \rangle}{\langle x_0 x_\tau \rangle + \langle v_0 v_\tau \rangle}.
\end{align}
This condition is simultaneously satisfied by the angle $\varphi_1$ of $\hat{\bf w}_1$
and the angle $\varphi_2 = \varphi_1 + \pi/2$ of the perpendicular direction $\hat{\bf w}_2$.

Example correlation functions for underdamped dynamics with $\eta = 0.5$
are shown in Fig.~\ref{fig:signal_prediction}a.
There, as can be seen, the correlation terms exhibit damped oscillations, each with its own phase.
The backward and forward signal statistics derived from these pairwise correlations 
are illustrated in Fig.~\ref{fig:signal_prediction}b for different choices of the forecast interval $\tau$.
The ellipses corresponding to these statistics increase in area with increasing forecast interval $\tau$,
eventually converging to the circle of the marginal distribution (shown in gray) as $\tau \rightarrow \infty$.

The optimal encoding angles $\varphi_1$ and $\varphi_2$ correspond to the orientation angles 
of the minor and major axes, respectively, of the backward statistics ellipses
(Fig.~\ref{fig:signal_prediction}b).
One can show that, irrespective of the damping regime or the forecast interval,
the angle $\varphi_1$ always lies in the first quadrant.
This is reflected in the counterclockwise tilt of the backward statistics ellipses
(Fig.~\ref{fig:signal_prediction}b).
Thus, the first encoding component combines the current signal 
and its derivative constructively (with weights of the same sign), whereas 
the second component combines them destructively (with weights of opposite signs).
The angle $\varphi_1$ for different damping regimes is given by
\begin{align}
\label{eqn:phi1_cases}
\varphi_1 = 
\begin{cases}
\frac{1}{2}\arctan \left( \frac{\tanh(\kappa \tau)}{\kappa} \right), &{\rm if} \, \, \eta > 2 \\[1.2ex]
\frac{1}{2}\arctan \left( \frac{\tan(\omega \tau)}{\omega} \right), &{\rm if} \, \, \eta < 2 \\[1.2ex]
\frac{1}{2}\arctan \tau, &{\rm if} \, \, \eta = 2
\end{cases}
\end{align}
Here, $\kappa = \sqrt{\eta^2/4 - 1}$ is defined for the overdamped regime ($\eta > 2$),
and $\omega = \sqrt{1 - \eta^2/4}$ is defined for the underdamped regime ($\eta < 2$).
Importantly, we restrict the output of the arctan function to the range $(0, \pi)$,
which ensures that the angle $\varphi_1$ lies within $(0, \pi/2)$ and thus remains
in the first quadrant, while the angle $\varphi_2 = \varphi_1 + \pi/2$ 
remains in the second quadrant.

We focus our further analysis of the optimal strategy on the underdamped case, 
which is arguably the more interesting regime, deferring the discussion of 
overdamped and critically damped cases to Appendix~E3.
Fig.~13a shows the encoding angles $\varphi_1$ and $\varphi_2 = \varphi_1 + \pi/2$
as functions of the forecast interval $\tau$ in the underdamped regime with $\eta = 0.5$.
The nearly linear dependence on $\tau$ can be easily deduced from Eq.~\ref{eqn:phi1_cases}
where $\omega = \sqrt{1 - \eta^2/4} \approx 1$ for $\eta = 0.5$, which results in
\begin{align}
\label{eqn:phi1_approx}
\varphi_1(\tau) \approx \frac{\tau}{2} \,\,   {\rm mod} \,\,  \frac{\pi}{2}.
\end{align}
For short forecast intervals, the autocorrelation $\langle x_0 x_\tau \rangle$  decays less than
$\langle v_0 v_\tau \rangle$ (Fig.~\ref{fig:signal_prediction}a). Thus, for small $\tau$, the dominant 
encoding component $z_1$ prioritizes the current signal value $x_0$ over 
the current derivative $v_0$. 
When the forecast interval exceeds a quarter period, $T/4$ (with $T = 2\pi/\omega$),
the autocorrelation $\langle v_0 v_\tau \rangle$ exceeds $\langle x_0 x_\tau \rangle$ in magnitude
(Fig.~\ref{fig:signal_prediction}a), and 
the priority shifts from $x_0$ to $v_0$.
As $\tau$ approaches half a period $T/2$,
$z_1$ becomes predominantly $v_0$-based.
This behavior repeats every half-period.
Indeed, as the dominant component, $z_1$ prioritizes the more informative signal feature at the given forecast interval.
The second component $z_2$, encoding in an orthogonal direction, does the opposite, 
reversing the prioritization between $x_0$ and $v_0$.

\begin{figure}[!t]
\centering
\includegraphics{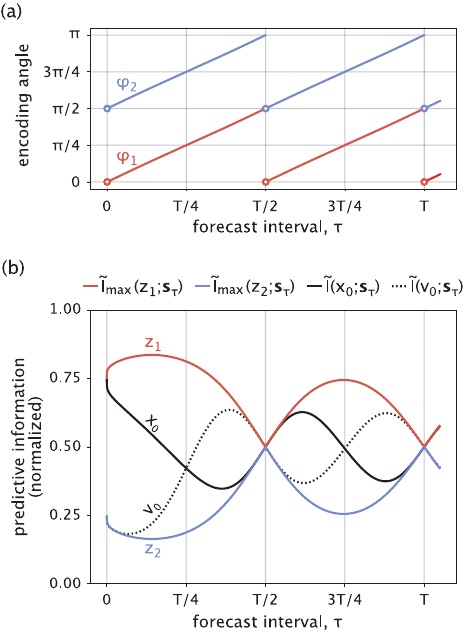}
\caption{
Features of the optimal encoding strategy for predicting signals arising from underdamped dynamics ($\eta = 0.5$).
(a) Optimal encoding angles $\varphi_1$ and $\varphi_2 = \varphi_1 + \pi/2$ for the components $z_1$ and $z_2$,
respectively, shown as a function of the forecast interval $\tau$. 
The angle $\varphi_1$ lies in the range $(0, \pi/2)$, while $\varphi_2$ lies in $(\pi/2, \pi)$.
(b) 
Normalized values of predictive information plotted against the forecast interval for different encoding strategies.
All strategies are evaluated in the zero-noise limit, $I(z_i; {\bf s}_0) \rightarrow \infty$.
Normalization is performed by the maximum achievable predictive information, namely $I({\bf s}_0; {\bf s}_\tau)$,
as determined by the signal statistics.
The upper (red) and lower (blue) curves represent the normalized predictive information 
available to the $z_1$ and $z_2$ components, respectively, in the limit of infinite encoding capacity.
Their sum equals one, namely $\tilde{I}_{\rm max}(z_1; {\bf s}_\tau) + \tilde{I}_{\rm max}(z_2; {\bf s}_\tau) = 1$.
The two curves in between correspond to purely $x_0$-based (solid black line) and 
purely $v_0$-based (dotted black line) strategies.
Analogous plots for the overdamped regime are shown in Fig.~A3.
}
\label{fig:signal_prediction_bottleneck}
\end{figure}

Additional insights about the predictive capacity of the two encoding components can be gained 
by examining the information metrics.
Specifically, we consider the maximum predictive information available to $z_1$ and $z_2$ components,
since the optimal number of
encoding dimensions (1D or 2D)
at finite encoding capacities depends on how different these maximum values 
are (e.g., see Fig.~\ref{fig:2d_information_curve}, with $I_a = I_{\rm max}(z_1;  {\bf s}_\tau)$
and $I_b = I_{\rm max}(z_2; {\bf s}_\tau)$).
Fig.~13b shows the normalized values of these information measures as a function of the forecast interval.
Each measure is normalized by the upper bound on predictive information set by the signal statistics, i.e.
$I({\bf s}_0; {\bf s}_\tau) = I_{\rm max}(z_1;  {\bf s}_\tau) +  I_{\rm max}(z_2; {\bf s}_\tau)$.
The plot thus shows how the relative importance of the two components varies with the forecast interval.
In the same figure, we also plot the normalized predictive information values corresponding to purely $x_0$-based
and purely $v_0$-based strategies.

Fig.~\ref{fig:signal_prediction_bottleneck}b reveals several important aspects of the two components' behavior.
As already recognized from Fig.~\ref{fig:signal_prediction_bottleneck}a,
the dominant component $z_1$ is $x_0$-based and the second component $z_2$ is $v_0$-based 
in the limit $\tau \rightarrow 0$.
In this limit, the first component always contains three times as much predictive information 
as the second component, irrespective of the damping regime (see Appendix~E3).
Interestingly, the strategy based on $z_1$ significantly outperforms the purely $x_0$-based strategy 
already at short forecast intervals.
This demonstrates the importance of incorporating the current signal derivative $v_0$ in the encoding, 
despite its relatively small weight 
($\hat{w}_{1,2} = \sin \varphi_1 \ll 1$ for small $\tau$, 
see Eq.~\ref{eqn:z1_z2_prediction}a and Eq.~\ref{eqn:phi1_approx}).
At a forecast interval equal to a quarter period ($\tau = T/4$), $x_0$ and $v_0$ become equally predictive 
of ${\bf s}_\tau$ ($\langle x_0 x_\tau \rangle = - \langle v_0 v_\tau \rangle$
and $\langle x_0 v_\tau \rangle = -\langle v_0 x_\tau \rangle$, see Fig.~\ref{fig:signal_prediction}a), 
and the optimal strategy combines them with equal weights 
($\hat{w}_{1,1} = \hat{w}_{1,2} = 1/\sqrt{2}$ since $\varphi_1(T/4) = \pi/4$).
As $\tau$ approaches half a period, $z_1$ becomes predominantly $v_0$-based, 
while $z_2$ becomes predominantly $x_0$-based. 
The predictive capacities of the two components also converge, becoming equal
at $\tau = T/2$, as reflected in the circular shape of the backward statistics ellipse at that forecast interval
(Fig.~\ref{fig:signal_prediction}b).
A similar pattern repeats every half-period.
In subsequent cycles, the relative predictive capacity of the $z_1$ component reaches its maximum 
at $\tau_k \approx T/4 + k (T/2)$, for $k \ge 1$. For sufficiently long forecast intervals ($\eta \tau_k \gg 1$),
this peak value is given by (see Appendix~E3)
\begin{align}
\tilde{I}_{\rm max}(z_1; {\bf s}_{\tau_k}) \approx \frac{1}{2} + \frac{\eta/2}{1 + (\eta/2)^2}.
\end{align}
The monotonic dependence of $\tilde{I}_{\rm max}(z_1; {\bf s}_{\tau_k})$ on the damping coefficient $\eta$
shows that the gap in the predictive capacities of the $z_1$ and $z_2$ components decreases 
with weaker damping (lower $\eta$).
Thus, the weaker the damping, the more important it becomes to incorporate 
the second component for accurate prediction.

Altogether, our reexamination of the signal prediction problem studied earlier by 
Sachdeva {\it et al.} \cite{Sachdeva2021} has 
led to a deeper understanding of the encoding features underlying optimal prediction.
First, the concise analytical expressions for the optimal encoding angle (Eq.~\ref{eqn:phi1_cases}),
derived from the standardized signal statistics,
provide direct, quantitative insights into how the optimal prioritization of $x_0$ over $v_0$ 
is determined in different settings.
Second, our extension of the study to multiple encoding components 
has revealed an intricate dependence of their predictive capacities on the forecast
interval, that, in turn, varies qualitatively across damping regimes 
(Fig.~\ref{fig:signal_prediction_bottleneck}, see also Fig.~A3).

\section{Discussion}

The information bottleneck method is an established framework for learning efficient signal 
representations that are maximally informative about a relevance variable, subject to limits 
on the encoding capacity \cite{Tishby1999}.
In a special, yet important case where the signal and relevance variables are jointly Gaussian, 
the optimal representation is also Gaussian, and an analytical solution 
to the bottleneck problem becomes available.
Since the original work by Chechik {\it et al.} \cite{Chechik2005}, 
the Gaussian bottleneck method has been 
applied to speaker recognition \cite{Hecht2009}, 
used in studies of cellular prediction \cite{Tjalma2023},
explored for its parallels with the renormalization group \cite{KlinePalmer2022},
and extended to more general problem settings
\cite{Rey2012, Mahvari2021, Ngampruetikorn2023}.
Despite its diverse applications,
our current understanding of the Gaussian information bottleneck method 
and its emergent features remains largely mathematical.

The main aim of our work was to advance our understanding of the method 
from multiple, mutually enriching perspectives.
The initial standardization of the problem variables enabled us 
to extract geometric insights into the optimal strategy, 
without compromising the generality of the analysis.
In particular, by studying the less commonly considered decoding distribution $P({\bf y} | {\bf z})$
in the relevance space and its relationship to the distribution $P({\bf y} | {\bf s})$ of the stochastic 
${\bf s} \rightarrow {\bf y}$ mapping, 
we found distinct signatures of the optimal strategy in the geometric depictions  
of these distributions (ellipses in 2D, ellipsoids in higher dimensions).
First, their axes are aligned, with each aligned pair corresponding to the same rank in the ascending 
order of lengths (i.e., the shortest axis of $P({\bf y} | {\bf z})$ is aligned with the shortest axis 
of $P({\bf y} | {\bf s})$, and so on).
Second, and more importantly, transitions from lower- to higher-dimensional optimal representations 
occur when the aspect ratios of these axes become equal at critical levels of encoded information
(Fig.~\ref{fig:3d_geometric}).

The latter geometric insight then hinted at the existence of an information-theoretic criterion 
for the bifurcation points and the optimal navigation between them.
Specifically, a new encoding dimension is introduced when the amount of relevant information 
that each of the existing components can still store becomes equal to the maximum 
relevant information available to the new, unused component.
After the new component is introduced, the additional encoded information is allocated equally 
among all components, yielding identical amounts of additional relevant information from each component
(Fig.~\ref{fig:3d_information_curve}).
Interestingly, this principle is reminiscent of the thermodynamic problem of distributing particles among boxes 
of different sizes in such a way that the total free energy is minimized. The optimal strategy 
is to first fill the larger box and only begin filling the smaller box when the chemical potentials 
of particles in the two boxes become equal.

The canonical solution to the Gaussian information bottleneck problem is expressed 
in terms of statistically independent encoding components.
With the standardization procedure applied in our treatment 
($\boldsymbol{\Sigma}_{\bf s} = r^2 {\bf I}_{\bf s}$, $\boldsymbol{\Sigma}_{\bf y} = r^2 {\bf I}_{\bf y}$),
these components encode along a set of orthonormal directions (eigenvectors of
$\boldsymbol{\Sigma}_{{\bf s} | {\bf y}}$)
and have independent sources of encoding noise,
with the more informative directions assigned lower noise levels or, equivalently,
higher encoded information.
The space of optimal solutions, however, is degenerate, composed of alternative 
strategies that involve correlated encoding components.
One noteworthy case is when the total encoding capacity 
is distributed equally among non-orthogonal encoding directions, 
leading to components correlated via the signal, 
with each component assigned the same intermediate level of independent noise.
If the resource cost of encoding the signal into a component does not scale linearly
with the amount of information encoded \cite{Tjalma2023}, and/or if the high-precision (low $\sigma$)
encoding of the dominant component in the canonical setup becomes prohibitive,
then an alternative strategy with correlated components may, in fact,
be advantageous.

As a practical demonstration of our perspective on the bottleneck method, 
we next revisited the previously studied problem of predicting signals 
coming from stochastically driven harmonic oscillator dynamics
\cite{Creutzig2009, Becker2015, Palmer2015, Sachdeva2021, Tjalma2023}.
This problem conveniently reduces to optimally encoding the current signal value
and its derivative, as these two features of the past signal trajectory 
fully specify the future signal statistics.
Our use of standardized signal statistics enabled concise analytical expressions 
for the orientation angles of the $P({\bf s} | {\bf y})$ ellipse, which directly 
indicate the relative weighting of the current signal and its derivative 
in the two encoding components.
Furthermore, our extended analysis of the second encoding component 
revealed the complex dependence of its predictive capacity 
on the forecast interval as well as on the damping regime of the signal dynamics.

The prevalence of linear models with Gaussian statistics in natural and engineered systems 
makes the Gaussian information bottleneck a practically applicable framework, 
in addition to serving as an analytically tractable baseline against which more 
sophisticated numerical frameworks can be compared.
Developing a deep, intuitive understanding of its various features is therefore essential
-- an aim that our present work has contributed to.
Looking forward, it would be interesting to explore the possible implications of our results 
for settings beyond the canonical Gaussian information bottleneck.
For instance, Ngampruetikorn and Schwab recently showed that 
similar structural transitions in the optimal representation occur 
when generalized correlation measures, such as Rényi and Jeffreys divergences,
are used instead of the Shannon mutual information \cite{Ngampruetikorn2023}.
Whether the geometric and component-wise information-theoretic criteria we derived for the transition points 
and the optimal navigation between them also hold for this generalized problem formulation
remains an open question.
It would also be fruitful to understand how and to what extent these new perspectives
on structural transitions 
continue to hold approximately when the variables are no longer jointly Gaussian
\cite{Rey2012, Bauer2023, Wu2020Phase}, 
and what insights can be gained from the nature of the resulting deviations.
When signal and relevance variables in such nonlinear systems represent entire trajectories rather than 
low-dimensional vectors, advanced simulation schemes may be required for the exact computation of 
information metrics \cite{Reindardt2023}.

\section{Acknowledgements}

This work was supported by the Dutch Research Council (NWO) and 
the European Research Council (ERC) under 
the European Union's Horizon 2020 research and innovation program (grant No. 885065). 
It was performed at the research institute AMOLF.

\bibliography{papers.bib}

\newpage
\onecolumngrid

\newpage

\begin{center}
{\large \bf Intuitive dissection of the Gaussian information bottleneck method
with an application to optimal prediction \\ {\phantom \,} \\ Supporting information}    
\end{center}

\begin{center}
Vahe Galstyan, Age Tjalma, Pieter Rein ten Wolde
\end{center}

\phantom{\,}
\newline
\phantom{\,}

\thispagestyle{empty}

\noindent
\hyperref[sec:AppendixA]{A. Standardization procedures} \dotfill \pageref{sec:AppendixA} \\
\hyperref[sec:AppendixA1]{\qquad 1. Standardization of ${\bf s}$ and ${\bf y}$} \dotfill \pageref{sec:AppendixA1} \\
\hyperref[sec:AppendixA2]{\qquad 2. Implications for the shapes of $P({\bf s} | {\bf y})$ and $P({\bf y} | {\bf s})$ distributions} \dotfill \pageref{sec:AppendixA2} \\
\hyperref[sec:AppendixA3]{\qquad 3. Standardization of encoding variables} \dotfill \pageref{sec:AppendixA3} \\

\noindent
\hyperref[sec:AppendixB]{B. One-dimensional encoding} \dotfill \pageref{sec:AppendixB} \\
\hyperref[sec:AppendixB1]{\qquad 1. Optimal encoding direction} \dotfill \pageref{sec:AppendixB1} \\
\hyperref[sec:AppendixB2]{\qquad 2. Slope of the $I(z; {\bf y})$ vs. $I(z; {\bf s})$ curve} \dotfill \pageref{sec:AppendixB2} \\
\hyperref[sec:AppendixB3]{\qquad 3. ``Decoding'' perspective} \dotfill \pageref{sec:AppendixB3} \\
\hyperref[sec:AppendixB4]{\qquad 4. Optimality in the ``decoding'' perspective} \dotfill \pageref{sec:AppendixB4} \\

\noindent
\hyperref[sec:AppendixC]{C. Two-dimensional encoding} \dotfill \pageref{sec:AppendixC} \\
\hyperref[sec:AppendixC1]{\qquad 1. Statistics of $P({\bf s} | {\bf z})$ and $P({\bf y} | {\bf z})$ distributions} \dotfill \pageref{sec:AppendixC1} \\
\hyperref[sec:AppendixC2]{\qquad 2. Transition from scalar to two-dimensional optimal encoding} \dotfill \pageref{sec:AppendixC2} \\
\hyperref[sec:AppendixC3]{\qquad 3. Behavior beyond the transition point} \dotfill \pageref{sec:AppendixC3} \\

\noindent
\hyperref[sec:AppendixD]{D. General multi-dimensional case} \dotfill \pageref{sec:AppendixD} \\
\hyperref[sec:AppendixD1]{\qquad 1. Solution to the Gaussian information bottleneck problem} \dotfill \pageref{sec:AppendixD1} \\
\hyperref[sec:AppendixD2]{\qquad 2. Degenerate space of optimal solutions} \dotfill \pageref{sec:AppendixD2} \\
\hyperref[sec:AppendixD3]{\qquad 3. Thermodynamic analogy for the optimal allocation of encoding capacity} \dotfill \pageref{sec:AppendixD3} \\

\noindent
\hyperref[sec:AppendixE]{E. Signal prediction problem} \dotfill \pageref{sec:AppendixE} \\
\hyperref[sec:AppendixE1]{\qquad 1. Stochastically driven damped harmonic oscillator model} \dotfill \pageref{sec:AppendixE1}\\
\hyperref[sec:AppendixE2]{\qquad 2. Solution of the stochastic model} \dotfill \pageref{sec:AppendixE2}\\
\hyperref[sec:AppendixE3]{\qquad 3. Optimal prediction strategy} \dotfill \pageref{sec:AppendixE3}\\

\newpage

\appendix

\setcounter{page}{1}
\renewcommand{\thepage}{\arabic{page}}

\setcounter{figure}{0}
\renewcommand{\thefigure}{A\arabic{figure}}

\newpage

\section{Standardization procedures}
\label{sec:AppendixA}

In this Appendix section, we provide details of the standardization procedure applied to signal and relevance variables, 
as well as to the encoding variables. We show that this procedure does not affect the generality of problem treatment. 
Also, we discuss the link between the shapes of $P({\bf s} | {\bf y})$ and $P({\bf y} | {\bf s})$ distributions  
after the standardization procedure. 
The contents of this section largely concern the claims made in {\bf Sec. II} of the main text.

\subsection{Standardization of ${\bf s}$ and ${\bf y}$}
\label{sec:AppendixA1}

In the general formulation of the problem, the covariance matrices of signal (${\bf s}$) and relevance (${\bf y}$) variables 
may not be diagonal, containing equal entries. This makes many of the geometric interpretations offered in the main text 
infeasible. Below we show that a linear transformation that standardizes the signal variable does not alter the information 
metrics associated with it. The same line of arguments can then be applied to the relevance variable.

We start with the assumption that the original covariance matrix $\boldsymbol{\Sigma}_{\bf s}$ is full-ranked. 
If any component of ${\bf s}$ were a deterministic linear combination of other components, we would assume that 
this component was eliminated before formulating the information bottleneck problem. 
Provided that $\boldsymbol{\Sigma}_{\bf s}$ is full-ranked, it can be diagonalized by an orthogonal matrix ${\bf P}$, namely
\begin{align}
\label{eqnA:SigmaS_diagonalization}
\boldsymbol{\Sigma}_{\bf s} = {\bf P} {\bf D} {\bf P}^{\rm T},
\end{align}
where ${\bf P} {\bf P}^{\rm T} = {\bf I}_{\bf s}$ (condition of orthogonality), and ${\bf D}$ is a diagonal matrix containing 
positive eigenvalues of $\boldsymbol{\Sigma}_{\bf s}$.

Now consider the transformed signal variable of the form:
\begin{align}
\label{eqnA:s_transformation}
\tilde{\bf s} = r {\bf D}^{-1/2} {\bf P}^{\rm T} {\bf s}.
\end{align}
Note that since the diagonal entries of ${\bf D}$ are all positive, the matrix ${\bf D}^{-1/2}$ is well-defined. Let us calculate 
the covariance matrix of this transformed variable, assuming that the mean of the original variable ${\bf s}$ has been 
subtracted off, so that $\langle {\bf s} \rangle = {\bf 0}$. By definition, we have
\begin{subequations}
\begin{align}
\boldsymbol{\Sigma}_{\tilde{\bf s}} &= \mathbb{E} \left[ \tilde{\bf s} \tilde{\bf s}^{\rm T} \right] \\
&= r^2 {\bf D}^{-1/2} {\bf P}^{\rm T} 
\begingroup
      \color{gray}
\underbracket[0.140ex]{\color{black} \mathbb{E} \left[ {\bf s} {\bf s}^{\rm T} \right] }_{
\boldsymbol{\Sigma}_{\bf s}}
\endgroup
{\bf P} {\bf D}^{-1/2} \\
&= 
r^2 {\bf D}^{-1/2} 
\begingroup
      \color{gray}
\underbracket[0.140ex]{\color{black} {\bf P}^{\rm T} {\bf P} }_{{\bf I}_{\bf s}}
\endgroup
{\bf D} 
\begingroup
      \color{gray}
\underbracket[0.140ex]{\color{black} {\bf P}^{\rm T} {\bf P} }_{{\bf I}_{\bf s}}
\endgroup
{\bf D}^{-1/2} \\
&= r^2 {\bf D}^{-1/2} \, {\bf D} \, {\bf D}^{-1/2} \\
&= r^2 {\bf I}_{\rm s}.
\end{align}
\end{subequations}
This shows that the linear transformation in Eq.~\ref{eqnA:s_transformation} turns the covariance matrix of the signal 
variable into a diagonal matrix that contains equal entries set to $r^2$.

Let us now demonstrate that the transformation in Eq.~\ref{eqnA:s_transformation} does not change the mutual information 
between the signal and relevance variables. We write that mutual information in the following form:
\begin{align}
I({\bf s}; {\bf y}) = H({\bf y}) - H({\bf y} | {\bf s}).
\end{align}
Since $H({\bf y})$ stays the same when transforming the ${\bf s}$-variable, we need to show that the conditional entropy of the 
${\bf y}$-variable does not change, namely
\begin{align}
\label{eqnA:equal_entropies_s_tilde}
H({\bf y} | {\bf s}) = H({\bf y} | \tilde{\bf s}).
\end{align}
As the variables of the problem are Gaussian, their entropies are set through the covariance matrices. Specifically, 
the entropy of a $d$-dimensional multivariable Gaussian random variable is $H = \frac{1}{2} \log \left( (2\pi e)^d |\boldsymbol{\Sigma}| \right)$. To prove Eq.~\ref{eqnA:equal_entropies_s_tilde}, it would suffice to 
show that $\boldsymbol{\Sigma}_{{\bf y} | {\bf s}} = \boldsymbol{\Sigma}_{{\bf y} | \tilde{\bf s}}$.
Using the Schur complement 
formula, we can write the conditional variance matrix $\boldsymbol{\Sigma}_{{\bf y} | \tilde{\bf s}}$ as
\begin{align}
\label{eqnA:Schur_cond_tilde}
\boldsymbol{\Sigma}_{{\bf y} | \tilde{\bf s}} = \boldsymbol{\Sigma}_{\bf y} - 
\boldsymbol{\Sigma}_{ {\bf y} \tilde{\bf s}}^{\,} 
\boldsymbol{\Sigma}_{\tilde{\bf s}}^{-1} \boldsymbol{\Sigma}_{{\bf y} \tilde{\bf s}}^{\rm T}.
\end{align}
From the definition of $\boldsymbol{\Sigma}_{{\bf y} \tilde{\bf s}}$, we have
\begin{subequations}
\begin{align}
\boldsymbol{\Sigma}_{{\bf y} \tilde{\bf s}} &= \mathbb{E} \left[ {\bf y} \tilde{\bf s}^{\rm T} \right] \\
&= \mathbb{E} \left[ {\bf y} \left( r {\bf D}^{-1/2} {\bf P}^{\rm T} {\bf s}\right)^{\rm T} \right] \\
&= \mathbb{E} \left[ 
{\bf y} {\bf s}^{\rm T} r {\bf P} {\bf D}^{-1/2}
\right] \\
&= \boldsymbol{\Sigma}_{{\bf y} {\bf s}} \, r {\bf P} {\bf D}^{-1/2}.
\end{align}
\end{subequations}
Substituting the above identity into Eq.~\ref{eqnA:Schur_cond_tilde}, we obtain
\begin{subequations}
\begin{align}
\boldsymbol{\Sigma}_{{\bf y} | \tilde{\bf s}} &= \boldsymbol{\Sigma}_{\bf y} - 
\left( \boldsymbol{\Sigma}^{\,}_{{\bf y} {\bf s}} r {\bf P} {\bf D}^{-1/2} \right) \, r^{-2} {\bf I}_{\bf s} \, 
\left(r {\bf D}^{-1/2} {\bf P}^{\rm T} 
\boldsymbol{\Sigma}_{{\bf y} {\bf s}}^{\rm T} \right) \\
\label{eqnA:S_cond_almost}
&= \boldsymbol{\Sigma}_{\bf y} - \boldsymbol{\Sigma}_{{\bf y} {\bf s}}^{\,} 
\, {\bf P} {\bf D}^{-1} {\bf P}^{\rm T} \, 
\boldsymbol{\Sigma}_{{\bf y} {\bf s}}^{\rm T}.
\end{align}
\end{subequations}
Now, from Eq.~\ref{eqnA:SigmaS_diagonalization} and the orthogonality condition ${\bf P} {\bf P}^{\rm T} = {\bf I}_{\bf s}$,
the inverse of the covariance matrix of the original ${\bf s}$-variable can be written as
\begin{subequations}
\begin{align}
\boldsymbol{\Sigma}_{\bf s}^{-1} &= \left( {\bf P}^{\rm T} \right)^{-1} {\bf D}^{-1} {\bf P}^{-1} \\
&= {\bf P} {\bf D}^{-1} {\bf P}^{\rm T}.
\end{align}
\end{subequations}
Noting the presence of this expression on the right-hand side of Eq.~\ref{eqnA:S_cond_almost}, we find
\begin{align}
\boldsymbol{\Sigma}_{ {\bf y} | \tilde{\bf s} } = 
\underbracket[0.140ex]{\boldsymbol{\Sigma}_{\bf y} 
- \boldsymbol{\Sigma}_{{\bf y} {\bf s}}^{\,} \boldsymbol{\Sigma}_{\bf s}^{-1} \boldsymbol{\Sigma}_{{\bf y} {\bf s}}^{\rm T}}_{
\boldsymbol{\Sigma}_{{\bf y} | {\bf s}}}.
\end{align}
The right-hand side represents the expanded form of $\boldsymbol{\Sigma}_{{\bf y} | {\bf s}}$, demonstrating that indeed 
$\boldsymbol{\Sigma}_{ {\bf y} | \tilde{\bf s} } = \boldsymbol{\Sigma}_{{\bf y} | {\bf s}}$. This shows that the standardization 
procedure does not change the mutual information between signal and relevance variables. 
An identical line of arguments can be made to justify the standardization of the relevance variable ${\bf y}$.

We note that an encoding rule ${\bf z} = \tilde{\bf W}^{\rm T} \tilde{\bf s} + \boldsymbol{\xi}$ 
written for the normalized variable $\tilde{\bf s}$ can be equivalently written for the original variable 
(${\bf z} = {\bf W}^{\rm T} {\bf s} + \boldsymbol{\xi}$)
by making the correspondence ${\bf W} = r {\bf P} {\bf D}^{-1/2} \tilde{\bf W}$.
Also, the original variable ${\bf s}$ can 
be recovered from the transformed variable via
\begin{align}
{\bf s} = r^{-1} {\bf P} {\bf D}^{1/2} \tilde{\bf s},
\end{align}
which follows from Eq.~\ref{eqnA:s_transformation} after multiplying 
both sides by $r^{-1}\left( {\bf P}^{\rm T} \right)^{-1} {\bf D}^{1/2}$
and noting that $\left( {\bf P}^{\rm T} \right)^{-1} = {\bf P}$ 
from the orthogonality condition (${\bf P} {\bf P}^{\rm T} = {\bf I}_{\bf s}$).

\subsection{Implications for the shapes of $P({\bf s} | {\bf y})$ and $P({\bf y} | {\bf s})$ distributions}
\label{sec:AppendixA2}

Next, we discuss the shapes of the $P({\bf s} | {\bf y})$ and $P({\bf y} | {\bf s})$ distributions once the signal and relevance 
variables have been standardized. For convenience, we will use the ${\bf s}$ and ${\bf y}$ notation for the standardized variables,
with $\boldsymbol{\Sigma}_{\bf s} = r^2 {\bf I}_{\bf s}$ and $\boldsymbol{\Sigma}_{\bf y} = r^2 {\bf I}_{\bf y}$.
Applied to the conditional covariance matrix $\boldsymbol{\Sigma}_{{\bf s} | {\bf y}}$, the Schur complement formula yields
\begin{align}
\boldsymbol{\Sigma}_{{\bf s} | {\bf y}} = \boldsymbol{\Sigma}_{\bf s} - 
\boldsymbol{\Sigma}_{{\bf s} {\bf y}}^{\,} \, \boldsymbol{\Sigma}_{\bf y}^{-1} \, \boldsymbol{\Sigma}_{{\bf s} {\bf y}}^{\rm T}.
\end{align}
Substituting $\boldsymbol{\Sigma}_{\bf s}$ and $\boldsymbol{\Sigma}_{\bf y}$, and using the notation 
$\tilde{\boldsymbol{\Sigma}} = r^{-2} \boldsymbol{\Sigma}$, we obtain
\begin{align}
\label{eqnA:S_s_given_y_schur_tilded}
\boldsymbol{\Sigma}_{{\bf s} | {\bf y}} = r^2 \left( {\bf I}_{\bf s} - \tilde{\boldsymbol{\Sigma}}_{{\bf s} {\bf y}}
\tilde{\boldsymbol{\Sigma}}_{{\bf y} {\bf s}} \right).
\end{align}
In an analogous way, we can write the conditional covariance matrix $\boldsymbol{\Sigma}_{{\bf y} | {\bf s}}$ as
\begin{align}
\label{eqnA:S_y_given_s_schur_tilded}
\boldsymbol{\Sigma}_{{\bf y} | {\bf s}} = r^2 \left(
{\bf I}_{\bf y} - \tilde{\boldsymbol{\Sigma}}_{{\bf y} {\bf s}} \tilde{\boldsymbol{\Sigma}}_{{\bf s} {\bf y}}
\right).
\end{align}

Let us first discuss the case where the signal and relevance variables have identical dimensions, 
implying ${\bf I}_{\bf s} = {\bf I}_{\bf y} \equiv {\bf I}$. 
Suppose $\hat{\bf w}$ is an eigenvector of $\boldsymbol{\Sigma}_{{\bf s} | {\bf y}}$ with a corresponding eigenvalue $\lambda$,
satisfying $\boldsymbol{\Sigma}_{{\bf s} | {\bf y}} \hat{\bf w} = \lambda \hat{\bf w}$. 
Using the form in Eq.~\ref{eqnA:S_s_given_y_schur_tilded}, we have
\begin{align}
r^2 \left( {\bf I} - \tilde{\boldsymbol{\Sigma}}_{{\bf s} {\bf y}}
\tilde{\boldsymbol{\Sigma}}_{{\bf y} {\bf s}} \right) \hat{\bf w} = \lambda \hat{\bf w}.
\end{align}
Multiplying both sides by $\tilde{\boldsymbol{\Sigma}}_{{\bf y} {\bf s}}$, we then write
\begin{align}
\begingroup
      \color{gray}
\underbracket[0.140ex]{\color{black} r^2 \left( {\bf I} - \tilde{\boldsymbol{\Sigma}}_{{\bf y} {\bf s}}
\tilde{\boldsymbol{\Sigma}}_{{\bf s} {\bf y}}
\right) }_{\boldsymbol{\Sigma}_{{\bf y} | {\bf s}}}
\endgroup
\begingroup
      \color{gray}
\underbracket[0.140ex]{\color{black} \tilde{\boldsymbol{\Sigma}}_{{\bf y} {\bf s}} \hat{\bf w} }_{\bf v}
\endgroup
= 
\lambda 
\begingroup
      \color{gray}
\underbracket[0.140ex]{\color{black} \tilde{\boldsymbol{\Sigma}}_{{\bf y} {\bf s}} \hat{\bf w} }_{\bf v}
\endgroup
\end{align}
We see that $\lambda$, being an eigenvalue of $\boldsymbol{\Sigma}_{{\bf s} | {\bf y}}$, is also 
an eigenvalue of $\boldsymbol{\Sigma}_{{\bf y} | {\bf s}}$ with a corresponding eigenvector 
${\bf v} = \tilde{\boldsymbol{\Sigma}}_{{\bf y} {\bf s}} \hat{\bf w}$.
Since $\boldsymbol{\Sigma}_{{\bf s} | {\bf y}}$ and $\boldsymbol{\Sigma}_{{\bf y} | {\bf s}}$ have the same number of 
eigenvalues when the dimensions of ${\bf s}$ and ${\bf y}$ are the same, the equality of these eigenvalues implies 
that the constant density contours of $P({\bf s} | {\bf y})$ and $P({\bf y} | {\bf s})$ distributions (ellipses if ${\rm dim}({\bf s}) = 2$
and ellipsoids if ${\rm dim}({\bf s}) = 3$) are identical, except for their orientations. If $\{ \hat{\bf w}_i \}$ are the principal axes 
corresponding to $P({\bf s} | {\bf y})$, then $\{ {\bf v}_i = \tilde{\boldsymbol{\Sigma}}_{{\bf y} {\bf s}} \hat{\bf w}_i \}$ are the 
principal axis orientations corresponding to $P({\bf y} | {\bf s})$.

Now let us discuss the case where the dimensions of ${\bf s}$ and ${\bf y}$ variables do not match. For concreteness, 
let's assume that $n_{\bf y} < n_{\bf s}$, where $n_{\bf y} = {\rm dim}({\bf y})$ and $n_{\bf s} = {\rm dim}({\bf s})$. 
The rank of the cross-correlation matrix $\tilde{\boldsymbol{\Sigma}}_{{\bf s} {\bf y}}$ equals the lower of the two dimensions,
namely $n_{\bf y}$. And since 
${\rm rank} \left( \tilde{\boldsymbol{\Sigma}}_{{\bf s} {\bf y}}^{\,} \tilde{\boldsymbol{\Sigma}}_{{\bf s} {\bf y}}^{\rm T} \right)
= {\rm rank} \left( \tilde{\boldsymbol{\Sigma}}_{{\bf s} {\bf y}} \right)$, the rank of the conditional covariance matrix 
$\boldsymbol{\Sigma}_{{\bf s} | {\bf y}}$ (Eq.~\ref{eqnA:S_s_given_y_schur_tilded}) will also be equal to $n_{\bf y}$.
Thus, when $n_{\bf y} < n_{\bf s}$, 
$\boldsymbol{\Sigma}_{{\bf s} | {\bf y}}$ will have as its eigenvalues all $n_{\bf y}$ eigenvalues of 
$\boldsymbol{\Sigma}_{{\bf y} | {\bf s}}$, and additionally, it will have $n_{\bf s} - n_{\bf y}$ eigenvalues 
equal to $r^2$.

This means that the constant-density contour of $P({\bf s} | {\bf y})$ will match in size with the $P({\bf y} | {\bf s})$ contour 
along $n_{\bf y}$ directions, and it will have the size of the marginal $P({\bf s})$ contour along the remaining 
$n_{\bf s} - n_{\bf y}$ dimensions.
As an example, if $n_{\bf y} = 2$ and $n_{\bf s} = 3$, then $P({\bf y} | {\bf s})$ will be represented by 
an ellipse on the ${\bf y}$-plane, and $P({\bf s} | {\bf y})$ will be represented by an ellipsoid in the ${\bf s}$-space. 
The $P({\bf s} | {\bf y})$ ellipsoid will be compressed along two directions $\hat{\bf w}_1$ and $\hat{\bf w}_2$ 
corresponding to the principal directions ${\bf v}_1 = \tilde{\boldsymbol{\Sigma}}_{{\bf y} {\bf s}} \hat{\bf w}_1$ and 
${\bf v}_2 = \tilde{\boldsymbol{\Sigma}}_{{\bf y} {\bf s}} \hat{\bf w}_2$ of the $P({\bf y} | {\bf s})$ ellipse, 
and it will not be compressed along the third perpendicular direction with a corresponding eigenvalue $r^2$.

\subsection{Standardization of encoding variables}
\label{sec:AppendixA3}

In the Gaussian information bottleneck method, the general form considered for the encoding variable ${\bf z}$ is that 
of noisy linear compression with Gaussian encoding noise, namely
\begin{align}
{\bf z} = {\bf W}^{\rm T} {\bf s} + \boldsymbol{\xi}.
\end{align}
For clarity, we also write the same expression in an expanded form below:
\begin{align}
\begin{bmatrix}
z_1 \\
\vdots \\
z_{n_{\bf z}}
\end{bmatrix}
=
\begin{bmatrix}
\rule[0.5ex]{2em}{0.25pt} \,\, {\bf w}^{\rm T}_1 \, \rule[0.5ex]{2em}{0.25pt} \\
\vdots\\
\rule[0.5ex]{2em}{0.25pt} \,\, {\bf w}^{\rm T}_{n_{\bf z}} \, \rule[0.5ex]{2em}{0.25pt}
\end{bmatrix}
\begin{bmatrix}
s_1 \\
s_2 \\
\vdots \\
s_{n_{\bf s}}
\end{bmatrix}
+
\begin{bmatrix}
\xi_1 \\
\vdots\\
\xi_{n_{\bf z}}
\end{bmatrix}.
\end{align}
Here, $n_{\bf z}$ is the number of encoding components, each individually given by $z_i = {\bf w}_i \cdot {\bf s} + \xi_i$.
In general, encoding noises $\xi_i$ can be correlated, with their correlations specified via the covariance matrix 
$\boldsymbol{\Sigma}_{\xi}$.

In their original work, Chechik {\it et al.} demonstrated that for a given pair of encoding variables,
$({\bf W}, \boldsymbol{\Sigma}_{\boldsymbol{\xi}})$, there exists a corresponding pair $(\tilde{\bf W}, {\bf I}_{n_{\bf z}})$ 
that yields the same value for the objective functional 
$\mathcal{L} = I({\bf z}; {\bf y}) - \gamma I({\bf z}; {\bf s})$ \cite{Chechik2005}.
This corresponding pair has the covariance matrix of encoding noises set equal to an identity matrix ${\bf I}_{n_{\bf z}}$,
implying a fixed variance of $1$ for each noise component and absence of correlations between them.
The goal of optimal encoding was then to identify the directions and magnitudes of the encoding vectors $\{ \tilde{\bf w}_i \}$
for different values of the Lagrange multiplier $\gamma$.

In our treatment of the problem, however, we consider encoding vectors as unit vectors $\{ \hat{\bf w}_i \}$ 
and encoding noises $\{ \xi_i \}$ as having distinct variances $\{ \sigma_i^2 \}$. Here, we demonstrate 
the correspondence between our choice for the $( {\bf W}, \boldsymbol{\Sigma}_{\boldsymbol{\xi}})$ pair 
and that of Chechik {\it et al.} \cite{Chechik2005}.

Specifically, suppose the encoding scheme ${\bf z} = {\bf W}^{\rm T} {\bf s} + {\boldsymbol{\xi}}$ 
uses uncorrelated noises that have a fixed variance of 1, i.e. $\boldsymbol{\Sigma}_{\boldsymbol{\xi}} = {\bf I}_{n_{\bf z}}$.
Let us show that a corresponding scheme $\tilde{\bf z} = \tilde{\bf W}^{\rm T} {\bf s} + \tilde{\boldsymbol{\xi}}$ 
consisting of unit encoding vectors $\hat{\bf w}_i = {\bf w}_i / || {\bf w}_i ||$ and distinct encoding noise strengths 
$\{ \tilde{\sigma}_i^2 = 1/|| {\bf w}_i ||^2\}$ yields an encoding that is equally informative about the signal and relevance variables 
as the original scheme. That is, we want to show that $I({\bf z}; {\bf x}) = I(\tilde{\bf z}; {\bf x})$ where ${\bf x}$ can 
be the signal or the relevance variable.

As done previously in this section, we write the mutual information $I({\bf z}; {\bf x})$ as
\begin{align}
I({\bf z}; {\bf x}) = H({\bf x}) - H({\bf x} | {\bf z}).
\end{align}
The equality $I({\bf z}; {\bf x}) = I(\tilde{\bf z}; {\bf x})$ will then hold if 
$H({\bf x} | {\bf z}) = H({\bf x} | \tilde{\bf z})$. And since the considered variables are Gaussian, it would be sufficient 
to show that $\boldsymbol{\Sigma}_{{\bf x} | {\bf z}} = \boldsymbol{\Sigma}_{{\bf x} | \tilde{\bf z}}$.

Applying the Schur complement formula, we can write
\begin{align}
\boldsymbol{\Sigma}_{{\bf x} | \tilde{\bf z}} = \boldsymbol{\Sigma}_{\bf x} - 
\boldsymbol{\Sigma}_{{\bf x}  \tilde{\bf z}}^{\,} \boldsymbol{\Sigma}_{ \tilde{\bf z}}^{-1} \boldsymbol{\Sigma}_{{\bf x} 
 \tilde{\bf z}}^{\rm T}.
\end{align}
Then, from the definition of $\boldsymbol{\Sigma}_{{\bf x}  \tilde{\bf z}}$, we have
\begin{subequations}
\begin{align}
\boldsymbol{\Sigma}_{{\bf x}  \tilde{\bf z}} &= \mathbb{E} \left[ {\bf x} \left( \tilde{\bf W}^{\rm T} {\bf s} 
+ \tilde{\boldsymbol{\xi}} \right)^{\rm T} \right] \\
&= \mathbb{E} \left[ {\bf x} {\bf s}^{\rm T} \tilde{\bf W} \right] \\
&= \boldsymbol{\Sigma}_{{\bf x} {\bf s}} \tilde{\bf W}.
\end{align}
\end{subequations}
Similarly, for $\boldsymbol{\Sigma}_{\tilde{\bf z}}$ we obtain
\begin{subequations}
\begin{align}
\boldsymbol{\Sigma}_{\tilde{\bf z}} &= \mathbb{E} \left[ 
\left( \tilde{\bf W}^{\rm T} {\bf s} + \tilde{\boldsymbol{\xi}} \right)
\left( \tilde{\bf W}^{\rm T} {\bf s} + \tilde{\boldsymbol{\xi}} \right)^{\rm T}
 \right] \\
&= \tilde{\bf W}^{\rm T} \mathbb{E} \left[ {\bf s} {\bf s}^{\rm T} \right] \tilde{\bf W} 
+ \mathbb{E} \left[ \tilde{\boldsymbol{\xi}} \tilde{\boldsymbol{\xi}}^{\rm T}\right] \\
&= \tilde{\bf W}^{\rm T} \boldsymbol{\Sigma}_{\bf s} \tilde{\bf W} + \boldsymbol{\Sigma}_{\tilde{\boldsymbol{\xi}}}.
\end{align}
\end{subequations}
The conditional covariance matrix $\boldsymbol{\Sigma}_{{\bf x} | \tilde{\bf z}}$ then takes the following expanded form:
\begin{align}
\label{eqnA:sigma_cond_x_given_ztilde}
\boldsymbol{\Sigma}_{{\bf x} | \tilde{\bf z}} = 
\boldsymbol{\Sigma}_{\bf x} - \boldsymbol{\Sigma}_{{\bf x} {\bf s}} \, \tilde{\bf W}
\left( \tilde{\bf W}^{\rm T} \boldsymbol{\Sigma}_{\bf s} \tilde{\bf W} + \boldsymbol{\Sigma}_{\tilde{\boldsymbol{\xi}}} \right)^{-1}
\tilde{\bf W}^{\rm T} \, \boldsymbol{\Sigma}_{ {\bf s} {\bf x}}.
\end{align}
Now, the ${\bf W} \rightarrow \tilde{\bf W}$ transformation that normalizes the encoding vectors can be written in a matrix form as
\begin{align}
\tilde{\bf W} = {\bf W} {\bf C},
\end{align}
where ${\bf C} = {\rm diag} (||{\bf w}_1||^{-1}, ..., ||{\bf w}_{n_{\bf z}}||^{-1})$ is a diagonal matrix composed of the inverse 
lengths of the original encoding vectors $\{ {\bf w}_i \}$. Similarly, the form $\tilde{\sigma}_i^2 = 1/||{\bf w}_i||^2$ for 
the transformed noise elements translates to the following matrix expression for the noise covariance matrix:
\begin{align}
\boldsymbol{\Sigma}_{\tilde{\boldsymbol{\xi}}} = {\bf C}^2.
\end{align}

Substituting $\tilde{\bf W}$ and $\boldsymbol{\Sigma}_{\tilde{\boldsymbol{\xi}}}$ into the expression 
for $\boldsymbol{\Sigma}_{{\bf x} | \tilde{\bf z}}$ (Eq.~\ref{eqnA:sigma_cond_x_given_ztilde}) and using 
the fact that ${\bf C} = {\bf C}^{\rm T}$, we find
\begin{subequations}
\begin{align}
\boldsymbol{\Sigma}_{{\bf x} | \tilde{\bf z}} &= \boldsymbol{\Sigma}_{\bf x} 
- \boldsymbol{\Sigma}_{{\bf x} {\bf s}} {\bf W} {\bf C} 
\left( {\bf C} {\bf W} \boldsymbol{\Sigma}_{\bf s} {\bf W} {\bf C} + {\bf C}^2 \right)^{-1}
{\bf C} {\bf W}^{\rm T} \boldsymbol{\Sigma}_{{\bf s} {\bf x}} \\
&= \boldsymbol{\Sigma}_{\bf x} 
- 
\begingroup
      \color{gray}
\underbracket[0.140ex]{\color{black} \boldsymbol{\Sigma}_{{\bf x} {\bf s}} {\bf W}  }_{ 
\boldsymbol{\Sigma}_{{\bf x} {\bf z}}}
\endgroup
\begingroup
      \color{gray}
\underbracket[0.140ex]{\color{black} {\bf C} {\bf C}^{-1} }_{ {\bf I} }
\endgroup
(
\begingroup
      \color{gray}
\underbracket[0.140ex]{\color{black} {\bf W} \boldsymbol{\Sigma}_{\bf s} {\bf W} + {\bf I}_{n_{\bf z}}   }_{ 
\boldsymbol{\Sigma}_{\bf z} }
\endgroup
)^{-1}
\begingroup
      \color{gray}
\underbracket[0.140ex]{\color{black} {\bf C}^{-1} {\bf C}  }_{ {\bf I} }
\endgroup
\begingroup
      \color{gray}
\underbracket[0.140ex]{\color{black} {\bf W}^{\rm T} \boldsymbol{\Sigma}_{{\bf s} {\bf x}} }_{ \boldsymbol{\Sigma}_{{\bf z} {\bf x}} }
\endgroup
\\
&= \boldsymbol{\Sigma}_{\bf x} - \boldsymbol{\Sigma}_{{\bf x} {\bf z}}^{\,} \boldsymbol{\Sigma}_{\bf z}^{-1} 
\boldsymbol{\Sigma}_{{\bf z} {\bf x}}^{\,}.
\end{align}
\end{subequations}
Recognizing the right-hand side as the expanded form of $\boldsymbol{\Sigma}_{{\bf x} | {\bf z}}$, we see that indeed 
$\boldsymbol{\Sigma}_{{\bf x} | {\bf z}} = \boldsymbol{\Sigma}_{{\bf x} | \tilde{\bf z}}$, from which it follows that
$I({\bf z}; {\bf x}) = I(\tilde{\bf z}; {\bf x})$. This shows that the form for the pair 
$({\bf W}, \boldsymbol{\Sigma}_{\boldsymbol{\xi}} )$ considered in our work does not reduce the generality of the problem 
treatment, just like the form considered in Chechik {\it et al.} does not \cite{Chechik2005}.

\newpage

\section{One-dimensional encoding}
\label{sec:AppendixB}

In this Appendix section, 
we derive the results presented in {\bf Sec. III} of the main text related to the scalar encoding scenario.
We first provide details on the optimal encoding direction $\hat{\bf w}_1^*$, 
discuss features of the $I(z; {\bf y})$ vs. $I(z; {\bf s})$ curve,
and then present calculations 
concerning the alternative ``decoding'' perspective on optimality.

\subsection{Optimal encoding direction}
\label{sec:AppendixB1}

In the main text, we considered the scalar encoding of a two-dimensional signal in the form 
$z = \hat{\bf w} \cdot {\bf s} + \xi$, where the unit vector $\hat{\bf w}$ is the encoding direction.
The information encoded in $z$ about the signal ${\bf s}$ is given by
\begin{align}
\label{eqnA:I_zs_1d}
I(z; {\bf s}) = \frac{1}{2} \log \left( \frac{\sigma^2_z}{\sigma^2_{z | {\bf s}}} \right) = 
\frac{1}{2} \log \left( 1 + \frac{r^2}{\sigma^2} \right),
\end{align}
where $r^2$ is the variance of each signal component, and $\sigma^2$ is the variance of encoding noise 
(Eq.~6 of the main text).

Similarly, the relevant information preserved in $z$ is of the form
\begin{align}
\label{eqnA:I_rel_scalar_general}
I(z; {\bf y}) = \frac{1}{2} \log \left( \frac{\sigma^2_z}{\sigma^2_{z | {\bf y}}} \right).
\end{align}
For the variance $\sigma^2_z$, we have
\begin{subequations}
\label{eqnA:sigma_z_step_by_step}
\begin{align}
\sigma^2_z &= \mathbb{E} [z^2] \\
&= \mathbb{E} \left[ \left( \hat{\bf w}^{\rm T} {\bf s} + \xi \right) \left( {\bf s}^{\rm T} \hat{\bf w} + \xi \right) \right] \\
&= \hat{\bf w}^{\rm T} \mathbb{E} \left[ {\bf s} {\bf s}^{\rm T} \right] \hat{\bf w} + \mathbb{E}[\xi^2] \\
\label{eqnA:sigma_z_genericish}
&= \hat{\bf w}^{\rm T} \boldsymbol{\Sigma}_{\bf s} \hat{\bf w} + \sigma^2 \\
&= \hat{\bf w}^{\rm T} \, r^2 {\bf I}_{\bf s} \, \hat{\bf w} + \sigma^2 \\
&= r^2 + \sigma^2.
\end{align}
\end{subequations}
Note that the independence of $\sigma^2_z$ from the encoding direction $\hat{\bf w}$ comes from the standardization 
procedure that results in $\boldsymbol{\Sigma}_{\bf s} = r^2 {\bf I}_{\bf s}$; without it, $\sigma^2_z$ would, in general, 
depend on $\hat{\bf w}$.

Next, following similar steps for $\sigma^2_{z | {\bf y}}$, we write
\begin{subequations}
\label{eqnA:sigma_z_given_y_step_by_step}
\begin{align}
\sigma^2_{z | {\bf y}} &= 
\mathbb{E} \left[ z^2 | {\bf y}\right] - \mathbb{E} \left[ z | {\bf y}\right] ^ 2 \\
&= \mathbb{E} \left[ \left( \hat{\bf w}^{\rm T} {\bf s} + \xi \right) \left( {\bf s}^{\rm T} \hat{\bf w} + \xi \right) \big| \, {\bf y} \right]
- \mathbb{E} \left[ \hat{\bf w}^{\rm T} {\bf s} + \xi \, | \, {\bf y} \right] ^ 2 \\
&= \hat{\bf w}^{\rm T} \left( \mathbb{E} \left[ {\bf s} {\bf s}^{\rm T} | \, {\bf y} \right] 
- \mathbb{E} \left[ {\bf s} | {\bf y} \right] \mathbb{E} \left[ {\bf s}^{\rm T} | {\bf y} \right]
\right)\hat{\bf w} + \sigma^2 \\
&= \hat{\bf w}^{\rm T} \boldsymbol{\Sigma}_{{\bf s} | {\bf y}} \hat{\bf w} + \sigma^2.
\end{align}
\end{subequations}
Substituting the results of Eq.~\ref{eqnA:sigma_z_step_by_step} and Eq.~\ref{eqnA:sigma_z_given_y_step_by_step} 
into the expression of relevant information (Eq.~\ref{eqnA:I_rel_scalar_general}), we obtain
\begin{align}
\label{eqnA:Iz_vecy_via_what}
I(z; {\bf y}) = \frac{1}{2} \log \left( \frac{r^2 + \sigma^2}{
\hat{\bf w}^{\rm T}  \boldsymbol{\Sigma}_{{\bf s} | {\bf y}} \hat{\bf w} + \sigma^2
} \right).
\end{align}
Now, since $\hat{\bf w}$ is a unit vector, 
fixing the encoded information $I(z; {\bf s})$ fixes the encoding noise $\sigma$ as per Eq.~\ref{eqnA:I_zs_1d}.
Maximizing the relevant information $I(z; {\bf y})$ for given $I(z; {\bf s})$ then means minimizing 
$\hat{\bf w}^{\rm T} \boldsymbol{\Sigma}_{{\bf s} | {\bf y} } \hat{\bf w}$ by an optimal choice of the encoding direction $\hat{\bf w}$.
Requiring that $\hat{\bf w}$ is a unit vector ($||\hat{\bf w}|| = 1$), we can write the minimization condition formally as
\begin{align}
\label{eqnA:minimization_1d_Lagrangian}
\frac{\partial}{\partial \hat{\bf w}} \left( 
\hat{\bf w}^{\rm T} \boldsymbol{\Sigma}_{{\bf s} | {\bf y}} \hat{\bf w} - \lambda^2 ||\hat{\bf w}||^2
\right) \Big|_{\hat{\bf w} = \hat{\bf w}^*} = 0,
\end{align}
where $\lambda^2$ is a (positive) Lagrange multiplier that constrains the magnitude of $\hat{\bf w}$ to be unity. 
Performing the differentiation, we obtain
\begin{align}
\left( \boldsymbol{\Sigma}_{{\bf s} | {\bf y}} + \boldsymbol{\Sigma}_{{\bf s} | {\bf y}}^{\rm T} \right) \hat{\bf w}^*
- 2\lambda^2 \hat{\bf w}^* = 0.
\end{align}
Noting that $\boldsymbol{\Sigma}_{{\bf s} | {\bf y}} = \boldsymbol{\Sigma}_{{\bf s} | {\bf y}}^{\rm T}$, we simplify 
this result further into
\begin{align}
\label{eqnA:1d_encoding_evec_condition}
\boldsymbol{\Sigma}_{{\bf s} | {\bf y}} \hat{\bf w}^* = \lambda^2 \hat{\bf w}^*.
\end{align}
Using Eq.~\ref{eqnA:1d_encoding_evec_condition}, 
the minimization objective $\hat{\bf w}^{\rm T} \boldsymbol{\Sigma}_{{\bf s} | {\bf y} } \hat{\bf w}$
in Eqs.~\ref{eqnA:Iz_vecy_via_what} and \ref{eqnA:minimization_1d_Lagrangian} 
evaluated at the optimum $\hat{\bf w} = \hat{\bf w}^*$ becomes
\begin{align}
\label{eqnA:1d_objective_eval_form}
\hat{\bf w}^{\rm T} \boldsymbol{\Sigma}_{{\bf s} | {\bf y} } \hat{\bf w} \, \big|_{\hat{\bf w} = \hat{\bf w}^*} = \lambda^2.
\end{align}
We can thus see that the optimal encoding direction $\hat{\bf w}^*$ must be an eigenvector of 
$\boldsymbol{\Sigma}_{{\bf s} | {\bf y}}$ (Eq.~\ref{eqnA:1d_encoding_evec_condition})
with the smaller corresponding eigenvalue in order minimize 
$\hat{\bf w}^{\rm T} \boldsymbol{\Sigma}_{{\bf s} | {\bf y} } \hat{\bf w}$ (Eq.~\ref{eqnA:1d_objective_eval_form}).
Denoting this smaller eigenvalue as $a^2$, where $a$ is the length of the semi-minor axis of the 
normalized $P({\bf s} | {\bf y})$ ellipse (hence $a < r$),
we write the relevant information under optimal encoding as
\begin{align}
I^{\rm opt}(z; {\bf y}) = \frac{1}{2} \log \left( \frac{r^2 + \sigma^2}{a^2 + \sigma^2} \right),
\end{align}
which is Eq.~10 presented in the main text.

\subsection{Slope of the $I(z; {\bf y})$ vs. $I(z; {\bf s})$ curve}
\label{sec:AppendixB2}

Considering Eq.~\ref{eqnA:I_zs_1d} for the encoded information, we can also write the relevant information as
\begin{align}
I^{\rm opt}(z; {\bf y}) = I(z; {\bf s}) - \frac{1}{2} \log \left( 1 + \frac{a^2}{\sigma^2} \right).
\end{align}
This form makes it clear that $I^{\rm opt} (z; {\bf y}) < I(z; {\bf s})$, 
and that the slope $\partial I^{\rm opt} (z; {\bf y}) / \partial I(z; {\bf s})$ 
of the $I^{\rm opt}(z; {\bf y})$ vs. $I(z; {\bf s})$ curve is always less than 1 (see Fig.~4 of the main text). This slope is
the largest in the $I(z; {\bf s}) \rightarrow 0$ limit of vanishingly small encoded information. To derive it, we approximate 
both information terms under the condition $\sigma \gg r$, obtaining
\begin{align}
I(z; {\bf s}) &\approx \frac{1}{2} \frac{r^2}{\sigma^2}, \\
I^{\rm opt}(z; {\bf y}) &\approx I(z; {\bf s}) - \frac{1}{2} \frac{a^2}{\sigma^2} \nonumber\\
&= (1-\tilde{a}^2) I(z; {\bf s}),
\end{align}
where $\tilde{a} = a/r$. Then for the slope we find
\begin{align}
\frac{\partial I^{\rm opt}(z; {\bf y})}{\partial I(z; {\bf s})} \bigg|_{I(z; {\bf s}) \rightarrow 0} = 1 - \tilde{a}^2.
\end{align}
If instead of the most optimal condition, the least optimal condition were considered (with $\lambda^2 = b^2$), 
the initial slope would be $1 - \tilde{b}^2$, which does not exceed $1 - \tilde{a}^2$ due to $a \le b$.

\subsection{``Decoding'' perspective}
\label{sec:AppendixB3}

Next, we derive the results presented in Sec. III of the main text concerning the alternative ``decoding'' perspective on the problem.
The two key distributions in this perspective are $P({\bf s} | z)$ and $P({\bf y} | z)$. We will start off by deriving expressions 
for the means and variances of these two distributions.

To derive the means, we will use the known result for jointly Gaussian variables ${\bf x}_1$ and ${\bf x}_2$ that states
\begin{align}
\label{eqnA:conditional_mean_general}
\langle {\bf x}_1 | {\bf x}_2 \rangle = \langle {\bf x}_1 \rangle + \boldsymbol{\Sigma}_{{\bf x}_1 {\bf x}_2}^{\,}
\boldsymbol{\Sigma}_{{\bf x}_2}^{-1} \left( {\bf x}_2 - \langle {\bf x}_2 \rangle \right).
\end{align}
This is akin to the Schur complement formula for the covariance matrices:
\begin{align}
\label{eqnA:Schur_general}
\boldsymbol{\Sigma}_{{\bf x}_1 | {\bf x}_2} = \boldsymbol{\Sigma}_{{\bf x}_1} - 
\boldsymbol{\Sigma}_{{\bf x}_1 {\bf x}_2}^{\,} \boldsymbol{\Sigma}_{{\bf x}_2}^{-1} 
\boldsymbol{\Sigma}_{{\bf x}_2 {\bf x}_1}^{\,}.
\end{align}

Starting with the mean for $P({\bf s} | z)$, we write
\begin{align}
\label{eqnA:s_cond_z_1d}
\langle {\bf s} | z \rangle &= \langle {\bf s} \rangle + \boldsymbol{\Sigma}_{{\bf s} z} \sigma^{-2}_z (z - \langle z \rangle).
\end{align}
Now, we know that $\langle {\bf s} \rangle = {\bf 0}$, $\langle z \rangle = 0$, and $\sigma_z^2 = r^2 + \sigma^2$ 
(Eq.~\ref{eqnA:sigma_z_step_by_step}). By definition, we also have
\begin{subequations}
\label{eqnA:Sigma_sz}
\begin{align}
\boldsymbol{\Sigma}_{{\bf s} z} &= \mathbb{E} [{\bf s} \, ({\bf s}^{\rm T} \hat{\bf w} + \xi)] \\
&= \boldsymbol{\Sigma}_{\bf s} \hat{\bf w} \\
&= r^2 \hat{\bf w}.
\end{align}
\end{subequations}
Substituting all these identities into Eq.~\ref{eqnA:s_cond_z_1d}, we obtain
\begin{align}
\label{eqnA:s_mean_given_z_final}
\langle {\bf s} | z \rangle = \frac{r^2}{r^2 + \sigma^2} \hat{\bf w} z.
\end{align}

Similarly, for the covariance of the $P({\bf s} | z)$ distribution, we apply the Schur complement formula and write
\begin{align}
\boldsymbol{\Sigma}_{{\bf s} | z} = \boldsymbol{\Sigma}_{\bf s} - 
\boldsymbol{\Sigma}_{{\bf s} z} \, \sigma^{-2}_z \, \boldsymbol{\Sigma}_{z {\bf s}}.
\end{align}
Substituting $\boldsymbol{\Sigma}_{\bf s} = r^2 {\bf I}_{\bf s}$ and the expressions for 
$\boldsymbol{\Sigma}_{{\bf s} z}$ (Eq.~\ref{eqnA:Sigma_sz})
and $\sigma^2_z$ (Eq.~\ref{eqnA:sigma_z_step_by_step}), we find
\begin{align}
\boldsymbol{\Sigma}_{{\bf s} | z} = r^2 \left( {\bf I}_{\bf s} - \frac{r^2}{r^2 + \sigma^2} \hat{\bf w} \hat{\bf w}^{\rm T} \right).
\end{align}
Together, $\langle {\bf s} | z \rangle$ and $\boldsymbol{\Sigma}_{{\bf s} | z}$ fully specify the distribution $P({\bf s} | z)$.

Note that $\hat{\bf w}$ is an eigenvector of $\boldsymbol{\Sigma}_{{\bf s} | z}$, since
\begin{subequations}
\begin{align}
\boldsymbol{\Sigma}_{{\bf s} | z} \hat{\bf w} &=  
r^2 \left( \hat{\bf w} - \frac{r^2}{r^2 + \sigma^2} \hat{\bf w} 
\underbracket[0.140ex]{\hat{\bf w}^{\rm T} \hat{\bf w}}_{=1} \right)
\\
&= r^2 \left( 1 - \frac{r^2}{r^2 + \sigma^2} \right) \hat{\bf w} \\
&= \frac{r^2 \sigma^2}{r^2 + \sigma^2} \hat{\bf w}.
\end{align}
\end{subequations}
The ellipse corresponding to $P({\bf s} | z)$ is therefore compressed along $\hat{\bf w}$, 
with the length of the semi-minor axis along that direction given by
\begin{align}
\label{eqnA:ell_1d}
\ell = \frac{r \sigma}{\sqrt{r^2 + \sigma^2}}.
\end{align}
The other eigenvector of $\boldsymbol{\Sigma}_{{\bf s} | z}$ is $\hat{\bf w}_\perp$ with a corresponding eigenvalue $r^2$
(follows from a similar calculation involving the identity $\hat{\bf w}^{\rm T} \hat{\bf w}_\perp = 0$),
which implies that the $P({\bf s} | z)$ ellipse is not compressed along the perpendicular direction, as 
reflected in Fig.~5a of the main text.

We next proceed with deriving the mean and the variance of the $P({\bf y} | z)$ distribution in the relevance space.
Analogous to Eq.~\ref{eqnA:s_cond_z_1d} for $\langle {\bf s} | z \rangle$, we write
\begin{align}
\label{eqnA:y_given_z_general}
\langle {\bf y} | z \rangle = \langle {\bf y} \rangle + \boldsymbol{\Sigma}_{{\bf y} z} \sigma_z^{-2} (z - \langle z \rangle).
\end{align}
By definition, we have
\begin{subequations}
\begin{align}
\boldsymbol{\Sigma}_{{\bf y}z} &= \mathbb{E} \left[ {\bf y} \left( {\bf s}^{\rm T} \hat{\bf w} + \xi \right) \right] \\
&= \boldsymbol{\Sigma}_{{\bf y} {\bf s}} \hat{\bf w} \\
\label{eqnA:Sigma_yz_tilde_notation}
&= r^2 \tilde{\boldsymbol{\Sigma}}_{{\bf y} {\bf s}} \hat{\bf w},
\end{align}
\end{subequations}
where in the last step we used 
$\tilde{\boldsymbol{\Sigma}}_{{\bf y} {\bf s}} = \boldsymbol{\Sigma}_{{\bf y} {\bf s}} / r^2$.
Let us here introduce the vector ${\bf v}$ defined as
\begin{align}
\label{eqnA:v_notation}
{\bf v} = \tilde{\boldsymbol{\Sigma}}_{{\bf y} {\bf s}} \hat{\bf w}.
\end{align}
While $\hat{\bf w}$ is a unit vector, ${\bf v}$, in general, is not. To see it, we write its squared magnitude as
\begin{subequations}
\begin{align}
||{\bf v}||^2 &= {\bf v}^{\rm T} {\bf v} \\
&= \hat{\bf w}^{\rm T} \tilde{\boldsymbol{\Sigma}}_{{\bf s} {\bf y}}
\tilde{\boldsymbol{\Sigma}}_{{\bf y} {\bf s}} \hat{\bf w} \\
&= \hat{\bf w}^{\rm T} \big( {\bf I}_{\bf s} - \big( 
{\bf I}_{\bf s}
- \tilde{\boldsymbol{\Sigma}}_{{\bf s} {\bf y}} \tilde{\boldsymbol{\Sigma}}_{{\bf y} {\bf s}}
\big)
\big) \hat{\bf w} \\
&= 1 - \hat{\bf w}^{\rm T} \tilde{\boldsymbol{\Sigma}}_{{\bf s} | {\bf y}} \hat{\bf w},
\end{align}
\end{subequations}
where in the last step we identified 
$\tilde{\boldsymbol{\Sigma}}_{{\bf s} | {\bf y}} = {\bf I}_{\bf s} - 
\tilde{\boldsymbol{\Sigma}}_{{\bf s} {\bf y}} \tilde{\boldsymbol{\Sigma}}_{{\bf y} {\bf s}}$.
Now the square form 
$\hat{\bf w}^{\rm T} \tilde{\boldsymbol{\Sigma}}_{{\bf s} | {\bf y}} \hat{\bf w}$ obeys the following inequlity:
\begin{align}
0 \, \le \, \tilde{a}^2 \, \le \,
\hat{\bf w}^{\rm T} \tilde{\boldsymbol{\Sigma}}_{{\bf s} | {\bf y}} \hat{\bf w}
\, \le \, \tilde{b}^2 \, \le \, 1,
\end{align}
where $\tilde{a}^2 = (a/r)^2$ and $\tilde{b}^2 = (b/r)^2$ are the eigenvalues of the normalized 
covariance matrix $\tilde{\boldsymbol{\Sigma}}_{{\bf s} | {\bf y}}$. From this, it follows that 
$||{\bf v}^2|| = 1 - \hat{\bf w}^{\rm T} \tilde{\boldsymbol{\Sigma}}_{{\bf s} | {\bf y}} \hat{\bf w} \le 1$, 
which shows that indeed the magnitude of the vector ${\bf v}$ is, in general, not $1$.
We can also write tighter bounds on $||{\bf v}^2||$, namely
\begin{align}
\label{eqnA:v_squared_bounds_1d}
1 - \tilde{b}^2 \le ||{\bf v}||^2 \le 1 - \tilde{a}^2.
\end{align}
The upper bound is achieved when $\hat{\bf w} = \hat{\bf w}_1^*$, while the lower bound is reached when 
$\hat{\bf w} = \hat{\bf w}_2^*$.

Now, using the ${\bf v}$-notation (Eq.~\ref{eqnA:v_notation}), 
we write $\boldsymbol{\Sigma}_{{\bf y} z}$ from Eq.~\ref{eqnA:Sigma_yz_tilde_notation} as
\begin{align}
\boldsymbol{\Sigma}_{{\bf y} z} = r^2 {\bf v}.
\end{align}
Substituting $\langle {\bf y} \rangle = {\bf 0}$, $\langle z \rangle = 0$, $\sigma^2_z = r^2 + \sigma^2$, 
and $\boldsymbol{\Sigma}_{{\bf y} z} = r^2 {\bf v}$ into Eq.~\ref{eqnA:y_given_z_general}, we obtain
the final expression for $\langle {\bf y} | z\rangle$:
\begin{align}
\langle {\bf y} | z \rangle = \frac{r^2}{r^2 + \sigma^2} {\bf v} z.
\end{align}
Comparing with the analogous expression for $\langle {\bf s} | z \rangle$ (Eq.~\ref{eqnA:s_mean_given_z_final}), 
we note that $|| \langle {\bf y} | z \rangle || \le || \langle {\bf s} | z \rangle ||$ due to $||{\bf v}|| \le 1$.

Moving on to the covariance matrix of the $P({\bf y} | z)$ distribution, we write it as
\begin{align}
\boldsymbol{\Sigma}_{{\bf y} | z} = \boldsymbol{\Sigma}_{\bf y} 
- \boldsymbol{\Sigma}_{{\bf y} z} \sigma_z^{-2} \boldsymbol{\Sigma}_{z {\bf y}}.
\end{align}
Substituting $\boldsymbol{\Sigma}_{\bf y} = r^2 {\bf I}_{\bf y}$, $\boldsymbol{\Sigma}_{{\bf y} z} = r^2 {\bf v}$,
and $\sigma^2_z = r^2 + \sigma^2$, we find
\begin{align}
\boldsymbol{\Sigma}_{{\bf y} | z} = r^2 \left( {\bf I}_{\bf y} - \frac{r^2}{r^2 + \sigma^2} {\bf v} {\bf v}^{\rm T}\right).
\end{align}
The vectors ${\bf v}$ and ${\bf v}_\perp$ are the two eigenvectors of $\boldsymbol{\Sigma}_{{\bf y} | z}$.
We have $\boldsymbol{\Sigma}_{{\bf y} | z} {\bf v}_\perp = r^2 {\bf v}_\perp$, implying that 
the $P({\bf y} | z)$ distribution is not compressed along ${\bf v}_\perp$. This means that the encoding 
variable $z$ informs about the relevance variable ${\bf y}$ only along the direction ${\bf v}$.
For the eigenvector ${\bf v}$, we have
\begin{subequations}
\begin{align}
\boldsymbol{\Sigma}_{{\bf y} | z} {\bf v} &= r^2 \left( 
{\bf I}_{\bf y} - \frac{r^2}{r^2 + \sigma^2} {\bf v} {\bf v}^{\rm T}
\right) {\bf v} \\
&= r^2 \left( 1 - \frac{r^2}{r^2 + \sigma^2} ||{\bf v}||^2 \right) {\bf v},
\end{align}
\end{subequations}
where we used ${\bf v}^{\rm T} {\bf v} = ||{\bf v}||^2$. 
The eigenvalue form implies that the semi-minor axis length of the $P({\bf y} | z)$ ellipse along the direction ${\bf v}$
is given by 
\begin{align}
\label{eqnA:Lambda_general_via_v}
\Lambda = r \, \sqrt{1 - \frac{r^2}{r^2+ \sigma^2} ||{\bf v}||^2}.
\end{align}
Notably, this length $\Lambda$ depends on the encoding direction $\hat{\bf w}$ 
(via ${\bf v} = \tilde{\boldsymbol{\Sigma}}_{{\bf y} {\bf s}} \hat{\bf w}$), while the semi-minor axis length 
$\ell$ of the $P({\bf s} | z)$ ellipse does not (Eq.~\ref{eqnA:ell_1d}, also see Fig.~5a,b in the main text).

\subsection{Optimality in the ``decoding'' perspective}
\label{sec:AppendixB4}

Now, we discuss what the optimal encoding strategy means in the decoding perspective. 
Our starting point will be the optimality condition derived earlier (Eq.~\ref{eqnA:1d_encoding_evec_condition}), 
which states that the optimal encoding direction $\hat{\bf w}^*_1$ is the eigenvector of 
the conditional covariance matrix $\boldsymbol{\Sigma}_{{\bf s} | {\bf y}}$ with the lower corresponding 
eigenvalue ($a^2$), namely
\begin{align}
\boldsymbol{\Sigma}_{{\bf s} | {\bf y}} \hat{\bf w}_1^* = a^2 \hat{\bf w}_1^*.
\end{align}
Considering the expanded form 
$\boldsymbol{\Sigma}_{{\bf s} | {\bf y}} = r^2 ({\bf I}_{\bf s} - 
\tilde{\boldsymbol{\Sigma}}_{{\bf s} {\bf y}} \tilde{\boldsymbol{\Sigma}}_{{\bf y} {\bf s}} )$ (see Eq.~\ref{eqnA:S_s_given_y_schur_tilded}), 
we can rewrite the above condition as
\begin{align}
r^2 ({\bf I}_{\bf s} - 
\tilde{\boldsymbol{\Sigma}}_{{\bf s} {\bf y}} \tilde{\boldsymbol{\Sigma}}_{{\bf y} {\bf s}} ) 
\hat{\bf w}_1^* = a^2 \hat{\bf w}_1^*.
\end{align}
Then, we multiply both sides by $\tilde{\boldsymbol{\Sigma}}_{{\bf y} {\bf s}}$ and after some rearrangements find
\begin{align}
\label{eqnA:optimality_decoding_1d}
\begingroup
      \color{gray}
\underbracket[0.140ex]{\color{black} r^2 \big( {\bf I}_{\bf y} - \tilde{\boldsymbol{\Sigma}}_{{\bf y} {\bf s}} 
\tilde{\boldsymbol{\Sigma}}_{{\bf s} {\bf y}}  \big)}_{ \boldsymbol{\Sigma}_{{\bf y} | {\bf s}} }
\endgroup
\begingroup
      \color{gray}
\underbracket[0.140ex]{\color{black} \tilde{\boldsymbol{\Sigma}}_{{\bf y} {\bf s}} \hat{\bf w}_1^*  }_{ {\bf v}^*_1 }
\endgroup
= 
a^2 
\begingroup
      \color{gray}
\underbracket[0.140ex]{\color{black} \tilde{\boldsymbol{\Sigma}}_{{\bf y} {\bf s}} \hat{\bf w}_1^*  }_{ {\bf v}^*_1 }
\endgroup
.
\end{align}
Identifying familiar quantities (Eq.~\ref{eqnA:S_y_given_s_schur_tilded} for $\boldsymbol{\Sigma}_{{\bf y} | {\bf s}}$ and 
Eq.~\ref{eqnA:v_notation} for ${\bf v}_1$), we rewrite Eq.~\ref{eqnA:optimality_decoding_1d} in its final form as
\begin{align}
\boldsymbol{\Sigma}_{{\bf y} | {\bf s}} {\bf v}_1^* = a^2 {\bf v}_1^*.
\end{align} 
This form shows that ${\bf v}_1^*$ is the eigenvector of $\boldsymbol{\Sigma}_{{\bf y} | {\bf s}}$ 
with the lowest eigenvalue $a^2$, and hence implies that, under the optimal encoding strategy, 
the direction ${\bf v}_1^*$ along which 
the encoding variable $z$ retains information about the relevance variable ${\bf y}$ 
matches the direction along which the ${\bf s} \rightarrow {\bf y}$ stochastic mapping has the least uncertainty
(see also Fig.~5c in the main text).

As argued earlier (see Eq.~\ref{eqnA:v_squared_bounds_1d} and the surrounding discussion), 
the squared magnitude of ${\bf v}_1^*$ is
\begin{align}
||{\bf v}_1^*||^2 = 1 - \tilde{a}^2.
\end{align}
Substituting $||{\bf v}_1^*||^2$ into Eq.~\ref{eqnA:Lambda_general_via_v} for $\Lambda$, we obtain
\begin{subequations}
\begin{align}
\Lambda_1 &= r \sqrt{1 - \frac{r^2}{r^2 + \sigma^2} ||{\bf v}_1^*||^2} \\
&= r \sqrt{1 - \frac{r^2}{r^2 + \sigma^2} (1 - \tilde{a}^2)} \\
&= r \sqrt{\frac{a^2 + \sigma^2}{r^2 + \sigma^2}}.
\end{align}
\end{subequations}
Note that in the limit of noiseless encoding ($\sigma \rightarrow 0$), the $P({\bf y} | z)$ ellipse is maximally 
compressed along the minor axis of the $P({\bf y} | {\bf s})$ ellipse, with its semi-axis length along that 
direction being $\Lambda_1 \rightarrow a$ (see Fig.~5c of the main text).
In this limit, the relevant information is the highest for a scalar encoding strategy and equals 
$I_{\rm max}(z; {\bf y}) = -\log \tilde{a}$.

With a similar set of arguments, we can also consider the least optimal encoding scenario where the 
encoding direction $\hat{\bf w}_2^*$ is the eigenvector of $\boldsymbol{\Sigma}_{{\bf s} | {\bf y}}$ 
with the {\it larger} corresponding eigenvalue $b^2$, namely
\begin{align}
\boldsymbol{\Sigma}_{{\bf s} | {\bf y}} \hat{\bf w}_2^* = b^2 \hat{\bf w}_2^*.
\end{align}
The corresponding ${\bf v}_2$ vector is similarly an eigenvector of $\boldsymbol{\Sigma}_{{\bf y} | {\bf s}}$
with the larger corresponding eigenvalue:
\begin{align}
\boldsymbol{\Sigma}_{{\bf y} | {\bf s}} {\bf v}_2^* = b^2 {\bf v}_2^*.
\end{align}
It is perpendicular to ${\bf v}_1^*$ and has a squared magnitude given by
\begin{align}
||{\bf v}_2^*||^2 = 1 - \tilde{b}^2.
\end{align}
Substituting $||{\bf v}_2^*||^2$ into Eq.~\ref{eqnA:Lambda_general_via_v}, we obtain 
the semi-minor axis length of the $P({\bf y} | z)$ distribution under the least optimal strategy to be
\begin{subequations}
\begin{align}
\Lambda_2 &= r \sqrt{1 - \frac{r^2}{r^2 + \sigma^2} ||{\bf v}_2^*||^2} \\
&= r \sqrt{1 - \frac{r^2}{r^2 + \sigma^2} (1 - \tilde{b}^2)} \\
&= r \sqrt{ \frac{b^2 + \sigma^2}{r^2 + \sigma^2} }.
\end{align}
\end{subequations}
In the limit $\sigma \rightarrow 0$, the $P({\bf y} | z)$ ellipse is maximally compressed along the {\it least} 
informative direction ${\bf v}_2^*$ of the stochastic ${\bf s} \rightarrow {\bf y}$ mapping, with $\Lambda_2$
approaching $b$ -- the semi-major axis length of the $P({\bf y} | {\bf s})$ ellipse (Fig.~5c of the main text).

\newpage
\section{Two-dimensional encoding}
\label{sec:AppendixC}

In this Appendix section, we provide details on the scenario where the signal is encoded in two distinct components
({\bf Sec. IV} of the main text).
Specifically, we derive the statistics of the decoding distributions $P({\bf s} | {\bf z})$ and $P({\bf y} | {\bf z})$, 
derive the condition of transitioning from scalar to 2D optimal encoding, 
and discuss the behavior beyond the transition point.

\subsection{Statistics of $P({\bf s} | {\bf z})$ and $P({\bf y} | {\bf z})$ distributions}
\label{sec:AppendixC1}

To compute the means and covariance matrices of the two conditional distributions
$P({\bf s} | {\bf z})$ and $P({\bf y} | {\bf z})$, we will use the same general formulae 
introduced in the scalar encoding case (Eq.~\ref{eqnA:Schur_general} and 
Eq.~\ref{eqnA:conditional_mean_general}). Starting with the mean of $P({\bf s} | {\bf z})$, we write
\begin{align}
\label{eqnA:s_cond_z_vector_general}
\langle {\bf s} | {\bf z} \rangle &= \langle {\bf s} \rangle + \boldsymbol{\Sigma}_{{\bf s} {\bf z}}^{\,} 
\boldsymbol{\Sigma}_{\bf z}^{-1} ({\bf z} - \langle {\bf z} \rangle).
\end{align}
Recalling the general form ${\bf z} = {\bf W}^{\rm T} {\bf s} + \boldsymbol{\xi}$, we write the covariance 
matrix $\boldsymbol{\Sigma}_{\bf z}$ as
\begin{subequations}
\begin{align}
\boldsymbol{\Sigma}_{\bf z} &= \mathbb{E} \left[ {\bf z} {\bf z}^{\rm T} \right] \\
&= \mathbb{E} \left[ 
\big( {\bf W}^{\rm T} {\bf s} + \boldsymbol{\xi} \big) \big( {\bf s}^{\rm T} {\bf W} + \boldsymbol{\xi}^{\rm T} \big) 
\right] \\
&= {\bf W}^{\rm T} \mathbb{E} \left[ {\bf s} {\bf s}^{\rm T} \right] {\bf W} + 
\mathbb{E} \left[ \boldsymbol{\xi} \boldsymbol{\xi}^{\rm T} \right] \\
&= {\bf W}^{\rm T} \boldsymbol{\Sigma}_{\bf s} {\bf W} + \boldsymbol{\Sigma}_{\boldsymbol{\xi}} \\
&= r^2 {\bf W}^{\rm T} {\bf W} + \boldsymbol{\Sigma}_{\boldsymbol{\xi}}.
\end{align}
\end{subequations}
Similarly, for the cross-correlation matrix $\boldsymbol{\Sigma}_{{\bf s} {\bf z}}$ we have
\begin{subequations}
\begin{align}
\boldsymbol{\Sigma}_{{\bf s} {\bf z}} &= \mathbb{E} \left[  {\bf s} {\bf z}^{\rm T} \right] \\
&= \mathbb{E} \left[ {\bf s} \, \big( {\bf s}^{\rm T} {\bf W} + \boldsymbol{\xi}^{\rm T} \big) \right] \\
&= \mathbb{E} \left[ {\bf s} {\bf s}^{\rm T} \right] {\bf W} \\
&= r^2 {\bf W}.
\end{align}
\end{subequations}
Substituting these two matrix expressions into Eq.~\ref{eqnA:s_cond_z_vector_general} 
and using $\langle {\bf s} \rangle = {\bf 0}$, $\langle {\bf z} \rangle = {\bf 0}$, we obtain
\begin{align}
\label{eqnA:s_given_z_vector_intermediate}
\langle {\bf s} | {\bf z} \rangle = {\bf W} \left( {\bf W}^{\rm T} {\bf W} + 
r^{-2} \boldsymbol{\Sigma}_{\boldsymbol{\xi}} \right)^{-1} {\bf z}.
\end{align}
Then, for the conditional covariance matrix $\boldsymbol{\Sigma}_{{\bf s} | {\bf z}}$, we apply the Schur 
complement formula ($\boldsymbol{\Sigma}_{{\bf s} | {\bf z}} = 
\boldsymbol{\Sigma}_{\bf s} - \boldsymbol{\Sigma}_{{\bf s} {\bf z}}^{\,} \boldsymbol{\Sigma}_{\bf z}^{-1}
\boldsymbol{\Sigma}_{{\bf z} {\bf s}}^{\,}$) and again substituting the expressions of matrices, find
\begin{align}
\label{eqnA:Sigma_s_given_z_vec_intermediate}
\boldsymbol{\Sigma}_{{\bf s} | {\bf z}} = r^2 \left( 
{\bf I}_{\bf s} - {\bf W} \left( 
{\bf W}^{\rm T} {\bf W} + r^{-2} \boldsymbol{\Sigma}_{\boldsymbol{\xi}}
\right)^{-1} {\bf W}^{\rm T}
\right).
\end{align}

Let us now derive similar general expressions for the statistics of the $P({\bf y} | {\bf z})$ distribution. 
For the mean, we have
\begin{align}
\label{eqnA:y_given_z_vector_general}
\langle {\bf y} | {\bf z} \rangle = \langle {\bf y} \rangle + 
\boldsymbol{\Sigma}_{{\bf y} {\bf z}}^{\,} \boldsymbol{\Sigma}_{\bf z}^{-1} ({\bf z} - \langle {\bf z} \rangle).
\end{align}
From the definition, 
the cross-correlation matrix $\boldsymbol{\Sigma}_{{\bf y} {\bf z}}$ is
\begin{subequations}
\begin{align}
\boldsymbol{\Sigma}_{{\bf y} {\bf z}} &= \mathbb{E} \left[ {\bf y} {\bf z}^{\rm T} \right] \\
&= \mathbb{E} \left[ 
{\bf y} \big( {\bf s}^{\rm T} {\bf W} + \boldsymbol{\xi}^{\rm T} \big)
\right] \\
&= \mathbb{E} \left[ {\bf y} {\bf s}^{\rm T} \right] {\bf W} \\
&= r^2 \tilde{\boldsymbol{\Sigma}}_{{\bf y} {\bf s}} {\bf W},
\end{align}
\end{subequations}
where in the last line we used the identity $\boldsymbol{\Sigma}_{{\bf y} {\bf s}} = r^2 \tilde{\boldsymbol{\Sigma}}_{{\bf y} {\bf s}}$.
Substituting $\boldsymbol{\Sigma}_{{\bf y} {\bf z}}$ and $\boldsymbol{\Sigma}_{\bf z}$
into Eq.~\ref{eqnA:y_given_z_vector_general}, 
together with $\langle {\bf y} \rangle = {\bf 0}$ and $\langle {\bf z} \rangle = {\bf 0}$, we obtain
\begin{align}
\label{eqnA:y_given_z_vector_intermediate}
\langle {\bf y} | {\bf z} \rangle = 
\tilde{\boldsymbol{\Sigma}}_{{\bf y} {\bf s}} {\bf W}
\left( 
{\bf W}^{\rm T} {\bf W} + r^{-2} \boldsymbol{\Sigma}_{\boldsymbol{\xi}}
\right)^{-1} {\bf z}.
\end{align}
Then for the covariance matrix, again using the Schur complement formula
($\boldsymbol{\Sigma}_{{\bf y} | {\bf z}} = 
\boldsymbol{\Sigma}_{\bf y} - \boldsymbol{\Sigma}_{{\bf s} {\bf z}}^{\,} \boldsymbol{\Sigma}_{\bf z}^{-1}
\boldsymbol{\Sigma}_{{\bf z} {\bf y}}^{\,}$), we find
\begin{align}
\label{eqnA:Sigma_y_given_z_vector_intermediate}
\boldsymbol{\Sigma}_{{\bf y} | {\bf z}} = r^2 \left( 
{\bf I}_{\bf y} - \tilde{\boldsymbol{\Sigma}}_{{\bf y} {\bf s}} {\bf W} \left( 
{\bf W}^{\rm T} {\bf W} + r^{-2} \boldsymbol{\Sigma}_{\boldsymbol{\xi}}
\right)^{-1} {\bf W}^{\rm T} \tilde{\boldsymbol{\Sigma}}_{{\bf s} {\bf y}}
\right).
\end{align}

These expressions for the different statistics are generic and hold for any choice of 
the pair $({\bf W}, \boldsymbol{\Sigma}_{\boldsymbol{\xi}})$. Recall that in our formulation of the problem
(section II of the main text),
${\bf W} = [\hat{\bf w}_1, \hat{\bf w}_2, ...]$ is set to be a collection of distinct encoding directions, 
while 
$\boldsymbol{\Sigma}_{\boldsymbol{\xi}}$ is considered to be a diagonal matrix of encoding noises, i.e. 
$\boldsymbol{\Sigma}_{\boldsymbol{\xi}}^{ij} = \sigma_i^2 \, \delta_{ij}$.
To focus on the question of assigning the noise strengths $\{ \sigma_i^2 \}$ under optimal encoding, 
in section IV of the main text we set the encoding directions $\hat{\bf w}_1$ and $\hat{\bf w}_2$ to be, 
respectively, 
the direction $\hat{\bf w}_1^*$ optimal in the 1D encoding setting, and the direction $\hat{\bf w}_2^*$ 
perpendicular to it (Fig.~3b).
While there are other choices of $\{ \hat{\bf w}_i \}$ that yield an equally optimal solution to the bottleneck problem, 
the choice $\hat{\bf w}_i = \hat{\bf w}_i^*$ under which $\{ z_i \}$ encode independent components of the 
signal is the most intuitive for illustrating how information from these multiple components is combined 
and how noise should be optimally assigned to each component.
We therefore proceed with this section having $\hat{\bf w}_i = \hat{\bf w}_i^*$ as a starting point, 
leaving the more general discussion of the degenerate space of solutions to Appendix D.

The $ij^{\rm th}$ component of the matrix product ${\bf W}^{\rm T} {\bf W}$, which is generally given by  
$\hat{\bf w}_i \cdot \hat{\bf w}_j$, reduces to $\delta_{ij}$ under the choice $\hat{\bf w}_i = \hat{\bf w}_i^*$
because the principal direction $\hat{\bf w}_1^*$ and $\hat{\bf w}_2^*$ are perpendicular. This implies that
\begin{align}
\label{eqnA:W_orthonormality}
{\bf W}^{\rm T} {\bf W} = {\bf I}_{\bf s}.
\end{align}
We now revisit the general expressions of the means and covariance matrices, having the above identity in mind.
The term $({\bf W}^{\rm T} {\bf W} + r^{-2} \boldsymbol{\Sigma}_{\boldsymbol{\xi}})^{-1}$ appearing in all 
of the four expressions 
(Eqs.~
\ref{eqnA:s_given_z_vector_intermediate}, \ref{eqnA:Sigma_s_given_z_vec_intermediate}, \ref{eqnA:y_given_z_vector_intermediate}, and
\ref{eqnA:Sigma_y_given_z_vector_intermediate})
can be rewritten as
\begin{subequations}
\label{eqnA:the_inverse_term}
\begin{align}
\left( {\bf W}^{\rm T} {\bf W} + r^{-2} \boldsymbol{\Sigma}_{\boldsymbol{\xi}} \right)^{-1} &=  
({\bf I}_{\bf s} + r^{-2} \boldsymbol{\Sigma}_{\boldsymbol{\xi}})^{-1} \\
&= 
\begin{bmatrix}
1 + \dfrac{\sigma_1^2}{r^2} & 0 \\
0 & 1 + \dfrac{\sigma_2^2}{r^2}
\end{bmatrix}^{-1} \\
&= 
\begin{bmatrix}
\dfrac{r^2}{r^2 + \sigma_1^2} & 0 \\
0 & \dfrac{r^2}{r^2 + \sigma_2^2}
\end{bmatrix}.
\end{align}
\end{subequations}
Substituting it into the expression for $\langle {\bf s} | {\bf z} \rangle$ 
(Eq.~\ref{eqnA:s_given_z_vector_intermediate}), we obtain
\begin{subequations}
\begin{align}
\langle {\bf s} | {\bf z} \rangle &= 
[\hat{\bf w}_1, \hat{\bf w}_2] 
\begin{bmatrix}
\dfrac{r^2}{r^2 + \sigma_1^2} & 0 \\
0 & \dfrac{r^2}{r^2 + \sigma_2^2}
\end{bmatrix}
\begin{bmatrix}
z_1 \\
z_2
\end{bmatrix} \\
&= 
\sum_{i \in \{1, 2\}} 
\frac{r^2}{r^2 + \sigma_i^2} \hat{\bf w}_i z_i.
\end{align}
\end{subequations}
Similarly, substituting Eq.~\ref{eqnA:the_inverse_term} into Eq.~\ref{eqnA:Sigma_s_given_z_vec_intermediate} for 
$\boldsymbol{\Sigma}_{{\bf s} | {\bf z}}$, we find
\begin{subequations}
\begin{align}
\boldsymbol{\Sigma}_{{\bf s} | {\bf z}} &= r^2 \left( {\bf I}_{\bf s}
- [\hat{\bf w}_1, \hat{\bf w}_2] 
\begin{bmatrix}
\dfrac{r^2}{r^2 + \sigma_1^2} & 0 \\
0 & \dfrac{r^2}{r^2 + \sigma_2^2}
\end{bmatrix}
\begin{bmatrix}
\hat{\bf w}_1^{\rm T} \\
\hat{\bf w}_2^{\rm T}
\end{bmatrix}
\right) \\
&= r^2 \bigg( {\bf I}_{\bf s} 
- \sum_{i \in \{ 1, 2\}} \frac{r^2}{r^2 + \sigma_i^2} \hat{\bf w}_i^{\,}  \hat{\bf w}_i^{\rm T}
\bigg).
\end{align}
\end{subequations}
Note that $\hat{\bf w}_1$ and $\hat{\bf w}_2$ 
(satisfying the orthonormality condition $\hat{\bf w}_i^{\rm T} \hat{\bf w}_j^{\,} = \delta_{ij}$ (Eq.~\ref{eqnA:W_orthonormality}))
are the two eigenvectors of $\boldsymbol{\Sigma}_{{\bf s} | {\bf z}}$; hence
$\boldsymbol{\Sigma}_{{\bf s} | {\bf z}} \hat{\bf w}_i = \ell_i^2 \hat{\bf w}_i$.
Here, $\{ \ell_i \}$ represent
the semi-axis lengths of the $P({\bf s} | {\bf z})$ ellipse, which is now compressed along both $\hat{\bf w}_1$
and $\hat{\bf w}_2$ (see Fig.~6a in the main text). The length $\ell_i$ is given by
\begin{align}
\label{eqnA:ell_i_general_expression}
\ell_i = \frac{r \sigma_i}{\sqrt{r^2 + \sigma_i^2}}.
\end{align}

We next derive the final expressions for the statistics of $P({\bf y} | {\bf z})$. In their general formulae
(Eq.~\ref{eqnA:y_given_z_vector_intermediate} for $\langle {\bf y} | {\bf z} \rangle$
and Eq.~\ref{eqnA:Sigma_y_given_z_vector_intermediate} for $\boldsymbol{\Sigma}_{{\bf y} | {\bf z}}$), 
the term $\tilde{\boldsymbol{\Sigma}}_{{\bf y} {\bf s}} {\bf W}$ appears. Denoting it by ${\bf V}$, we write 
\begin{subequations}
\begin{align}
{\bf V} &= \tilde{\boldsymbol{\Sigma}}_{{\bf y} {\bf s}} {\bf W} \\
&= 
[\tilde{\boldsymbol{\Sigma}}_{{\bf y} {\bf s}} \hat{\bf w}_1, \tilde{\boldsymbol{\Sigma}}_{{\bf y} {\bf s}} \hat{\bf w}_2] \\
&= [{\bf v}_1, {\bf v}_2],
\end{align}
\end{subequations}
where ${\bf v}_i = \tilde{\boldsymbol{\Sigma}}_{{\bf y} {\bf s}} \hat{\bf w}_i$. Using this notation
in Eq.~\ref{eqnA:y_given_z_vector_intermediate}, we obtain
\begin{subequations}
\begin{align}
\langle {\bf y} | {\bf z} \rangle &= 
[{\bf v}_1, {\bf v}_2] 
\begin{bmatrix}
\dfrac{r^2}{r^2 + \sigma_1^2} & 0 \\
0 & \dfrac{r^2}{r^2 + \sigma_2^2}
\end{bmatrix}
\begin{bmatrix}
z_1 \\
z_2
\end{bmatrix} \\
&= 
\sum_{i \in \{1, 2\}} 
\frac{r^2}{r^2 + \sigma_i^2} {\bf v}_i z_i.
\end{align}
\end{subequations}
Similarly, the covariance matrix $\boldsymbol{\Sigma}_{{\bf y} | {\bf z}}$ becomes
\begin{subequations}
\begin{align}
\boldsymbol{\Sigma}_{{\bf y} | {\bf z}} &= 
r^2 \left( 
{\bf I}_{\bf y} - {\bf V} \left( 
{\bf W}^{\rm T} {\bf W} + r^{-2} \boldsymbol{\Sigma}_{\boldsymbol{\xi}}
\right)^{-1} {\bf V}^{\rm T} \right)
\\
&= r^2 \left( {\bf I}_{\bf y}
- [{\bf v}_1, {\bf v}_2] 
\begin{bmatrix}
\dfrac{r^2}{r^2 + \sigma_1^2} & 0 \\
0 & \dfrac{r^2}{r^2 + \sigma_2^2}
\end{bmatrix}
\begin{bmatrix}
{\bf v}_1^{\rm T} \\
{\bf v}_2^{\rm T}
\end{bmatrix}
\right) \\
&= r^2 \bigg( {\bf I}_{\bf y} 
- \sum_{i \in \{ 1, 2\}} \frac{r^2}{r^2 + \sigma_i^2} {\bf v}_i^{\,}  {\bf v}_i^{\rm T}
\bigg).
\end{align}
\end{subequations}
Under optimal encoding ($\hat{\bf w}_i^{\,} = \hat{\bf w}_i^*$), the vectors $\{ {\bf v}_i \}$ are 
orthogonal (${\bf v}_i^{\rm T} {\bf v}_j^{\,} = ||{\bf v}_i ||^2 \delta_{ij}$) and represent the eigenvectors 
of the covariance matrix $\boldsymbol{\Sigma}_{{\bf y} | {\bf s}}$ (see Appendix B4, also Fig.~5c in the main text).
They are also the eigenvectors of the matrix $\boldsymbol{\Sigma}_{{\bf y} | {\bf z}}$ above, satisfying 
$\boldsymbol{\Sigma}_{{\bf y} | {\bf z}} {\bf v}_i = \Lambda_i^2 {\bf v}_i$, where
$\{ \Lambda_i \}$ are the semi-axis lengths of the $P({\bf y} | {\bf z})$ ellipse (Fig.~6b), 
with their expressions given by
\begin{subequations}
\label{eqnA:Lambda_i_final_expressions}
\begin{align}
\Lambda_i &= r \sqrt{1 - \frac{r^2}{r^2 + \sigma^2} ||{\bf v}_i||^2} \\
& = r \sqrt{ \frac{\lambda_i^2 + \sigma_i^2}{r^2 + \sigma_i^2}}.
\end{align}
\end{subequations}
Here we used the identity $||{\bf v}_i||^2 = 1 - (\lambda/r)^2$ (with $\lambda_i \in \{a, b\}$), 
already derived in Appendix~B4.

\subsection{Transition from scalar to two-dimensional optimal encoding}
\label{sec:AppendixC2}

We now proceed to discuss the question of optimally assigning the encoding noises 
$\sigma_1$ and $\sigma_2$ for different amounts of fixed encoded information.
As we will show, the optimal assignment will indicate a transition from scalar to two-dimensional 
encoding when the amount of encoded information becomes sufficiently large.

Fixing the encoded information $I({\bf z}; {\bf s})$ amounts to fixing the area of the $P({\bf s} | {\bf z})$ ellipse
(see Fig.~6a), which, up to a constant multiple, is given by $\ell_1 \times \ell_2$, where $\ell_1$ and $\ell_2$
are the semi-axis lengths of the ellipse (Eq.~\ref{eqnA:ell_i_general_expression}). The exact relation between 
the encoded information and these lengths is
\begin{subequations}
\begin{align}
I({\bf z}; {\bf s}) &= \log \left( \frac{r^2}{\ell_1 \times \ell_2} \right) \\
&= \frac{1}{2} \log \left( 1 + \frac{r^2}{\sigma_1^2} \right) \left( 1 + \frac{r^2}{\sigma_2^2} \right),
\end{align}
\end{subequations}
where in the last step we substituted the expressions of $\ell_i$ (Eq.~\ref{eqnA:ell_i_general_expression}).

Similarly, the relevant information $I({\bf z}; {\bf y})$ is set by the area of the $P({\bf z}; {\bf y})$ ellipse
(Fig.~6b), the semi-axis lengths of which are $\Lambda_1$ and $\Lambda_2$ 
(Eq.~\ref{eqnA:Lambda_i_final_expressions}). The exact relation between $I({\bf z}; {\bf y})$ and these lengths is
\begin{subequations}
\begin{align}
I({\bf z}; {\bf y}) &= \log \left( \frac{r^2}{\Lambda_1 \times \Lambda_2} \right) \\
&= \frac{1}{2} \log \left( \frac{r^2 + \sigma_1^2}{a^2 + \sigma_1^2} \right)
\left( \frac{r^2 + \sigma_2^2}{b^2 + \sigma_2^2} \right),
\end{align}
\end{subequations}
where in the last step we substituted the expressions of $\Lambda_i$ (Eq.~\ref{eqnA:Lambda_i_final_expressions})
with $\lambda_i \in \{ a, b \}$ representing the dimensions of the $P({\bf y} | {\bf s})$ ellipse.

Now, fixing the encoded information $I({\bf z}; {\bf s})$ means fixing the product
\begin{align}
\label{eqnA:Omega_definition}
\Omega^2 = \left( 1 + \frac{r^2}{\sigma_1^2} \right) \left( 1 + \frac{r^2}{\sigma_2^2} \right),
\end{align}
where we introduced $\Omega= 2^{I({\bf z}; {\bf s})}$ for convenience. Note that since both terms on the right-hand
side are greater than or equal to $1$, the variable $\Omega$ must satisfy the constraint
\begin{align}
\Omega \ge 1.
\end{align}
For given $\Omega$, we want to maximize $I({\bf z}; {\bf y})$ which we can write as
\begin{subequations}
\begin{align}
I({\bf z}; {\bf y}) &= 
\frac{1}{2} \log \left( 
\frac{1 + r^2/\sigma_1^2}{1 + a^2 / \sigma_1^2} 
\right)
\left(
\frac{1 + r^2/\sigma_2^2}{1 + b^2 / \sigma_2^2}
\right) \\
\label{eqnA:I_rel_given_Omega}
&= \frac{1}{2} \log \frac{\Omega^2}{(1 + a^2/\sigma_1^2)(1 + b^2/\sigma_2^2)}.
\end{align}
\end{subequations}
For convenience, we now introduce
\begin{align}
\label{eqnA:theta_definition}
\theta_i = r^2/\sigma_i^2.
\end{align}
The fixed $\Omega$ constraint
in Eq.~\ref{eqnA:Omega_definition} can be rewritten as
\begin{align}
\label{eqnA:Omega2_simple}
\Omega^2 = \left(1 + \theta_1 \right) \left(1 + \theta_2 \right).
\end{align}
Maximizing $I({\bf z}; {\bf y})$ for given $\Omega$ (see Eq.~\ref{eqnA:I_rel_given_Omega})
then becomes equivalent to minimizing 
\begin{align}
\label{eqnA:h_theta1_theta2}
h(\theta_1, \theta_2) = \big( 1 + \tilde{a}^2 \theta_1 \big) \big( 1 + \tilde{b}^2 \theta_2 \big),
\end{align}
where $\tilde{a} = a/r$ and $\tilde{b} = b/r$. Expressing $\theta_2$ in Eq.~\ref{eqnA:Omega2_simple} as
\begin{align}
\label{eqnA:theta2_intermsof_Omega_theta1}
\theta_2 = \frac{\Omega^2}{1 + \theta_1} -1
\end{align}
and substituting into Eq.~\ref{eqnA:h_theta1_theta2}, we obtain
\begin{align}
h(\theta_1 | \Omega) = \tilde{a}^2 \tilde{b}^2 \Omega^2 +
\big( 1 - \tilde{b}^2  \big) \big( 1 + \tilde{a}^2 \theta_1 \big)
+ \Omega^2 \frac{\tilde{b}^2 (1 - \tilde{a}^2)}{1 + \theta_1}.
\end{align}

The optimal $\theta_1$ can be found by setting $\partial h_1/\partial \theta_1 = 0$ and solving for 
$\theta_1$, which yields
\begin{align}
\theta_1^* = \Omega \sqrt{\frac{\tilde{b}^2 (1 - \tilde{a}^2)}{\tilde{a}^2(1 - \tilde{b}^2)}} - 1.
\end{align}
Note that since $\partial^2 h/\partial \theta_1^2 > 0$, $\theta_1^*$ minimizes $h(\theta_1 | \Omega)$, or, 
equivalently, maximizes $I({\bf z}; {\bf y})$ for the given $\Omega$.
Substituting $\theta_1^*$ into Eq.~\ref{eqnA:theta2_intermsof_Omega_theta1}, 
we find the corresponding optimal $\theta_2^*$ as
\begin{align}
\theta_2^* = \Omega \sqrt{\frac{\tilde{a}^2 (1 - \tilde{b}^2)}{\tilde{b}^2(1 - \tilde{a}^2)}} - 1.
\end{align}
While $\theta_1^*$ is guaranteed to be non-negative (due to $a \le b$), the same condition will hold for $\theta_2^*$ 
only when $\Omega \ge \Omega_c$, where the critical value $\Omega_c$ is 
\begin{align}
\label{eqnA:Omega_c_expression}
\Omega_c = \sqrt{ \frac{\tilde{b}^2 ( 1 - \tilde{a}^2 )}{\tilde{a}^2 (1 - \tilde{b}^2 )}  }.
\end{align}
The value $\Omega_c$ corresponds to the critical amount of encoded information 
$I^\dagger({\bf z}; {\bf s})$ via
\begin{align}
\label{eqnA:I_thresh_1st}
I^\dagger({\bf z}; {\bf s}) = \log \Omega_c.
\end{align}
Using two encoding components becomes preferred over scalar encoding only when the amount of encoded 
information exceeds the threshold $I^\dagger({\bf z}; {\bf y})$. 

With the $\Omega_c$ notation, the optimal $\theta_1$ and $\theta_2$ expressions 
in the $\Omega \ge \Omega_c$ regime can be written as
\begin{subequations}
\label{eqnA:theta12_star}
\begin{align}
\label{eqnA:theta1_star}
\theta_1^* &= \Omega \times \Omega_c - 1, \\
\label{eqnA:theta2_star}
\theta_2^* &= \Omega \div \Omega_c - 1.
\end{align}
\end{subequations}
Recalling that $\theta_i = r^2/\sigma_i^2$ (Eq.~\ref{eqnA:theta_definition}),
we then obtain the optimal noises $\sigma_1^*$ and $\sigma_2^*$ as
\begin{subequations}
\begin{align}
\label{eqnA:sigma_star_1}
\big( \sigma_1^* \big)^2 &= \frac{r^2}{\Omega \times \Omega_c - 1}, \\
\label{eqnA:sigma_star_2}
\big( \sigma_2^* \big)^2 &= \frac{r^2}{\Omega \div \Omega_c - 1}.
\end{align}
\end{subequations}

At the transition point ($\Omega = \Omega_c$), we have $\sigma_2^\dagger \rightarrow \infty$, meaning that 
the second component is not yet used, and for the noise of the first component we find
\begin{subequations}
\begin{align}
\label{eqnA:sigma1_dagger}
\sigma_1^\dagger &= \frac{r}{\sqrt{\Omega_c^2 - 1}} \\
&= \sqrt{\frac{r^2 - b^2}{(b/a)^2 - 1}},
\end{align}
\end{subequations}
where in the last step we substituted the expression of $\Omega_c$ from Eq.~\ref{eqnA:Omega_c_expression}.

Let us now compute the semi-minor axis lengths of $P({\bf s} | {\bf z})$ and $P({\bf y} | {\bf z})$ ellipses 
at the first transition point. Starting with $\ell_1^\dagger$ for $P({\bf s} | {\bf z})$ (see Eq.~\ref{eqnA:ell_i_general_expression}), we have
\begin{subequations}
\begin{align}
\ell_1^\dagger &= \frac{r \sigma_1^\dagger}{\sqrt{r^2 + (\sigma_1^\dagger)^2}} \\
&= \frac{r}{\sqrt{1 + \theta_1^\dagger}} \\
\label{eqnA:ell1_dagger}
&= \frac{r}{\Omega_c}.
\end{align}
\end{subequations}
Here we used the definition $\theta_1 = r^2/\sigma_1^2$ (Eq.~\ref{eqnA:theta_definition}) and the identity $\theta_1^\dagger = \Omega_c^2 - 1$
that follows from Eq.~\ref{eqnA:theta1_star} evaluated at $\Omega = \Omega_c$. We can also rewrite the above result in terms of the threshold 
encoded information (Eq.~\ref{eqnA:I_thresh_1st}) as
\begin{align}
\ell_1^\dagger = r \, 2^{-I^\dagger({\bf z}; {\bf s})}.
\end{align}
Next, we derive an expression for $\Lambda_1^\dagger$ -- the semi-minor axis of $P({\bf y} | {\bf z})$ at 
the transition point. From Eq.~\ref{eqnA:Lambda_i_final_expressions}, we have
\begin{subequations}
\begin{align}
\Lambda_1^\dagger &= r \sqrt{\frac{a^2 + (\sigma_1^\dagger)^2}{r^2 + (\sigma_1^\dagger)^2}} \\
\label{eqnA:Lambda_1_dagger}
&= r \, \frac{a}{b},
\end{align}
\end{subequations}
where we skipped the simplification steps that follow after substituting $\sigma_1^\dagger$ 
(Eq.~\ref{eqnA:sigma1_dagger}).
Recalling that at the transition point the $P({\bf s} | {\bf z})$ and $P({\bf y} | {\bf z})$ ellipses are not compressed
along their major axes, i.e. $\ell_2^\dagger = \Lambda_2^\dagger = r$, we can rewrite the two geometric 
conditions (Eq.~\ref{eqnA:ell1_dagger} and Eq.~\ref{eqnA:Lambda_1_dagger}) in their final forms:
\begin{align}
\ell_1^\dagger / \ell_2^\dagger &= 2^{-I^\dagger({\bf z}; {\bf s})}, \\
\Lambda_1^\dagger / \Lambda_2^\dagger &= a / b.
\end{align}
As we will show next, these ratios remain unchanged when $I({\bf z}; {\bf s})$ is 
increased beyond the critical value $I^\dagger({\bf z}; {\bf s})$.

\subsection{Behavior beyond the transition point}
\label{sec:AppendixC3}

We now discuss the implications of optimal noise allocation beyond the transition point where 
$\Omega > \Omega_c$ or, equivalently, $I({\bf z}; {\bf s}) > I^\dagger({\bf z}; {\bf s})$.
Starting with the dimensions $\ell_1^*$ and $\ell_2^*$ of the $P({\bf s} | {\bf z})$ ellipse (see Eq.~\ref{eqnA:ell_i_general_expression}), 
we write
\begin{subequations}
\begin{align}
\ell_i^* &= \frac{r \sigma_i^*}{\sqrt{r^2 + (\sigma_i^*)^2}} \\
&= \frac{r}{\sqrt{1 + \theta_i^*}},
\end{align}
\end{subequations}
where we used the notation $\theta_i = r^2/\sigma_i^2$ (Eq.~\ref{eqnA:theta_definition}). 
Substituting the optimal $\theta_1^*$ and $\theta_2^*$
from Eq.~\ref{eqnA:theta1_star} and Eq.~\ref{eqnA:theta2_star}, respectively, we obtain
\begin{subequations}
\begin{align}
\ell_1^* / r &= \frac{1}{\sqrt{\Omega \times \Omega_c}} = 
2^{\displaystyle -0.5 \big( I({\bf z}; {\bf s}) + I^\dagger({\bf z}; {\bf s}) \big)}, \\
\ell_2^* / r &= \frac{1}{\sqrt{\Omega \div \Omega_c}} = 
2^{\displaystyle -0.5 \big( I({\bf z}; {\bf s}) - I^\dagger({\bf z}; {\bf s}) \big)}.
\end{align}
\end{subequations}
As expected, both $\ell_1^*$ and $\ell_2^*$ decrease with increasing $I({\bf z}; {\bf s})$, 
indicating the finer decoding of the signal ${\bf s}$ (see Fig.~8a in the main text).
Their ratio, however, 
stays the same for all $I({\bf z}; {\bf s}) > I^\dagger({\bf z}; {\bf s})$, namely
\begin{align}
\frac{\ell_1^*}{\ell_2^*} = 2^{-I^\dagger({\bf z}; {\bf s})}.
\end{align}
Recalling that $I(z_i; {\bf s}) = \log (r/\ell_i)$, we can recast the above condition into
\begin{align}
I(z_2; {\bf s}) = I(z_1; {\bf s}) - I^\dagger({\bf z}; {\bf s}),
\end{align}
which shows that, under the optimal strategy, the additional encoded information beyond the transition point is allocated equally 
between the two components.

Next, we do a similar calculation for the dimensions $\Lambda_1^*$ and $\Lambda_2^*$ of the $P({\bf y} | {\bf z})$
ellipse. Starting with $\Lambda_1^*$ (see Eq.~\ref{eqnA:Lambda_i_final_expressions}), we have
\begin{subequations}
\begin{align}
\Lambda_1^* &=  r \sqrt{\frac{a^2 + (\sigma_1^*)^2}{r^2 + (\sigma_1^*)^2}} \\
&= r \sqrt{\frac{1 + \tilde{a}^2 \theta_1^*}{1 + \theta_1^*}} \\
&= r \sqrt{\tilde{a}^2  + \frac{1 - \tilde{a}^2}{\Omega / \Omega_c}} \\
&= a \sqrt{1 + \frac{1}{\Omega} \frac{\sqrt{ (1 - \tilde{a}^2)(1 - \tilde{b}^2)} }{\tilde{a} \, \tilde{b}}},
\end{align}
\end{subequations}
where in the first step we used the definition $\theta_1 = r^2/\sigma_1^2$ (Eq.~\ref{eqnA:theta_definition}) 
and the notation $\tilde{a} = a/r$,
in the second step we used the result $1 + \theta_1^* = \Omega/\Omega_c$ (Eq.~\ref{eqnA:theta1_star}),
and then in the third step substituted the expression for $\Omega_c$ (Eq.~\ref{eqnA:Omega_c_expression}).
An analogous set of calculations yields a similar expression for $\Lambda_2^*$, namely
\begin{align}
\Lambda_2^* = b \sqrt{1 + \frac{1}{\Omega} \frac{\sqrt{ (1 - \tilde{a}^2)(1 - \tilde{b}^2)} }{\tilde{a} \, \tilde{b}}}.
\end{align}
Again, $\Lambda_1^*$ and $\Lambda_2^*$ both decrease with increasing $\Omega$ (or $I({\bf z}; {\bf s})$), 
eventually converging to $a$ and $b$, respectively, in the limit $\Omega \rightarrow \infty$. However, 
their ratio, once established at the transition point, 
remains the same for $I({\bf z}; {\bf s}) > I^\dagger({\bf z}; {\bf s})$:
\begin{align}
\frac{\Lambda_1^*}{\Lambda_2^*} = \frac{a}{b}.
\end{align}
This property is illustrated in Fig.~8b of the main text.

\newpage
\section{General multi-dimensional case}
\label{sec:AppendixD}

In this Appendix section, we derive the general solution to the Gaussian information bottleneck problem,
discuss the degenerate space of optimal solutions, derive the results pertaining the geometric 
interpretation of optimality in the general case, and provide details on the thermodynamic analogy for 
the optimal allocation of encoding capacity.
The contents of this section support claims made in {\bf Sec. V} of the main text.

\subsection{Solution to the Gaussian information bottleneck problem}
\label{sec:AppendixD1}

The problem of maximizing the relevant information $I({\bf z}; {\bf y})$ for a given amount of encoded information 
$I({\bf z}; {\bf s})$ can be formulated generally as a constrained optimization problem, where the functional 
\begin{align}
\label{eqnA:lagr_most_general}
\mathcal{L}[p({\bf z} | {\bf s})] = I({\bf z}; {\bf y}) - \gamma I({\bf z}; {\bf s})
\end{align}
is maximized over all possible encoding strategies $p({\bf z} | {\bf s})$, with 
$\gamma$ representing a Lagrange multiplier that controls the level of compression.
Interpreting the two mutual information terms as reductions in the entropy of the variable ${\bf z}$
when ${\bf s}$ and ${\bf y}$, respectively, are known, we write them as
\begin{align}
\label{eqnA:Izs_reminder}
I({\bf z}; {\bf s}) &= \frac{1}{2} \log \left( | \boldsymbol{\Sigma}_{\bf z} | / | \boldsymbol{\Sigma}_{ {\bf z} | {\bf s} } | \right), \\
I({\bf z}; {\bf y}) &= \frac{1}{2} \log \left( | \boldsymbol{\Sigma}_{\bf z} | / | \boldsymbol{\Sigma}_{ {\bf z} | {\bf y} } | \right).
\end{align}
Recalling the noisy encoding form ${\bf z} = {\bf W}^{\rm T} {\bf s} + \boldsymbol{\xi}$,
the three covariance matrices appearing above can be expressed as
\begin{align}
\label{eqnA:covariance_matrices_reminder_a}
\boldsymbol{\Sigma}_{{\bf z} | {\bf s}} &= \boldsymbol{\Sigma}_{\boldsymbol{\xi}}, \\
\label{eqnA:covariance_matrices_reminder_b}
\boldsymbol{\Sigma}_{\bf z} &= {\bf W}^{\rm T} \boldsymbol{\Sigma}_{\bf s} {\bf W} 
+ \boldsymbol{\Sigma}_{\boldsymbol{\xi}}, \\
\label{eqnA:covariance_matrices_reminder_c}
\boldsymbol{\Sigma}_{{\bf z} | {\bf y}} &= {\bf W}^{\rm T} \boldsymbol{\Sigma}_{{\bf s} | {\bf y}} {\bf W} 
+ \boldsymbol{\Sigma}_{\boldsymbol{\xi}}.
\end{align}

Now, in general, ${\bf W}$ may represent a set of non-perpendicular encoding directions, and a non-diagonal
$\boldsymbol{\Sigma}_{\boldsymbol{\xi}}$ may represent correlated noises corresponding to these encoding directions.
For any choice of the $\{ {\bf W}, \boldsymbol{\Sigma}_{\boldsymbol{\xi}} \}$ pair, however, 
there is a corresponding $P({\bf s} | {\bf z})$ ellipse (or ellipsoid in higher dimensions)
in the ${\bf s}$-space.
When the signal ${\bf s}$ is standardized, i.e. $\boldsymbol{\Sigma}_{\bf s} = r^2 {\bf I}$,  
the axes of this ellipse and the degree to which the ellipse is compressed along its perpendicular axes 
correspond to an effective orthonormal set of encoding directions and independent noises for each direction.
For example, if the ellipse is compressed along the direction $\hat{\bf w}$ and has a half-length $\ell$ along that 
direction, then, from the $\ell(\sigma)$ relation (see Eq.~\ref{eqnA:ell_i_general_expression}), 
the independent noise corresponding to that direction $\hat{\bf w}$ can be obtained as $\sigma = r \ell / \sqrt{r^2 - \ell^2}$.
No compression $(\ell \rightarrow r)$ means infinite encoding noise, and maximal compression $(\ell \rightarrow 0)$ means 
zero encoding noise, as expected.

Because of this equivalence, we will formulate the general information bottleneck problem as that of obtaining 
the optimal set of orthonormal encoding directions and independent noises corresponding to these directions
which maximize the functional in Eq.~\ref{eqnA:lagr_most_general}.
This principal solution will then dictate the family of degenerate solutions which we will discuss later in this section.
The orthonormality condition on the encoding directions means that the matrix 
${\bf W} = [\hat{\bf w}_1, \hat{\bf w}_2, ...]$ must satisfy the constraint
\begin{align}
{\bf W}^{\rm T} {\bf W} = {\bf I}.
\end{align}
And the condition of independent noises means that 
$\boldsymbol{\Sigma}_{\boldsymbol{\xi}}$ is diagonal, i.e.
\begin{align}
\boldsymbol{\Sigma}_{\boldsymbol{\xi}} = {\rm diag} \left( \sigma_1^2, \sigma_2^2, ... \right).
\end{align}
For convenience, we will assume a convention where $\{ \sigma^2_i \}$ have an ascending order 
(i.e. $\sigma_1 \le \sigma_2 \le ...$).

With such choices for ${\bf W}$ and $\boldsymbol{\Sigma}_{\boldsymbol{\xi}}$, and with the standardized
${\bf s}$-variable ($\boldsymbol{\Sigma}_{\bf s} = r^2 {\bf I}$), the covariance matrices 
$\boldsymbol{\Sigma}_{{\bf z} | {\bf s}}$ and $\boldsymbol{\Sigma}_{\bf z}$ 
(Eq.~\ref{eqnA:covariance_matrices_reminder_a} and Eq.~\ref{eqnA:covariance_matrices_reminder_b})
become diagonal, namely
\begin{align}
\boldsymbol{\Sigma}_{{\bf z} | {\bf s}} &= {\rm diag} \left( \sigma_1^2, \sigma_2^2, ... \right), \\
\label{eqnA:Sigma_z_diagonal}
\boldsymbol{\Sigma}_{\bf z} &= {\rm diag} \left( r^2 + \sigma_1^2, r^2 + \sigma_2^2, ... \right).
\end{align}
Their determinants are then simply given by the product of the diagonal elements. 
Applying this property, we can write the encoded information $I({\bf z}; {\bf s})$ 
(see Eq.~\ref{eqnA:Izs_reminder}) as
\begin{subequations}
\label{eqnA:Izs_general}
\begin{align}
I({\bf z}; {\bf s}) &= \frac{1}{2} \log \left( \frac{\prod_i (r^2 + \sigma_i^2)}{\prod_i \sigma_i^2} \right) \\
&= \frac{1}{2} \left( \sum_i \log \left( r^2 + \sigma_i^2 \right) - \sum_i \log \sigma_i^2 \right).
\end{align}
\end{subequations}
Notably, due to the choices made for ${\bf W}$, $\boldsymbol{\Sigma}_{\boldsymbol{\xi}}$ and $\boldsymbol{\Sigma}_{\bf s}$,
the encoded information does not depend on ${\bf W} = [\hat{\bf w}_1, \hat{\bf w}_2, ...]$
and is instead controlled by the independent noise magnitudes 
$\{ \sigma_i^2 \}$ assigned to the orthonormal set of encoding directions.\\

Now, our objective is to maximize the Lagrangian $\mathcal{L} = I({\bf z}; {\bf y}) - \gamma I({\bf z}; {\bf s})$
over all possible collections of orthonormal encoding directions ${\bf W} = [\hat{\bf w}_1, \hat{\bf w}_2, ...]$
and noise strengths $\{\sigma_1^2, \sigma_2^2, ... \}$ assigned to these directions.
We start with the optimization with respect to ${\bf W}$ for a given choice of noise strengths. 
Since $I({\bf z}; {\bf s})$ is independent of ${\bf W}$,
maximizing the Lagrangian is equivalent to maximizing the relevant information $I({\bf z}; {\bf y})$.
This optimization problem can be formulated mathematically as
\begin{align}
\label{eqnA:dLagr_dW}
\frac{\delta}{\delta {\bf W}} \left[ I({\bf z}; {\bf y}) - {\rm tr} \left( {\bf \Gamma} ( {\bf W}^{\rm T} {\bf W} - {\bf I} ) \right) \right]
= {\bf 0}.
\end{align}
The trace term subtracted from $I({\bf z}; {\bf y})$ is there to ensure that ${\bf W}$ is an orthonormal matrix.
There, ${\bf \Gamma}$ is an $n \times n$ symmetric Lagrange multiplier matrix with $n(n+1)/2$ independent terms.
This is exactly the number of conditions that the encoding directions $\{ \hat{\bf w}_i \}_{i=1}^n$ must satisfy because 
the orthonormality requirement $\hat{\bf w}_i \cdot \hat{\bf w}_j = \delta_{ij}$ must hold for all pairs of $i$ and $j$,
of which there are $n(n+1)/2$ in total. To see how the trace term accounts for these requirement, we write it out
in an expanded form, namely
\begin{subequations}
\begin{align}
{\rm tr} \left( {\bf \Gamma} ({\bf W}^{\rm T} {\bf W} - {\bf I} )\right) &= 
\sum_i \left( {\bf \Gamma} ({\bf W}^{\rm T} {\bf W} - {\bf I} )\right)_{ii} \\
&= \sum_i \sum_j \Gamma_{ij} ({\bf W}^{\rm T} {\bf W} - {\bf I})_{ji}  \\
&= \sum_{i,j} {\bf \Gamma}_{ij} (\hat{\bf w}_i \cdot \hat{\bf w}_j - \delta_{ij}),
\end{align}
\end{subequations}
where we used $({\bf W}^{\rm T} {\bf W})_{ij} = \hat{\bf w}_i \cdot \hat{\bf w}_j$ and $I_{ji} = \delta_{ij}$.
Due to the symmetry $\hat{\bf w}_i \cdot \hat{\bf w}_j = \hat{\bf w}_j \cdot \hat{\bf w}_i$, the matrix ${\bf \Gamma}$
also needs to be symmetric, i.e. $\Gamma_{ij} = \Gamma_{ji}$ or
\begin{align}
\label{eqnA:Gamma_symmetric}
{\bf \Gamma} = {\bf \Gamma}^{\rm T}.
\end{align}

Of the two covariance matrices entering the expression of the relevant information 
$I({\bf z}; {\bf y}) = \frac{1}{2} \log \left( |\boldsymbol{\Sigma}_{\bf z}| / |\boldsymbol{\Sigma}_{{\bf z} | {\bf y}}| \right)$
only $\boldsymbol{\Sigma}_{{\bf z} | {\bf y}}$ depends on the orthonormal matrix ${\bf W}$.
Substituting its expression (Eq.~\ref{eqnA:covariance_matrices_reminder_c}), we write the optimization problem in 
Eq.~\ref{eqnA:dLagr_dW} as
\begin{align}
\label{eqnA:Lagr_minimization_wrt_W}
\frac{\delta}{\delta {\bf W}} \left[ - \frac{1}{2} \log \left| {\bf W}^{\rm T} \boldsymbol{\Sigma}_{{\bf s} | {\bf y}} {\bf W} 
+ \boldsymbol{\Sigma}_{\boldsymbol{\xi}} \right| 
- {\rm tr} \left( {\bf \Gamma} ({\bf W}^{\rm T} {\bf W} - {\bf I} ) \right)
\right] = 0.
\end{align}
Using the matrix identities 
$\frac{\delta}{\delta {\bf X}} \log |{\bf X}^{\rm T} \boldsymbol{\Sigma} {\bf X} + {\bf A}| = 
2 \boldsymbol{\Sigma} {\bf X} 
({\bf X}^{\rm T} \boldsymbol{\Sigma} {\bf X} + {\bf A})^{-1}$
and $\frac{\delta}{\delta {\bf X}} {\rm tr} ({\bf A} {\bf X}^{\rm T} {\bf X}) = {\bf X} ({\bf A} + {\bf A}^{\rm T})$,
we perform the differentiation with respect to ${\bf W}$ and obtain
\begin{align}
\boldsymbol{\Sigma}_{{\bf s} | {\bf y}} {\bf W} \left( {\bf W}^{\rm T} \boldsymbol{\Sigma}_{{\bf s} | {\bf y}} {\bf W}
+ \boldsymbol{\Sigma}_{\boldsymbol{\xi}} \right)^{-1} + 2 {\bf W} {\bf \Gamma} = {\bf 0},
\end{align}
where we used the symmetry of ${\bf \Gamma}$ to write ${\bf \Gamma} + {\bf \Gamma}^{\rm T} = 2 {\bf \Gamma}$.
We then multiply the above equation by ${\bf W}^{\rm T}$ from the left, 
use the orthonormality condition ${\bf W}^{\rm T} {\bf W} = {\bf I}$
and solve for the Lagrange multiplier matrix ${\bf \Gamma}$, arriving at
\begin{align}
\label{eqnA:Gamma_expression}
{\bf \Gamma} = - \frac{1}{2} {\bf W}^{\rm T} \boldsymbol{\Sigma}_{{\bf s} | {\bf y}} {\bf W}
\left( {\bf W}^{\rm T} \boldsymbol{\Sigma}_{{\bf s} | {\bf y}} {\bf W} + \boldsymbol{\Sigma}_{\boldsymbol{\xi}} \right)^{-1}.
\end{align}
Next, we exploit the symmetry condition on ${\bf \Gamma}$ (Eq.~\ref{eqnA:Gamma_symmetric}).
Denoting ${\bf K} = {\bf W}^{\rm T} \boldsymbol{\Sigma}_{{\bf s} | {\bf y}} {\bf W}$ for succinctness, 
we can write ${\bf \Gamma} = -\frac{1}{2} {\bf K} ({\bf K} + \boldsymbol{\Sigma}_{\boldsymbol{\xi}})^{-1}$.
Applying the matrix identities $\left( {\bf A} {\bf B} \right)^{\rm T} = {\bf B}^{\rm T} {\bf A}^{\rm T}$ 
and $\left( {\bf A}^{\rm T} \right)^{-1} = \left( {\bf A}^{-1} \right)^{\rm T}$,
and using the fact that ${\bf K}$ and $\boldsymbol{\Sigma}_{\boldsymbol{\xi}}$ are both symmetric matrices, 
we impose the condition ${\bf \Gamma} = {\bf \Gamma}^{\rm T}$, finding
\begin{align}
{\bf K} ({\bf K} + \boldsymbol{\Sigma}_{\boldsymbol{\xi}})^{-1} = ({\bf K} + \boldsymbol{\Sigma}_{\boldsymbol{\xi}})^{-1} {\bf K}.
\end{align}
Multiplying both sides from the left and right with ${\bf K} + \boldsymbol{\Sigma}_{\boldsymbol{\xi}}$, we get
\begin{align}
({\bf K} + \boldsymbol{\Sigma}_{\boldsymbol{\xi}}) {\bf K} = {\bf K} ({\bf K} + \boldsymbol{\Sigma}_{\boldsymbol{\xi}}).
\end{align}
Simplifying further, we finally arrive at
\begin{align}
\boldsymbol{\Sigma}_{\boldsymbol{\xi}} {\bf K} = {\bf K} \boldsymbol{\Sigma}_{\boldsymbol{\xi}} .
\end{align}
Now, we know that the noise covariance matrix is diagonal, 
$\boldsymbol{\Sigma}_{\boldsymbol{\xi}} = {\rm diag}(\sigma_1^2, \sigma_2^2, ...)$.
With this in mind, the $ij^{\rm th}$ element of the above identity becomes
\begin{align}
\sigma_i^2 K_{ij} = \sigma_j^2 K_{ij}.
\end{align}
Reorganizing, we find
\begin{align}
\left( \sigma_i^2 - \sigma_j^2 \right) K_{ij} = 0.
\end{align}
In general, the encoding noise strengths are distinct ($\sigma^2_i \ne \sigma^2_j$ when $i \ne j$), 
implying that $K_{ij} = 0$ when $i \ne j$ or,
equivalently, that ${\bf K} = {\bf W}^{\rm T} \boldsymbol{\Sigma}_{{\bf s} | {\bf y}} {\bf W}$ is a diagonal matrix.
The fact that the orthogonal matrix ${\bf W}$ diagonalizes the conditional covariance matrix 
$\boldsymbol{\Sigma}_{{\bf s} | {\bf y}}$ means that ${\bf W}$ consists of the eigenvectors 
of $\boldsymbol{\Sigma}_{{\bf s} | {\bf y}}$, and the diagonal entries of ${\bf K}$ are the 
eigenvalues of $\boldsymbol{\Sigma}_{{\bf s} | {\bf y}}$. 
However, the order in which these eigenvalues appear
as the diagonal entries of ${\bf K}$ is still not specified.

The different orders correspond to local extrema of the Lagrangian. 
The global maximum of the Lagrangian is reached by the ordering that 
minimizes $|\boldsymbol{\Sigma}_{{\bf z} | {\bf y}}| = |{\bf K}  + \boldsymbol{\Sigma}_{\boldsymbol{\xi}}|$
(see Eq.~\ref{eqnA:Lagr_minimization_wrt_W}).
Let us denote the eigenvalues of $\boldsymbol{\Sigma}_{{\bf s} | {\bf y}}$ by $\{ \lambda_i^2 \}$,
where $\lambda_i$ is the semi-axis length of the $P({\bf s} | {\bf y})$ ellipse
along its $i^{\rm th}$ axis. We assume a convention where these eigenvalues
have an ascending order (i.e. $\lambda^2_1 \le \lambda^2_2 \le ...$).
Now, if $q(i)$ is the index of the eigenvalue appearing at the $i^{\rm th}$ diagonal position of 
matrix ${\bf K}$,
then the determinant $|{\bf K} + \boldsymbol{\Sigma}_{\boldsymbol{\xi}}|$ will take the form
\begin{align}
\label{eqnA:Sigma_z_given_y_extrema}
|\boldsymbol{\Sigma}_{{\bf z} | {\bf y}}| = \prod_i \left( \lambda^2_{q(i)} + \sigma_i^2 \right).
\end{align}
Our goal is to find the mapping relation $q(i)$ that minimizes the above product. We first consider
the case where $n=2$. Then the two possible orderings are $\{ q(1) = 1, \, q(2) = 2 \}$
and $\{ q(1) = 2, \, q(2) = 1\}$. The product in these two cases is given by
\begin{subequations}
\label{eqnA:cases_extrema}
\begin{align}
\text{case 1:} \qquad \left( \lambda^2_1 + \sigma^2_1 \right) \left( \lambda^2_2 + \sigma^2_2 \right) &= 
\left( \lambda_1^2 + \sigma_1^2 \right) \left( \lambda_1^2 + 
{\color{darkgray} (\lambda_2^2 - \lambda_1^2)}
+ \sigma_1^2 + 
{\color{darkgray} (\sigma_2^2 - \sigma_1^2)} \right) \nonumber\\
&= (\lambda_1^2 + \sigma_1^2)^2 + (\lambda_1^2 + \sigma_1^2) \left( 
{\color{darkgray} (\lambda_2^2 - \lambda_1^2)}
+ {\color{darkgray} (\sigma_2^2 - \sigma_1^2)} \right), \\
\text{case 2:} \qquad \left( \lambda_2^2 + \sigma_1^2 \right) \left( \lambda_1^2 + \sigma_2^2 \right) &= 
\left( \lambda_1^2 + {\color{darkgray} (\lambda_2^2 - \lambda_1^2)} + \sigma_1^2 \right)
\left( \lambda_1^2 + \sigma_1^2 + {\color{darkgray} (\sigma_2^2 - \sigma_1^2)} \right) \nonumber\\
&= \left( \lambda_1^2 + \sigma_1^2 \right)^2 + \left( \lambda_1^2 + \sigma_1^2 \right)
\left( {\color{darkgray} (\lambda_2^2 - \lambda_1^2)} + {\color{darkgray} (\sigma_2^2 - \sigma_1^2)} \right) 
+ \underbracket[0.140ex]{{\color{darkgray} (\lambda_2^2 - \lambda_1^2)}}_{\ge 0}
\underbracket[0.140ex]{{\color{darkgray} (\sigma_2^2 - \sigma_1^2)}}_{\ge 0}.
\end{align}
\end{subequations}
As can be seen, the product in the second case contains an extra non-negative term because, by our chosen
convention, $\lambda_2 \ge \lambda_1$ and $\sigma_2 \ge \sigma_1$.
What this means is that in the case of optimal two-dimensional encoding, the smaller encoding noise $\sigma_1$
should be assigned to the direction $\hat{\bf w}_1$ corresponding to the smaller eigenvalue
of $\boldsymbol{\Sigma}_{{\bf s} | {\bf y}}$, and the larger encoding noise $\sigma_2$ should be assigned 
to the direction $\hat{\bf w}_2$ corresponding to the larger eigenvalue of 
 $\boldsymbol{\Sigma}_{{\bf s} | {\bf y}}$.
 
 This principle can be generalized to arbitrary dimensions. Namely, the lowest encoding noise is 
 assigned to the encoding direction $\hat{\bf w}_1$ with the lowest corresponding eigenvalue $\lambda_1$,
 the second lowest noise is assigned to the encoding direction with the second lowest eigenvalue, etc., 
 and the largest encoding noise is assigned to the encoding direction with the largest corresponding eigenvalue.
To see why this general principle must hold, let us considering an arbitrary initial ordering $q(i)$.
If for some $i < j$, $\lambda_{q(i)} > \lambda_{q(j)}$, i.e. the lower noise $\sigma_i$ is assigned 
to the larger eigenvalue $\lambda^2_{q(i)}$, then flipping the ordering of this pair is guaranteed 
to reduce the product, as per the derivation above. Repeatedly applying this procedure for all 
pairs that violate the principle, we arrive at the sequence where $\{ \lambda_i \}$ are paired 
with $\{ \sigma_i \}$ of the same sorting order, which proves the general principle.

With the optimal $\lambda_i \leftrightarrow \sigma_i$ pairing, we can then write the relevant information
$I({\bf z}; {\bf y}) = \frac{1}{2} \log \left( |\boldsymbol{\Sigma}_{\bf z}| / |\boldsymbol{\Sigma}_{{\bf z} | {\bf y}}| \right)$,
with $|\boldsymbol{\Sigma}_{\bf z}| = \prod_i (r^2 + \sigma_i^2)$ (Eq.~\ref{eqnA:Sigma_z_diagonal})
and $|\boldsymbol{\Sigma}_{{\bf z} | {\bf y}}| = \prod_i (\lambda_i^2 + \sigma_i^2)$ (Eq.~\ref{eqnA:Sigma_z_given_y_extrema}) 
for a given set of $\{ \sigma^2_i \}$ as
\begin{subequations}
\label{eqnA:Izy_general_optW}
\begin{align}
I({\bf z}; {\bf y}) &= \frac{1}{2} \log \left( \frac{\prod_i (r^2 + \sigma_i^2)}{\prod_i (\lambda_i^2 + \sigma_i^2)} \right) \\
&= \frac{1}{2} \left( \sum_i \log \left( r^2 + \sigma_i^2 \right) - \sum_i \log \left( \lambda_i^2 + \sigma_i^2 \right) \right).
\end{align}
\end{subequations}

\begin{figure}[!ht]
\centering
\includegraphics[width=0.75\textwidth]{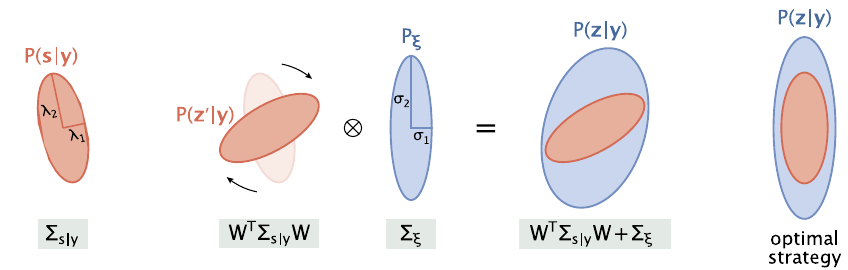}
\caption{
Geometric interpretation of the formation of the $P({\bf z} | {\bf y})$ distribution as 
a convolution between the reoriented $P({\bf z}' | {\bf y})$ distribution and the encoding noise distribution
$P_{\boldsymbol{\xi}}$. The optimal strategy minimizing the area of the $P({\bf z} | {\bf y})$ distribution 
aligns the $P({\bf z}' | {\bf y})$ and $P_{\boldsymbol{\xi}}$ ellipses.
}
\label{figA:optimal_orientation}
\end{figure}

Before we move on to the question of optimally choosing the encoding noise strengths $\{ \sigma_i^2 \}$, 
we provide additional geometric intuition on the optimal choice of orthonormal ${\bf W}$
that maximizes $I({\bf z}; {\bf y})$ for fixed $\{ \sigma_i^2 \}$, which, in turn, fix $I({\bf z}; {\bf s})$.
Recalling the definition of the encoding variable, namely 
${\bf z} = {\bf W}^{\rm T} {\bf s} + \boldsymbol{\xi}$, we first introduce ${\bf z}' = {\bf W}^{\rm T} {\bf s}$
as the deterministic part of the encoding. From the orthonormality condition ${\bf W}^{\rm T} {\bf W} = {\bf I}$,
we have ${\rm det}^2({\bf W}) = 1$ or ${\rm det}({\bf W}) = \pm 1$, 
which means that the variable ${\bf z}'$ rotates ($|{\bf W}| = 1$) or reflects ($|{\bf W}| = -1$)
the signal variable ${\bf s}$. Using this variable, 
we then write the formation of the distribution $P({\bf z} | {\bf y})$ as
\begin{align}
\label{eqnA:Pz_given_y_pre_convolution}
P({\bf z} | {\bf y}) = \int {\rm d} {\bf z}' \, P({\bf z} | {\bf z}') P({\bf z}' | {\bf y}).
\end{align}
Here, $P({\bf z} | {\bf z}')$ captures the addition of the encoding noise $\boldsymbol{\xi}$, 
while $P({\bf z}' | {\bf y})$ represents the rotation/reflection of the conditional distribution $P({\bf s} | {\bf y})$.
Now, since the probability $P({\bf z} | {\bf z}')$ depends on the difference $\boldsymbol{\xi} = {\bf z} - {\bf z}'$,
we can write it as $P({\bf z} | {\bf z}') = P_{\boldsymbol{\xi}} ({\bf z} - {\bf z}')$, where 
$P_{\boldsymbol{\xi}}$ is a multivariate Gaussian centered at the origin and having the covariance matrix 
$\boldsymbol{\Sigma}_{\boldsymbol{\xi}} = {\rm diag}(\sigma_1^2, \sigma_2^2, ...)$.
Substituting this identity into Eq.~\ref{eqnA:Pz_given_y_pre_convolution}, we obtain
\begin{align}
P({\bf z} | {\bf y}) = \int {\rm d} {\bf z}' P_{\boldsymbol{\xi}} ({\bf z} - {\bf z}') P({\bf z}' | {\bf y}).
\end{align}
The above formulation shows that $P({\bf z} | {\bf y})$ can be interpreted as the convolution
of the reoriented $P({\bf s} | {\bf y})$ distribution (represented via $P({\bf z}' | {\bf y})$)
with the encoding noise distribution $P_{\boldsymbol{\xi}}$. 
We illustrate this visually in Fig.~\ref{figA:optimal_orientation} 
for the case of two-dimensional signal and encoding variables.

Optimization of the relevant information $I({\bf z}; {\bf y})$
with respect to ${\bf W}$ aims to minimize the area of the $P({\bf z} | {\bf y})$ ellipse.
As captured in Fig.~\ref{figA:optimal_orientation}, this is achieved when the red $P({\bf s} | {\bf y})$
ellipse is oriented vertically so that its smaller axis with a half-length $\lambda_1$ aligns with 
the smaller axis of the blue $P_{\boldsymbol{\xi}}$ ellipse with a half-length $\sigma_1$ and, similarly,
the axes with longer lengths also align. The resulting optimal $P({\bf z} | {\bf y})$ ellipse has 
axis lengths $\sigma_{z_1 | {\bf y}} = \sqrt{\lambda_1^2 + \sigma_1^2}$ and 
$\sigma_{z_2 | {\bf y}} = \sqrt{\lambda_2^2 + \sigma_2^2}$, with components $z_1$ and $z_2$
encoding along the principal directions $\hat{\bf w}_1$ and $\hat{\bf w}_2$.
Also note that the least optimal strategy, which is ``case 2'' 
discussed in our earlier derivation (Eq.~\ref{eqnA:cases_extrema}),
reorients the $P({\bf s} | {\bf y})$ ellipse horizontally (instead of vertically), resulting
in a $P({\bf z} | {\bf y})$ ellipse with semi-axes $\sqrt{\lambda_2^2 + \sigma_1^2}$ and 
$\sqrt{\lambda_1^2 + \sigma_2^2}$, and a maximum corresponding area for given $\sigma_1$ and $\sigma_2$.\\

Having derived the optimal set of orthonormal encoding directions $\{ \hat{\bf w}_i \}$ for fixed encoding noises, 
we now proceed with deriving the optimal way of allocating these noises for a given amount of encoded 
information $I({\bf z}; {\bf s})$.
Previously, in our study of the two-dimensional encoding case (Appendix~\ref{sec:AppendixC}2),
we derived expressions for optimally assigned noise strengths as an explicit function of $I({\bf z}; {\bf s})$
(e.g. see Eq.~\ref{eqnA:theta12_star} where $I({\bf z}; {\bf s}) = \log \Omega$).
While such an approach is also possible in the general case, a more elegant way is to perform the optimization 
using a Lagrange multiplier, which we introduced 
in the general formulation of the problem at the beginning of the section (Eq.~\ref{eqnA:lagr_most_general}).

Substituting the encoded information $I({\bf z}; {\bf s})$ from Eq.~\ref{eqnA:Izs_general} and 
the relevant information $I({\bf z}; {\bf y})$ from Eq.~\ref{eqnA:Izy_general_optW}
into the expression for the Lagrangian $\mathcal{L} = I({\bf z}; {\bf y}) - \gamma I({\bf z}; {\bf s})$, we obtain\begin{align}
\mathcal{L} = \frac{1}{2} \sum_i \Big( (1 - \gamma) \log \left( r^2 + \sigma_i^2 \right) - \log \left( \lambda_i^2 + \sigma_i^2 \right) 
+ \gamma \log \sigma_i^2 \Big).
\end{align}
The factorized form of the Lagrangian makes the optimization with respect to $\{ \sigma_i \}$ very straightforward.
Specifically, requiring $\partial \mathcal{L} / \partial \sigma_i = 0$ leads to
\begin{align}
\frac{1 - \gamma}{r^2 + \sigma_i^2} - \frac{1}{\lambda_i^2 + \sigma_i^2} + \frac{\gamma}{\sigma_i^2} = 0.
\end{align}
Using the notation $\tilde{\lambda}_i = \lambda_i / r$ for succinctness, we solve for $\sigma_i^2$ and obtain
\begin{align}
\label{eqnA:sigma_i_sq_principal}
\sigma_i^2 = r^2 \frac{\gamma \tilde{\lambda}_i^2}{(1 - \gamma) - \tilde{\lambda}_i^2}.
\end{align}
Since the denominator cannot be negative, encoding along 
the direction $\hat{\bf w}_i$ is used only when the Lagrange multiplier $\gamma$ is below the critical value
\begin{align}
\gamma_i^c = 1 - \tilde{\lambda}_i^2.
\end{align}
Substituting $\sigma_i^2(\gamma)$ into the expressions for the encoded information (Eq.~\ref{eqnA:Izs_general})
and relevant information (Eq.~\ref{eqnA:Izy_general_optW}), and performing simplifications, we find these quantities 
as a function of $\gamma$ as
\begin{align}
\label{eqnA:I_enc_fn_gamma}
I({\bf z}; {\bf s}) &= \frac{1}{2} \sum_{i = 1}^{n(\gamma)} \log 
\left( \frac{1 - \gamma}{\gamma} \frac{1 - \tilde{\lambda}_i^2}{\tilde{\lambda}_i^2} \right), \\
\label{eqnA:I_rel_fn_gamma}
I({\bf z}; {\bf y}) &= \frac{1}{2} \sum_{i = 1}^{n(\gamma)} \log \left( \frac{1 - \gamma}{\tilde{\lambda}_i^2} \right).
\end{align}
Here, $n(\gamma)$ represents the number of components used at a given $\gamma$ value, 
which is equal to the number of $\tilde{\lambda}_i$ for which the condition $\gamma \le 1 - \tilde{\lambda}_i^2$ 
is satisfied.
The first component is introduced when $\gamma$ crosses the first threshold $\gamma_1^c = 1 - \tilde{\lambda}_1^2$.
And in the limit $\gamma \rightarrow 0$, which corresponds to the limit of infinite encoded information, 
all components are used. These results are analogous to those in the original work of Chechik {\it et al.} \cite{Chechik2005}.\\

At the end, to demonstrate the generality of our geometric interpretation of critical points and the behavior past them
(e.g. see Fig.~8 and Fig.~10 in the main text), we compute the semi-axis lengths $\ell_i$ and $\Lambda_i$ of the $P({\bf s} | {\bf z})$
and $P({\bf y} | {\bf z})$ ellipses (or ellipsoids), respectively, as a function of the Lagrange multiplier $\gamma$. 
As the encoded and relevant information amounts for the $i^{\rm th}$ component are of the form
$I(z_i; {\bf s}) = \log (r/\ell_i)$ and $I(z_i; {\bf y}) = \log (r / \Lambda_i)$, respectively, 
expressions for $\ell_i(\gamma)$ and $\Lambda_i(\gamma)$ follow directly from Eq.~\ref{eqnA:I_enc_fn_gamma}
and Eq.~\ref{eqnA:I_rel_fn_gamma} as
\begin{align}
\ell_i(\gamma) &= r \sqrt{\frac{\gamma}{1 - \gamma}} \sqrt{\frac{\tilde{\lambda}_i^2}{ 1 - \tilde{\lambda}_i^2}}, \\
\Lambda_i(\gamma) &= \frac{\lambda_i}{\sqrt{1 - \gamma}}.
\end{align}
An alternative route yielding the same final expressions 
would be to substitute the result for $\sigma_i^2(\gamma)$ (Eq.~\ref{eqnA:sigma_i_sq_principal})
into the definitions of $\ell_i$ (Eq.~\ref{eqnA:ell_i_general_expression})
and $\Lambda_i$ (Eq.~\ref{eqnA:Lambda_i_final_expressions}).
Note that for both $\ell_i(\gamma)$ and $\Lambda_i(\gamma)$, 
the $\gamma$-dependence is in the form of a multiplicative prefactor, 
suggesting that once the aspect ratios $\ell_i \div \ell_j$ and $\Lambda_i \div \Lambda_j$ are established
upon introducing the $i^{\rm th}$ and $j^{\rm th}$ encoding components, they remain the same when 
$\gamma$ is dialed down further.
Notably, the ratios of $\{ \Lambda_i \}$ under the optimal encoding strategy are simply the ratios of $\{\lambda_i \}$ -- the dimensions of 
${\bf s} \leftrightarrow {\bf y}$ mapping ellipses/ellipsoids, which offers a simple geometric interpretation 
of how $P({\bf y} | {\bf z})$ ellipse/ellipsoid is reshaped with increasing encoded information.

\subsection{Degenerate space of optimal solutions}
\label{sec:AppendixD2}

The general solution to the Gaussian information bottleneck problem presented in the first part of this Appendix section
consisted of an orthogonal set of encoding directions $\{ \hat{\bf w}_i \}$ and noise strengths $\{ \sigma_i^2(\gamma) \}$
associated with these directions; we will refer to this solution as the principal solution.
However, as mentioned earlier, there is a family of encoding strategies specified by the pair 
$\{ {\bf W}, \boldsymbol{\Sigma}_{\boldsymbol{\xi}} \}$ that yield the same 
decoding distribution $P({\bf s} | {\bf z})$ on the ${\bf s}$-space and hence, the same $P({\bf y} | {\bf z})$
distribution on the relevance space, thereby constituting equally optimal solutions. 
Here, we characterize this family of solutions.

Let's denote the orthonormal collection of principal encoding directions as ${\bf W}^* = \{ \hat{\bf w}_i^* \}$
and the associated diagonal encoding noise matrix as 
$\boldsymbol{\Sigma}_{\boldsymbol{\xi}}^* = {\rm diag} \left( \sigma_1^2(\gamma), \sigma_2^2(\gamma), ... \right)$, 
where $\sigma_i^2(\gamma)$ are those specified in Eq.~\ref{eqnA:sigma_i_sq_principal}. 
The encoding direction matrix ${\bf W}$ of an alternative strategy can be generally expressed in terms of ${\bf W}^*$ via
\begin{align}
\label{eqnA:W_mixing}
{\bf W} = {\bf W}^* {\bf M},
\end{align}
where ${\bf M}$ is the so-called mixing matrix. 
As the columns of ${\bf W}$ must be unit vectors, the choices of ${\bf M}$ are constrained. 
To derive the constraint, we first write the $j^{\rm th}$ column of ${\bf W}$ as
\begin{align}
{\bf W}_{:,j} = \sum_i M_{ij} \hat{\bf w}_i^*.
\end{align}
Then, using the orthogonality condition of the vector set $\{ \hat{\bf w}_i^* \}$, 
we compute the squared norm of the $j^{\rm th}$ column:
\begin{align}
||{\bf W}_{:,j} ||^2 = \sum_i M_{ij}^2.
\end{align}
Requiring $||{\bf W}_{:,j} ||^2 = 1$ implies that $\sum_i M_{ij}^2 = 1$, or, equivalently, 
the columns of the mixing matrix must be unit vectors. 
The columns of ${\bf M}$, however, do not generally represent an orthonormal set
(${\bf M}^{\rm T} {\bf M} \ne {\bf I}$).
This reflects on the fact that after mixing the principal directions $\{ \hat{\bf w}_i^* \}$, 
the alternative encoding directions $\{ \hat{\bf w}^{\rm alt}_i \}$ will generally be
non-perpendicular (${\bf W}^{\rm T} {\bf W} = {\bf M}^{\rm T} {\bf M} \ne {\bf I}$).

Now, if the decoding distribution $P({\bf s} | {\bf z})$ is imposed by the optimal strategy,
then the conditional covariance matrices $\boldsymbol{\Sigma}_{{\bf s} | {\bf z}}$ and 
$\boldsymbol{\Sigma}_{{\bf s} | {\bf z}^*}$ must match.
Recalling the general formula
$\boldsymbol{\Sigma}_{{\bf s} | {\bf z}} = r^2 ( 
{\bf I} - {\bf W} \left( 
{\bf W}^{\rm T} {\bf W} + r^{-2} \boldsymbol{\Sigma}_{\boldsymbol{\xi}}
\right)^{-1} {\bf W}^{\rm T} )$ (copied from Eq. \ref{eqnA:Sigma_s_given_z_vec_intermediate} for convenience),
we write $\boldsymbol{\Sigma}_{{\bf s} | {\bf z}^*}$ for the principal strategy as
\begin{align}
\label{eqnA:Sigma_s_given_z_principal}
\boldsymbol{\Sigma}_{{\bf s} | {\bf z}^*} = r^2 \left( {\bf I} - {\bf W}^* ({\bf I} + r^{-2} 
\boldsymbol{\Sigma}_{\boldsymbol{\xi}}^* )^{-1} {\bf W}^*{^{\rm T}} \right),
\end{align}
where we employed the orthogonality condition ${\bf W}^*{^{\rm T}} {\bf W}^* = {\bf I}$. 
Then, substituting ${\bf W} = {\bf W}^* {\bf M}$ for the alternative strategy, we find
\begin{subequations}
\label{eqnA:Sigma_s_given_z_alt}
\begin{align}
\boldsymbol{\Sigma}_{{\bf s} | {\bf z}} &= r^2 \left( {\bf I} - {\bf W}^* {\bf M} \, ({\bf I} + r^{-2} 
\boldsymbol{\Sigma}_{\boldsymbol{\xi}} )^{-1}  {\bf M}^{\rm T} {\bf W}^*{^{\rm T}} \right) \\
&= r^2 \left( {\bf I} - {\bf W}^* \, ({\bf I} + r^{-2} 
({\bf M}^{\rm T})^{-1} \boldsymbol{\Sigma}_{\boldsymbol{\xi}} {\bf M}^{-1} )^{-1}  {\bf W}^*{^{\rm T}} \right) .
\end{align}
\end{subequations}
The equality of Eq.~\ref{eqnA:Sigma_s_given_z_principal} and Eq.~\ref{eqnA:Sigma_s_given_z_alt} then implies
\begin{align}
\label{eqnA:Sigma_xi_mixing}
\boldsymbol{\Sigma}_{\boldsymbol{\xi}} = {\bf M}^{\rm T} \boldsymbol{\Sigma}^*_{\boldsymbol{\xi}} \, {\bf M}.
\end{align}
This result means that for any choice of the mixing matrix ${\bf M}$, one can obtain an equally optimal solution 
$\{ {\bf W}, \boldsymbol{\Sigma}_{\boldsymbol{\xi}} \}$ 
from the principal solution $\{ {\bf W}^*, \boldsymbol{\Sigma}^*_{\boldsymbol{\xi}} \}$
by setting the encoding direction according to Eq.~\ref{eqnA:W_mixing}
and the encoding noises according to Eq.~\ref{eqnA:Sigma_xi_mixing}.
This choice keeps $P({\bf s} | {\bf z})$ invariant.

Notably, while the different encoding components $\{ z_i^* \}$ of the principal solution have noises that are independent
(diagonal $\boldsymbol{\Sigma}^*_{\boldsymbol{\xi}}$), the same is generally not true for alternative strategies 
where the ``mixing'' of the diagonal matrix $\boldsymbol{\Sigma}^*_{\boldsymbol{\xi}}$ generally makes it 
non-diagonal (Eq.~\ref{eqnA:Sigma_xi_mixing}), whereby correlating the noises of the encoding components $\{ z_i \}$.\\

A particularly interesting class of alternative strategies is one that assigns {\it identical} and {\it independent}
noises to non-perpendicular directions. This is in contrast to the principal strategy that assigns non-identical
independent noises, $\boldsymbol{\Sigma}^*_{\boldsymbol{\xi}} = {\rm diag}(\sigma_1^2(\gamma), \sigma_2^2(\gamma), ...)$,
to perpendicular directions $\{ \hat{\bf w}_i^* \}$. This requirement on the alternative strategy turns 
the condition in Eq.~\ref{eqnA:Sigma_xi_mixing} into
\begin{align}
\label{eqnA:sigma_alt_condition_init}
{\bf M}^{\rm T}
\boldsymbol{\Sigma}^*_{\boldsymbol{\xi}}
{\bf M}
= 
\sigma^2_{\rm alt} {\bf I},
\end{align}
where $\sigma_{\rm alt}^2$ is the identical noise strength assigned to the alternative encoding directions.
To obtain $\sigma_{\rm alt}$, we first note the identity
${\bf M}^{\rm T} \left( \boldsymbol{\Sigma}^*_{\boldsymbol{\xi}} \right)^{1/2} = \sigma_{\rm alt} {\bf Q}^{\rm T}$
that follows from Eq.~\ref{eqnA:sigma_alt_condition_init} for some orthogonal matrix ${\bf Q}$ 
(i.e. ${\bf Q}^{\rm T} {\bf Q} =  {\bf Q} {\bf Q}^{\rm T} = {\bf I}$).\footnote{This general approach for finding $\sigma_{\rm alt}$
was suggested by user1551\\ on math.stackexchange.com (question \#5055826).\\}
Isolating ${\bf M}$, we find 
${\bf M}^{\rm T} = \sigma_{\rm alt} {\bf Q}^{\rm T} \left( \boldsymbol{\Sigma}^*_{\boldsymbol{\xi}} \right)^{-1/2}$.
Then, we compute ${\bf M}^{\rm T} {\bf M}$ as
\begin{align}
\label{eqnA:MTM_condition_sigma_alt}
{\bf M}^{\rm T} {\bf M} = \sigma^2_{\rm alt} {\bf Q}^{\rm T}  \left( \boldsymbol{\Sigma}^*_{\boldsymbol{\xi}} \right)^{-1} {\bf Q}.
\end{align}
Now, since the columns of ${\bf M}$ are unit vectors, the diagonal of ${\bf M}^{\rm T} {\bf M}$ is made of all ones and hence,
${\rm tr}({\bf M}^{\rm T} {\bf M}) = n$, where $n$ is the number of encoding components.
Next, using the cyclic property of matrix traces, namely ${\rm tr}({\bf ABC}) = {\rm tr}({\bf BCA})$, 
and the orthogonality condition of the matrix ${\bf Q}$ (i.e. ${\bf Q} {\bf Q}^{\rm T} = {\bf I}$), we can write 
\begin{align}
{\rm tr} \left( {\bf Q}^{\rm T}  \left( \boldsymbol{\Sigma}^*_{\boldsymbol{\xi}} \right)^{-1} {\bf Q} \right)
=
{\rm tr} \left(  \left( \boldsymbol{\Sigma}^*_{\boldsymbol{\xi}} \right)^{-1} {\bf Q} {\bf Q}^{\rm T} \right)
=
{\rm tr} \left( \left( \boldsymbol{\Sigma}^*_{\boldsymbol{\xi}} \right)^{-1} \right).
\end{align}
Taking the trace of both sides of Eq.~\ref{eqnA:MTM_condition_sigma_alt} then yields
\begin{align}
n = \sigma^2_{\rm alt} {\rm tr} \left( \left( \boldsymbol{\Sigma}^*_{\boldsymbol{\xi}} \right)^{-1} \right).
\end{align}
Solving for $\sigma^2_{\rm alt}$, we obtain
\begin{align}
\frac{1}{\sigma^2_{\rm alt}} = \frac{1}{n} {\rm tr} \left( \left( \boldsymbol{\Sigma}^*_{\boldsymbol{\xi}} \right)^{-1} \right).
\end{align}
In an expanded form, our final result for $\sigma^2_{\rm alt}$ becomes
\begin{align}
\frac{1}{\sigma^2_{\rm alt} (\gamma)} = \frac{1}{n} \left( \frac{1}{\sigma^2_1(\gamma)} + \frac{1}{\sigma^2_2(\gamma)} + \dots + 
\frac{1}{\sigma^2_n(\gamma)} \right).
\end{align}
This result shows that the identical noise strength $\sigma^2_{\rm alt}$ 
assigned to the alternative encoding directions falls between the lowest and highest noise strengths, 
namely $\sigma_1^2$ and $\sigma_n^2$, respectively, assigned to the most informative 
($\hat{\bf w}_1^*$) and the least informative ($\hat{\bf w}_n^*$) principal encoding directions.\\

The problem of finding the mixing matrix ${\bf M}$ from Eq.~\ref{eqnA:sigma_alt_condition_init}
that makes such an alternative strategy possible 
is generally underspecified for $n \ge 3$ dimensions and may therefore allow for multiple solutions.
In two dimensions, however, the solution for ${\bf M}$ can be worked out explicitly, yielding
\begin{align}
{\bf M} = 
\frac{1}{\sqrt{1/\sigma_1^2 + 1/\sigma_2^2}}
\begin{bmatrix}
\phantom{-}1/\sigma_1 & 1/\sigma_1  \\
-1/\sigma_2 & 1/\sigma_2
\end{bmatrix},
\end{align}
where, as a reminder, $\sigma_1^2(\gamma)$ and $\sigma_2^2(\gamma)$ are the noise 
strengths assigned to the principal directions $\hat{\bf w}_1^*$ and $\hat{\bf w}_2^*$ 
at a given choice of $\gamma < \gamma_c$ beyond the transition point
(see Eq.~\ref{eqnA:sigma_i_sq_principal}). 
This choice of the mixing matrix results in the following expressions for the 
two alternative encoding directions:
\begin{subequations}
\begin{align}
\hat{\bf w}_1^{\rm alt} &= \frac{\hat{\bf w}^*_1 / \sigma_1 - \hat{\bf w}^*_2 / \sigma_2}
{ \sqrt{1/\sigma_1^2 + 1/\sigma_2^2} }, \\
\hat{\bf w}_2^{\rm alt} &= \frac{\hat{\bf w}^*_1 / \sigma_1 + \hat{\bf w}^*_2 / \sigma_2}
{ \sqrt{1/\sigma_1^2 + 1/\sigma_2^2} }.
\end{align}
\end{subequations}
Importantly, while $\hat{\bf w}_1^*$ and $\hat{\bf w}_2^*$ are independent of $\gamma$
and are set solely by the ${\bf s} \leftrightarrow {\bf y}$ mapping statistics, 
the encoding directions in the alternative strategy vary along the information curve 
via the $\gamma$-dependence of $\sigma^2_1(\gamma)$ and $\sigma^2_2 (\gamma)$.

The identical noise strength assigned to the alternative encoding directions is given by 
\begin{align}
\label{eqnA:sigma_alt}
\frac{1}{\sigma_{\rm alt}} = \sqrt{\frac{1/\sigma_1^{2} + 1/\sigma_2^2}{2}}.
\end{align}
Under this assignment, we have
\begin{align}
\sigma_1 \le \sigma_{\rm alt} \le \sigma_2,
\end{align}
which means that 
the low-noise encoding along $\hat{\bf w}_1$ and high-noise encoding along $\hat{\bf w}_2$
is substituted with an intermediate-noise encoding along both $\hat{\bf w}_1^{\rm alt}$ and $\hat{\bf w}_2^{\rm alt}$.

In Fig.~A2, we illustrate this alternative optimal encoding setting, showing how the same 
signal decoding distribution can be attained by combining information from non-perpendicular encoding
directions.
The key difference between the principal and the alternative strategies 
is that now the encoding components 
$z_1^{\rm alt} = \hat{\bf w}_1^{\rm alt} \cdot {\bf s} + \xi_1$ and 
$z_2^{\rm alt} = \hat{\bf w}_2^{\rm alt} \cdot {\bf s} + \xi_2 $ are correlated
($\langle z_1^{\rm alt} z_2^{\rm alt} \rangle = r^2 \, \hat{\bf w}_1^{\rm alt} \cdot \hat{\bf w}_2^{\rm alt} \ne 0$),
implying that the sum of individually encoded information amounts will be greater than the information 
jointly encoded about the signal, i.e.
\begin{align}
I(z_1^{\rm alt}; {\bf s}) + I(z_2^{\rm alt}; {\bf s}) \ge I({\bf z}^{\rm alt}; {\bf s}).
\end{align}
Here, $I(z_1^{\rm alt}; {\bf s})$ and $I(z_2^{\rm alt}; {\bf s})$ are identical and are given by
\begin{align}
I(z_1^{\rm alt}; {\bf s}) = I(z_2^{\rm alt}; {\bf s}) = \frac{1}{2} \log \left( 1 + \left( \frac{r}{\sigma_{\rm alt}}\right)^2 \right),
\end{align}
while the information encoded jointly is
\begin{subequations}
\begin{align}
I({\bf z}^{\rm alt}; {\bf s}) &= I({\bf z}^*; {\bf s}) \\
&= I(z_1^*; {\bf s}) + I(z_2^*; {\bf s}) \\
&= 
\frac{1}{2} \log \left( 1 + \Big(\frac{r}{\sigma_1} \Big)^2 \right)
+ \frac{1}{2} \log \left( 1 + \Big(\frac{r}{\sigma_2} \Big)^2 \right).
\end{align}
\end{subequations}

Functionally, this would imply that if the energetic cost of coding would scale linearly with the sum of 
component-wise encoding information amounts, i.e. ${\rm Cost} \propto I(z_1; {\bf s}) + I(z_2; {\bf s})$,
then the principal strategy with orthogonal encoding directions would be the optimal one.
However, with orthogonal encoding, the information stored in the first component is, in general,
higher than that stored in the second component.
This asymmetry means that if the cost of encoding scales non-linearly with 
the individual encoded information amounts \cite{MulderPRR2025}, 
then the strategy with non-orthogonal encoding might be more beneficial.

\begin{figure}[!ht]
\centering
\includegraphics[width=0.4\textwidth]{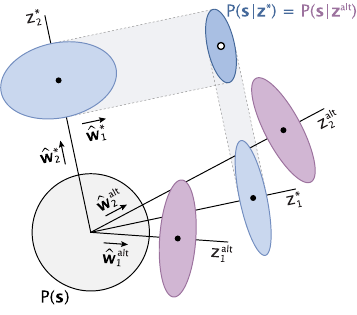}
\caption{
Illustration of how an alternative strategy that encodes the signal along non-perpendicular directions 
can be equally optimal compared to the scheme where the principal encoding directions are used.
The two purple ellipses represent signal decoding distributions $P({\bf s} | z_1^{\rm alt})$
and $P({\bf s} | z_2^{\rm alt})$ for the individual components $z_1^{\rm alt}$ and $z_2^{\rm alt}$.
Jointly, they yield a decoding distribution $P({\bf s} | {\bf z}^{\rm alt})$ that is identical in shape 
to $P({\bf s} | {\bf z}^*)$. 
In making the figure, the $z$-values were chosen as 
${\bf z}^* = \hat{\bf W}^*{^{\rm T}} {\bf s}$ and 
${\bf z}^{\rm alt} = \hat{\bf W}^{\rm T}_{\rm alt} \, {\bf s}$
for a fixed choice of the signal ${\bf s} = [s_1, s_2]^{\rm T}$. Such a choice, made for illustrative purposes, 
makes the centers of the distributions $P({\bf s} | {\bf z}^*)$ and $P({\bf s} | {\bf z}^{\rm alt})$ match.
}
\label{figA:alternative}
\end{figure}

\subsection{Thermodynamic analogy for the optimal allocation of encoding capacity}
\label{sec:AppendixD3}

Here, we explain why the thermodynamic analogy presented in Sec. V of the main text is valid.
We consider a lattice model for the boxes of varying sizes, with the $i^{\rm th}$ box having $N_i$ sites.
The different sizes correspond to the varying amounts of relevant information available to 
the principal encoding components, as reflected in different eigenvalues 
of the conditional covariance matrix $\boldsymbol{\Sigma}_{{\bf s} | {\bf y}}$.
The goal is to distribute a total of $n_{\rm tot}$ distinguishable particles among the boxes in such a way that 
the total free energy is minimized.

We can write the free energy as $F = U - TS$, where $U$ is the internal energy, $T$ is the temperature, 
and $S$ is the entropy. Suppose $n_i$ particles are distributed in the $i^{\rm th}$ box. The associated internal energy 
will be 
\begin{align}
U_i = n_i \varepsilon,
\end{align}
where $\varepsilon$ is the energy per particle. 
The associated entropy of distinguishable particles is set by the multiplicity of states via
\begin{align}
S_i = k_{\rm B} \log \frac{N_i!}{\left( N_i - n_i \right)!}.
\end{align}
Consider the case where $N_i - n_i \gg 1$, we apply the Sterling approximation to write
\begin{align}
S_i \approx k_{\rm B} \log N_i! - k_{\rm B}  \big( (N_i - n_i) \log (N_i - n_i) - (N_i - n_i) \big).
\end{align}
Substituting the expressions for $U_i$ and $S_i$ into $F_i = U_i - T S_i$, we obtain the free energy for the $i^{\rm th}$ box as
\begin{align}
\label{eqnA:F_i_free_energy}
F_i =  n_i \varepsilon - k_{\rm B} T \big( \log N_i! + (N_i - n_i) \log (N_i - n_i) - (N_i - n_i) \big).
\end{align}

Our goal is to distribute the particles in such a way that the total free energy $F_{\rm tot} = \sum_i F_i$ is minimized.
Accounting for the constraint $\sum_i n_i = n_{\rm tot}$ using a Lagrange multiplier $\mu$, we write the minimization condition as
\begin{align}
\frac{\partial}{\partial n_i} \left( \sum_i F_i - \mu \bigg(\sum_i n_i - n_{\rm tot} \bigg) \right) = 0.
\end{align}
Since $\mu_i = \partial F_i / \partial n_i$ is the chemical potential of particles in the $i^{\rm th}$ box, 
performing the partial differentiation in the above condition yields $\mu_i = \mu$,
which is a requirement of equal chemical potentials.

We can obtain $\mu_i$ by differentiating Eq.~\ref{eqnA:F_i_free_energy} with respect to $n_i$, which yields
\begin{align}
\mu_i = \varepsilon + k_{\rm B} T \log (N_i - n_i).
\end{align}
As we can see, equal chemical potentials imply an equal number of available sites, $N_i - n_i$. Thus, when all boxes are initially empty,
the optimal strategy is to first fill the largest box until the number of its available sites equals the total capacity of the second-largest box.
Afterward, both boxes are filled simultaneously until their available capacity equals the total capacity of the third-largest box, and so on.

\newpage
\section{Signal prediction problem}
\label{sec:AppendixE}

In this Appendix section, we derive the properties of the stochastic signal studied in {\bf Sec. VI} of the main text.
We start off by deriving the dimensionless form of the stochastic differential equations used for signal generation
(Eq.~41 and Eq.~42 in the main text), followed by the solution to these equations. Next, we derive conditions for the set of 
optimal encoding directions and study how amounts of maximum predictive information available to the different 
components vary with the forecast interval.

\subsection{Stochastically driven damped harmonic oscillator model}
\label{sec:AppendixE1}

The physically motivated model that we consider in the prediction problem is that of a stochastically driven 
damped harmonic oscillator. In it, an object of mass $m$ is connected via a spring to the origin and experiences 
stochastic kicks from the viscous environment. The dynamics of its one-dimensional motion are 
characterized via
\begin{subequations}
\label{eqnA:oscillator_SDE}
\begin{align}
\frac{{\rm d}x}{{\rm d}t} &= v, \\
m\frac{{\rm d}v}{{\rm d}t} &= - kx - \eta v + A \psi(t),
\end{align}
\end{subequations}
where $x$ and $v$ are the position and the velocity of the object, respectively, $k$ is the spring constant, 
$\eta$ is the drag coefficient, $A$ is a parameter that sets the strength of the stochastic kicks, 
and $\psi(t)$ is unit Gaussian white noise. 

While many parameters enter the generic description of the model, the stochastic structure of the dynamics 
depends only on a single effective parameter. To arrive at this parameter, we first introduce the dimensionless 
quantities $\tilde{x}$, $\tilde{v}$, and $\tilde{t}$ to stand for position, velocity, and time, respectively, 
and define them as
\begin{subequations}
\begin{align}
\tilde{x} &= \frac{x}{\, x^*}, \\
\tilde{v} &= \frac{v}{\, v^*}, \\
\tilde{t} &= \frac{t}{\, t^*}.
\end{align}
\end{subequations}
The ``starred" quantities are characteristic values of their respective dimensional pairs. Substituting 
$x = x^* \tilde{x}$, $v = v^* \tilde{v}$, and $t = t^* \tilde{t}$ into Eq.~\ref{eqnA:oscillator_SDE}
and regrouping the terms, we obtain
\begin{subequations}
\label{eqnA:nondim_system_interm}
\begin{align}
\frac{{\rm d} \tilde{x}}{{\rm d} \tilde{t}} &= \left( \frac{v^* t^*}{x^*} \right) \tilde{v}, \\
\frac{{\rm d} \tilde{v}}{{\rm d} \tilde{t}} &= - \frac{k}{m} \left(  \frac{x^* t^*}{v^*} \right) \tilde{x} 
- \left( \frac{\eta}{m} t^* \right) \tilde{v} + \left( \frac{A}{m} \frac{t^*}{v^*} \right) \psi (t^* \tilde{t}\,).
\end{align}
\end{subequations}
Now, $\psi$ is the derivative of a Wiener process $W$ which follows the scaling rule 
$W(t^* \tilde{t}\,) = \sqrt{t^*} W(\tilde{t}\,)$. Taking the derivative and using $W' = \psi$, we obtain
\begin{align}
\label{eqnA:white_noise_scaling}
\psi(t^* \tilde{t}\,) = \frac{1}{\sqrt{t^*}} \psi ( \tilde{t} \,).
\end{align}
We also introduce $\omega_0$ as the angular frequency of undamped oscillations and $\tau_\eta$ 
as a characteristic damping time scale, defining them via
\begin{align}
\label{eqnA:omega0_def}
\omega_0 &= \sqrt{k/m}, \\
\label{eqnA:tau_eta_def}
\tau_\eta &= m/\eta.
\end{align}
With this notation and the identity in Eq.~\ref{eqnA:white_noise_scaling}, the system of equations 
for the dimensionless variables (Eq.~\ref{eqnA:nondim_system_interm}) can be rewritten as
\begin{subequations}
\label{eqnA:nondim_system_interm2}
\begin{align}
\frac{{\rm d} \tilde{x}}{{\rm d} \tilde{t}} &= \left( \frac{v^* t^*}{x^*} \right) \tilde{v}, \\
\frac{{\rm d} \tilde{v}}{{\rm d} \tilde{t}} &= - \omega_0^2 \left(  \frac{x^* t^*}{v^*} \right) \tilde{x} 
- \left( \frac{t^*}{\tau_\eta} \right) \tilde{v} + \left( \frac{A}{m} \frac{t^*}{v^*} \right) \frac{1}{\sqrt{t^*}} \psi (\tilde{t}\,).
\end{align}
\end{subequations}

We next proceed with choosing the characteristic ``starred'' quantities, namely $x^*$, $v^*$, and $t^*$. 
First, to reduce Eq.~\ref{eqnA:nondim_system_interm2}a to a simple and intuitive form, we set
\begin{align}
\label{eqnA:nondim_cond1}
\frac{v^* t^*}{x^*} = 1.
\end{align}
We then make the coefficient in front of $\tilde{x}$ in Eq.~\ref{eqnA:nondim_system_interm2}b equal to unity
by requiring
\begin{align}
\label{eqnA:nondim_cond2}
\omega_0^2 \left( \frac{x^* t^*}{v^*} \right) = 1.
\end{align}
From these two conditions (Eq.~\ref{eqnA:nondim_cond1} and Eq.~\ref{eqnA:nondim_cond2}), 
the characteristic time scale $t^*$ can be obtained as
\begin{align}
\label{eqnA:t_star_choice}
t^* = \frac{1}{\omega_0}.
\end{align}
As we can see, $t^*$ is set by the time scale of undamped oscillations. This choice differs from that of  
Sachdeva {\it et al.} where the authors chose $\tau_\eta$ to be the characteristic time scale \cite{Sachdeva2021}.

Here, we denote the ratio of $t^*$ and $\tau_\eta$ as
\begin{align}
\label{eqnA:eta_tilde_def}
\tilde{\eta} = \frac{t^*}{\tau_\eta}
{\color{gray} = \frac{\omega_0^{-1}}{\tau_\eta}
= \frac{\eta}{\sqrt{mk}}
}.
\end{align}
The dimensionless effective damping coefficient $\tilde{\eta}$ captures the relative magnitude of 
inertial and damping influences on the motion of the object.

Next, considering a thermal source for the fluctuations, we write the fluctuation amplitude $A$ 
from the fluctuation-dissipation relation as
\begin{align}
A = \sqrt{2 k_{\rm B} T \eta},
\end{align}
where $k_{\rm B}$ is the Boltzmann constant, and $T$ is the temperature. With the above form for $A$, 
the coefficient in front of the noise term in Eq.~\ref{eqnA:nondim_system_interm2}b becomes
\begin{subequations}
\label{eqnA:noise_coeff_simplification}
\begin{align}
\left( \frac{A}{m} \frac{t^*}{v^*} \right) \frac{1}{\sqrt{t^*}} &= \frac{\sqrt{2 k_{\rm B} T (\eta / m) t^*}}{\sqrt{m {v^*}^2}} \\
&= \sqrt{2 \, \frac{t^*}{\tau_\eta}} \sqrt{\frac{k_{\rm B} T}{m {v^*}^2}} \\
&= \sqrt{2 \tilde{\eta}} \, \sqrt{\frac{k_{\rm B} T}{m {v^*}^2}},
\end{align}
\end{subequations}
where in the first step we identified $\eta/m = 1/\tau_\eta$ (Eq.~\ref{eqnA:tau_eta_def})
and in the second step $t^*/\tau_\eta = \tilde{\eta}$ (Eq.~\ref{eqnA:eta_tilde_def}).

A natural choice for the characteristic velocity $v^*$ that reduces the coefficient of the fluctuation term 
to $\sqrt{2 \tilde{\eta}}$ follows from the equipartition theorem, namely
\begin{align}
\frac{m {v^*}^2}{2} &= \frac{k_{\rm B} T}{2} \Rightarrow \\
v^* &= \sqrt{\frac{k_{\rm B} T}{m}}.
\end{align}
With this choice for $v^*$ and the choice we made for $t^*$ (Eq.~\ref{eqnA:t_star_choice}),
the characteristic position scale $x^* = v^* t^*$ (Eq.~\ref{eqnA:nondim_cond1}) becomes
\begin{subequations}
\begin{align}
x^* &= \sqrt{\frac{k_{\rm B} T}{m}} \frac{1}{\omega_0} \\
&= \sqrt{\frac{k_{\rm B} T}{k}},
\end{align}
\end{subequations}
where we made the substitution $\omega_0 = \sqrt{k/m}$ (Eq.~\ref{eqnA:omega0_def}).
Note that this choice for $x^*$ is also the natural one following from the equipartition theorem:
\begin{align}
\frac{k {x^*}^2}{2} = \frac{k_{\rm B} T}{2}.
\end{align}

Substituting Eqs.~\ref{eqnA:nondim_cond1}, \ref{eqnA:nondim_cond2}, \ref{eqnA:eta_tilde_def}, and \ref{eqnA:noise_coeff_simplification} into the system of equations for signal dynamics 
(Eq.~\ref{eqnA:nondim_system_interm2}), we obtain its final dimensionless form as
\begin{subequations}
\label{eqnA:SDE_system_dimensionless}
\begin{align}
\frac{{\rm d} \tilde{x}}{{\rm d} \tilde{t}} &= \tilde{v}, \\
\frac{{\rm d} \tilde{v}}{{\rm d} \tilde{t}} &= - \tilde{x} - \tilde{\eta} \tilde{v} + \sqrt{2 \tilde{\eta}} \, \psi(\tilde{t}\,).
\end{align}
\end{subequations}
For simplicity, in the main text we skipped the ``tilde'' symbol on top of the dimensionless variables.
We will also do the same in the subsequent parts of this Appendix section.

\subsection{Solution of the stochastic model}
\label{sec:AppendixE2}

We now proceed to solve the system of stochastic differential equations for the dimensionless position
and velocity (Eq.~\ref{eqnA:SDE_system_dimensionless}), skipping the ``tilde'' symbols for convenience. 
To that end, we first recast Eq.~\ref{eqnA:SDE_system_dimensionless} in a matrix form, namely
\begin{align}
\label{eqnA:SDE_system_matrix_interm}
\frac{{\rm d}}{{\rm d}t} 
\begin{bmatrix}
x \\
v
\end{bmatrix}
=
\underbracket[0.140ex]{
\begin{bmatrix}
0 & 1 \\
-1 & - \eta
\end{bmatrix}}_{\bf F}
\begin{bmatrix}
x \\
v
\end{bmatrix}
+ 
\underbracket[0.140ex]{
\begin{bmatrix}
0 \\
\sqrt{2 \eta}
\end{bmatrix}}_{\bf B}
\psi(t),
\end{align}
where the matrix ${\bf F}$ represents the coefficients in the deterministic part of the dynamics, 
while the non-zero entry in the matrix ${\bf B}$ is the coefficient of the stochastic term.

Because the dynamics is linear and the noise term is additive with $\langle \psi(t) \rangle = 0$, 
the dynamics of the mean is set solely by the deterministic term via
\begin{align}
\left \langle 
\begin{bmatrix}
x_t\\
v_t
\end{bmatrix}
\bigg|
\begin{bmatrix}
x_0 \\
v_0
\end{bmatrix}
\right \rangle
= 
e^{{\bf F}t}
\begin{bmatrix}
x_0 \\
v_0
\end{bmatrix},
\end{align}
where $e^{{\bf F}t}$ represents a matrix exponential and $t > 0$. Introducing the vector notation
\begin{align}
{\bf s} = [x, v]^{\rm T}
\end{align}
for succinctness, we can rewrite the above result for the mean dynamics as
\begin{align}
\label{eqnA:st_mean_given_s0}
\left\langle {\bf s}_t \, | \, {\bf s}_0 \right\rangle = e^{{\bf F}t} \, {\bf s}_0.
\end{align}

Before discussing the mean dynamics any further, we look at the autocorrelation of ${\bf s}$. Specifically, we consider 
the matrix $\boldsymbol{\Sigma}_{{\bf s}_t {\bf s}_0} = \langle {\bf s}_t {\bf s}_0^{\rm T} \rangle$, writing it from its definition as
\begin{subequations}
\label{eqnA:autorrelation_matrix_interm}
\begin{align}
\boldsymbol{\Sigma}_{{\bf s}_t {\bf s}_0} 
&= \int \int {\rm d}{\bf s}_t {\rm d}{\bf s}_0 \,  {\bf s}_t {\bf s}_0^{\rm T} P({\bf s}_t, {\bf s}_0) \\
&= \int {\rm d} {\bf s}_0 \left[ \int {\rm d}{\bf s}_t \,  {\bf s}_t  P({\bf s}_t \, | \, {\bf s}_0) \right] {\bf s}_0^{\rm T} P({\bf s}_0) \\
&= \int {\rm d} {\bf s}_0 \left \langle {\bf s}_t | {\bf s}_0 \right \rangle {\bf s}_0^{\rm T} P({\bf s}_0) \\
&= \int {\rm d} {\bf s}_0 \, e^{{\bf F} t} {\bf s}_0^{\,} {\bf s}_0^{\rm T} P({\bf s}_0) \\
&= e^{{\bf F}t} \left \langle {\bf s}_0^{\,} {\bf s}_0^{\rm T} \right \rangle \\
&= e^{{\bf F}t} \boldsymbol{\Sigma}_{{\bf s}_0}.
\end{align}
\end{subequations}
Here we substituted the expression for $\langle {\bf s}_t | {\bf s}_0 \rangle$ from Eq.~\ref{eqnA:st_mean_given_s0}
and introduced $\boldsymbol{\Sigma}_{{\bf s}_0} = \langle {\bf s}_0^{\,} {\bf s}_0^{\rm T} \rangle$ to represent the covariance  
matrix of ${\bf s}_0$. Considering the signal to be stationary, the covariance matrix $\boldsymbol{\Sigma}_{{\bf s}_0}$ 
can be obtained by solving the following Lyapunov equation:
\begin{align}
\label{eqnA:Sigma_s0_stationary}
{\bf F} \boldsymbol{\Sigma}_{{\bf s}_0} + \boldsymbol{\Sigma}_{{\bf s}_0}  {\bf F}^{\rm T} + {\bf B} {\bf B}^{\rm T} = {\bf 0}.
\end{align}
For details about the above claim, 
we refer the reader to section 6.5 of the book by Särkkä and Solin on stochastic processes \cite{Sarkka_Solin_2019}.
To briefly give an idea of how this condition emerges, we outline its derivation below. 
We write the time derivative of the product ${\bf s} {\bf s}^{\rm T}$ as
\begin{align}
\label{eqnA:Ito_applied_original}
\frac{{\rm d}}{{\rm d}t} \left( {\bf s} {\bf s}^{\rm T} \right) &= \frac{{\rm d} {\bf s}}{{\rm d}t} 
\, {\bf s}^{\rm T}  + {\bf s} \, \frac{{\rm d} {\bf s}^{\rm T}}{{\rm d}t\, } + {\bf B}{\bf B}^{\rm T}.
\end{align}
The first two terms on the right-hand side are what one would expect from ordinary calculus. 
The third term (${\bf B} {\bf B}^{\rm T}$), however, emerges when applying Itô's lemma 
to the nonlinear function ${\bf s}{\bf s}^{\rm T}$ of the stochastic variable ${\bf s}$ 
(see Särkkä and Solin \cite{Sarkka_Solin_2019}). 
Rewriting Eq.~\ref{eqnA:SDE_system_matrix_interm} as $\frac{\rm d}{{\rm d}t} {\bf s}  = {\bf F}{\rm d}{\bf s} + {\bf B} \psi(t)$
and substituting it into Eq.~\ref{eqnA:Ito_applied_original}, we find
\begin{align}
\frac{{\rm d}}{{\rm d}t} \left( {\bf s} {\bf s}^{\rm T} \right) = \left( {\bf F} {\bf s} + {\bf B} \psi(t) \right) {\bf s}^{\rm T}  + 
{\bf s} \left( {\bf s}^{\rm T} {\bf F}^{\rm T} + {\bf B}^{\rm T} \psi(t) \right) + {\bf B}{\bf B}^{\rm T}.
\end{align}
Next, we take the expectation of both sides. Noting that $\boldsymbol{\Sigma}_{{\bf s}} = \langle {\bf s} {\bf s}^{\rm T} \rangle$
and that $\psi(t)$ is uncorrelated with ${\bf s}(t)$, which implies $\langle \psi(t) \, {\bf s} (t) \rangle = {\bf 0}$, we obtain
\begin{align}
\label{eqnA:Sigma_s_time_dynamics}
\frac{{\rm d}}{{\rm d}t} \boldsymbol{\Sigma}_{\bf s} = {\bf F} \boldsymbol{\Sigma}_{\bf s} 
+ \boldsymbol{\Sigma}_{\bf s} {\bf F}^{\rm T} + {\bf B} {\bf B}^{\rm T}.
\end{align}
For a stationary signal, the covariance matrix is not a function of time. Setting the left-hand side of 
Eq.~\ref{eqnA:Sigma_s_time_dynamics} equal to zero results in Eq.~\ref{eqnA:Sigma_s0_stationary} 
that we claimed for the stationary $\boldsymbol{\Sigma}_{{\bf s}_0}$.

The substitution of ${\bf F}$ and ${\bf B}$ (see Eq.~\ref{eqnA:SDE_system_matrix_interm}) into the equation for
$\boldsymbol{\Sigma}_{{\bf s}_0}$ (Eq.~\ref{eqnA:Sigma_s0_stationary}) yields
\begin{align}
\begin{bmatrix}
0 & 1 \\
-1 & - \tilde{\eta}
\end{bmatrix}
\boldsymbol{\Sigma}_{{\bf s}_0}
+
\boldsymbol{\Sigma}_{{\bf s}_0}
\begin{bmatrix}
0 & -1 \\
1 & - \tilde{\eta}
\end{bmatrix}
+
\begin{bmatrix}
0 & 0 \\
0 & 2\tilde{\eta}
\end{bmatrix}
= {\bf 0}.
\end{align}
While solving such matrix equations is generally a non-trivial task, in our case the solution is simply 
\begin{align}
\label{eqnA:S_s0_identity}
\boldsymbol{\Sigma}_{{\bf s}_0} = {\bf I},
\end{align}
which can be easily confirmed by a visual inspection. Since the signal is stationary, ${\bf s}_t$ will have the same 
covariance matrix, namely
\begin{align}
\label{eqnA:S_st_identity}
\boldsymbol{\Sigma}_{{\bf s}_t} = {\bf I}.
\end{align}

In the information bottleneck problem, ${\bf s}_0$ and ${\bf s}_t$ serve as the signal and relevance variables, respectively.
The fact that their covariance matrices are unity conveniently shows that both variables are already standardized after 
the non-dimensionalization procedure, and that signal statistics is encoded in ${\bf s}_0 \leftrightarrow {\bf s}_t$
stochastic mappings. Substituting the identity
$\boldsymbol{\Sigma}_{{\bf s}_0} = {\bf I}$ into Eq.~\ref{eqnA:autorrelation_matrix_interm}f,
we find
\begin{align}
\label{eqnA:Sigma_st_s0}
\boldsymbol{\Sigma}_{{\bf s}_t {\bf s}_0} = 
\begin{bmatrix}
\langle x_0 x_t \rangle & \langle  v_0 x_t \rangle \\
\langle x_0 v_t \rangle & \langle  v_0 v_t \rangle
\end{bmatrix}
=
e^{{\bf F}t}.
\end{align}
With $\boldsymbol{\Sigma}_{{\bf s}_t {\bf s}_0}$ at hand, the covariance matrices of forward and backward mapping 
distributions can be obtained by applying the Schur complement formula (Eq.~\ref{eqnA:Schur_general}), namely
\begin{align}
\boldsymbol{\Sigma}_{{\bf s}_t | {\bf s}_0} &= {\bf I} - \boldsymbol{\Sigma}^{\,}_{{\bf s}_t {\bf s}_0} 
\boldsymbol{\Sigma}_{{\bf s}_0 {\bf s}_t}, \\
\label{eqnA:Sigma_s0_given_set_Schur}
\boldsymbol{\Sigma}_{{\bf s}_0 | {\bf s}_t} &= {\bf I} - \boldsymbol{\Sigma}^{\,}_{{\bf s}_0 {\bf s}_t} 
\boldsymbol{\Sigma}_{{\bf s}_t {\bf s}_0}.
\end{align}

The functional form of the matrix exponential $e^{{\bf F}t}$ that defines the auto- and cross-correlation functions
via Eq.~\ref{eqnA:Sigma_st_s0} depends on the choice of the dimensionless damping coefficient $\eta$ that 
enters the definition of the matrix ${\bf F} = \bigl[ \begin{smallmatrix} 0 & 1 \\ -1 & -\eta \end{smallmatrix} \bigr]$
(see Eq.~\ref{eqnA:SDE_system_matrix_interm}). Three distinct regimes can be identified: 
overdamped ($\eta > 2$), underdamped ($\eta < 2$), and critically damped ($\eta = 2$).
The expressions of the matrix exponential in these three distinct regimes are shown below:
\begin{subequations}
\label{eqnA:pairwise_correlations}
\begin{align}
\eta > 2: \quad e^{{\bf F}t} &= 
\begin{bmatrix}
\cosh(\kappa t) + \frac{\eta}{2\kappa}  \sinh(\kappa t) 
& \phantom{\dfrac{1}{1}} \frac{1}{\kappa} \sinh(\kappa t) \phantom{\dfrac{1}{1}} \\
- \frac{1}{\kappa} \sinh(\kappa t)  & 
\cosh(\kappa t) - \frac{\eta}{2\kappa}  \sinh(\kappa t) 
\end{bmatrix}
e^{-\eta t/2}, \\
\eta < 2: \quad e^{{\bf F}t} &= 
\begin{bmatrix}
\cos(\omega t) + \frac{\eta}{2\omega}  \sin(\omega t) 
& \phantom{\dfrac{1}{1}} \frac{1}{\omega} \sin(\omega t) \phantom{\dfrac{1}{1}} \\
- \frac{1}{\omega} \sin(\omega t)  & 
\cos(\omega t) - \frac{\eta}{2\omega}  \sin(\omega t) 
\end{bmatrix}
e^{-\eta t/2}, \\
\eta = 2: \quad e^{{\bf F}t} &= 
\begin{bmatrix}
1+t & t  \\
-t & 1-t
\end{bmatrix} e^{-t}.
\end{align}
\end{subequations}
Here $\kappa = \sqrt{\eta^2/4-1}$ is defined for the overdamped case ($\eta > 2$), while $\omega = \sqrt{1 - \eta^2/4}$ 
representing the angular frequency of oscillators is defined for the underdamped case ($\eta < 2$). Note that in all cases
the two off-diagonal elements are identical in magnitude but opposite in sign. This implies the relation
\begin{align}
\label{eqn:xv_correlation_identity}
\langle x_0 v_t \rangle = - \langle v_0 x_t \rangle,
\end{align}
since the two elements represent 
$\langle x_0 v_t \rangle$ and $\langle v_0 x_t \rangle$ (see Eq.~\ref{eqnA:Sigma_st_s0}).
In fact, Eq.~\ref{eqn:xv_correlation_identity} holds generally for all stationary signals. 
To see how it emerges, note that 
$\langle x_{t_0} x_{t_0 + t} \rangle$ is independent of $t_0$ due to stationarity. Taking the derivative with respect to $t_0$ yields 
$\frac{{\rm d}}{{\rm d}t_0} \langle x_{t_0} x_{t_0 + t} \rangle 
= \langle v_{t_0} x_{t_0 + t} \rangle + \langle x_{t_0} v_{t_0 + t} \rangle = 0$. In the limit $t_0 \rightarrow 0$, we arrive 
at Eq.~\ref{eqn:xv_correlation_identity}.

\subsection{Optimal prediction strategy}
\label{sec:AppendixE3}

Having obtained the signal statistics, we now turn to the question of optimal prediction. 
The key quantity that dictates the properties of the optimal strategy is the covariance matrix 
$\boldsymbol{\Sigma}_{{\bf s}_0 | {\bf s}_\tau}$, where $\tau$ stands for the prediction interval.
The principal encoding directions are the eigenvectors of this matrix. Representing a unit encoding vector 
via an angle, namely $\hat{\bf w} = [\cos \varphi, \sin \varphi]^{\rm T}$, we can write the eigenvector condition as
\begin{align}
\label{eqnA:eigen_varphi_condition_init}
\boldsymbol{\Sigma}_{{\bf s}_0 | {\bf s}_\tau} 
\begin{bmatrix}
\cos \varphi \\
\sin \varphi
\end{bmatrix}
=
\lambda^2 
\begin{bmatrix}
\cos \varphi \\
\sin \varphi
\end{bmatrix}.
\end{align}
We will use $\varphi_1$ as the direction of the eigenvector with the smaller corresponding eigenvalue $\lambda_1^2$,
where $\lambda_1$ is the semi-minor axis length of the $P({\bf s}_0 | {\bf s}_\tau)$ ellipse.
Similarly, we denote the semi-major axis length of the $P({\bf s}_0 | {\bf s}_\tau)$ ellipse by $\lambda_2$ 
and the corresponding angle by $\varphi_2$.

To obtain these two angles and the corresponding $\lambda$-values, we first multiply both sides of 
Eq.~\ref{eqnA:eigen_varphi_condition_init} (from the left) 
by the row vector $[\cos \varphi, \sin \varphi]$, which makes the 
right-hand side independent of $\varphi$ (since $\cos^2 \varphi + \sin^2 \varphi = 1$). 
With $\boldsymbol{\Sigma}_{{\bf s}_0 | {\bf s}_\tau}$ written 
in its matrix form, this multiplication results in
\begin{align}
\begin{bmatrix}
\cos \varphi & \sin \varphi
\end{bmatrix}
\begin{bmatrix}
\sigma^2_{x_0 | {\bf s}_\tau} & {\rm cov}(x_0, v_0 | {\bf s}_\tau) \\
{\rm cov}(x_0, v_0 | {\bf s}_\tau) & \sigma^2_{v_0 | {\bf s}_\tau}
\end{bmatrix}
\begin{bmatrix}
\cos \varphi \\
\sin \varphi
\end{bmatrix} = \lambda^2,
\end{align}
where ${\rm cov}(x_0, v_0 | {\bf s}_\tau)$ is the covariance between $x_0$ and $v_0$, conditional on ${\bf s}_\tau$.
Performing the matrix products, we find
\begin{align}
\left( \cos^2 \varphi \right) \sigma^2_{x_0 | {\bf s}_\tau} + 
\left( \sin^2 \varphi \right) \sigma^2_{v_0 | {\bf s}_\tau} + \left( 2 \sin \varphi \cos \varphi \right)
 {\rm cov}(x_0, v_0 | {\bf s}_\tau) = \lambda^2.
\end{align}
Next, we use the identities $\cos^2\varphi = \frac{1}{2}(1 + \cos 2 \varphi)$, $\sin^2\varphi = \frac{1}{2}(1 - \cos 2 \varphi)$
and $2 \sin \varphi \cos \varphi = \sin 2\varphi$ to rewrite the above equation as
\begin{align}
\label{eqnA:w_Sigma_w_varphi_interm}
\frac{1}{2}\left( \sigma^2_{x_0 | {\bf s}_\tau} + \sigma^2_{v_0 | {\bf s}_\tau} \right) 
+ \frac{\cos (2\varphi)}{2} \left( \sigma^2_{x_0 | {\bf s}_\tau} - \sigma^2_{v_0 | {\bf s}_\tau} \right)
+ \sin (2\varphi) \, {\rm cov}(x_0, v_0 | {\bf s}_\tau) = \lambda^2.
\end{align}
Now, the left-hand side of Eq.~\ref{eqnA:w_Sigma_w_varphi_interm}, which,
recalling that $\hat{\bf w} = [\cos \varphi, \sin \varphi]^{\rm T}$, is equal to
$\hat{\bf w}^{\rm T} \boldsymbol{\Sigma}_{{\bf s}_0 | {\bf s}_\tau} \hat{\bf w}$, achieves its smallest ($\lambda_1^2$)
and largest ($\lambda_2^2$) values at $\varphi = \varphi_1$ and $\varphi = \varphi_2$, respectively.
Therefore, to find these angles, we differentiate the left-hand side by $\varphi$ and set it equal to zero, obtaining
\begin{align}
- \sin(2 \varphi)  \left( \sigma^2_{x_0 | {\bf s}_\tau} - \sigma^2_{v_0 | {\bf s}_\tau} \right) 
+ 2 \cos(2 \varphi) \, {\rm cov} (x_0, v_0 | {\bf s}_\tau) = 0.
\end{align}
After some rearrangements, the condition for $\varphi \in \{\varphi_1, \varphi_2 \}$ becomes
\begin{align}
\label{eqnA:condition_tan_2phi_init}
\tan(2 \varphi) = \frac{2 \, {\rm cov} (x_0, v_0 | {\bf s}_\tau)}{\sigma^2_{x_0 | {\bf s}_\tau} - \sigma^2_{v_0 | {\bf s}_\tau}}.
\end{align}
Note that this condition on $\tan(2\varphi)$ specifies the angle up to a multiple of $\pi/2$ because the function $\tan(\cdot)$ has a period of $\pi$.
This explains why the angles $\varphi_1$ and $\varphi_2$ corresponding to the perpendicular directions $\hat{\bf w}_1$ and $\hat{\bf w}_2$
both satisfy Eq.~\ref{eqnA:condition_tan_2phi_init}. 

Before discussing the assignment of angles $\varphi_1$ and $\varphi_2$, let us express the terms appearing on the right-hand side 
of Eq.~\ref{eqnA:condition_tan_2phi_init} in terms of pairwise correlations. 
To that end, we use Eq.~\ref{eqnA:Sigma_s0_given_set_Schur} for $\boldsymbol{\Sigma}_{{\bf s}_0 | {\bf s}_\tau}$, 
the matrix form in Eq.~\ref{eqnA:Sigma_st_s0} and the identity $\langle x_0 v_\tau \rangle = -\langle v_0 x_\tau \rangle$
(Eq.~\ref{eqn:xv_correlation_identity}) to write
\begin{subequations}
\label{eqnA:Sigma_s0_given_stau_pairwise}
\begin{align}
\boldsymbol{\Sigma}_{{\bf s}_0 | {\bf s}_\tau} &= {\bf I}
-
\begin{bmatrix}
\langle x_0 x_\tau \rangle & \langle x_0 v_\tau \rangle \\
\langle v_0 x_\tau \rangle & \langle v_0 v_\tau \rangle
\end{bmatrix}
\begin{bmatrix}
\langle x_0 x_\tau \rangle & \langle v_0 x_\tau \rangle \\
\langle x_0 v_\tau \rangle & \langle v_0 v_\tau \rangle
\end{bmatrix} \\
&= {\bf I} 
- 
\begin{bmatrix}
\langle x_0 x_\tau \rangle & -\langle v_0 x_\tau \rangle \\
\langle v_0 x_\tau \rangle & \langle v_0 v_\tau \rangle
\end{bmatrix}
\begin{bmatrix}
\langle x_0 x_\tau \rangle & \langle v_0 x_\tau \rangle \\
-\langle v_0 x_\tau \rangle & \langle v_0 v_\tau \rangle
\end{bmatrix} \\
&= 
\begin{bmatrix}
\underbracket[0.140ex]{1 - \langle x_0 x_\tau \rangle^2 - \langle v_0 x_\tau \rangle^2}_{\sigma^2_{x_0 | {\bf s}_\tau}} & 
\underbracket[0.140ex]{-\langle v_0 x_\tau \rangle \big( \langle x_0 x_\tau \rangle - \langle v_0 v_\tau\rangle \big)}_{{\rm cov}(x_0, v_0 | {\bf s}_\tau)}\\
\underbracket[0.140ex]{-\langle v_0 x_\tau \rangle \big( \langle x_0 x_\tau \rangle - \langle v_0 v_\tau\rangle \big)}_{{\rm cov}(x_0, v_0 | {\bf s}_\tau)} & 
\underbracket[0.140ex]{1 - \langle v_0 v_\tau \rangle^2 - \langle v_0 x_\tau \rangle^2}_{\sigma^2_{v_0 | {\bf s}_\tau}}
\end{bmatrix}.
\end{align}
\end{subequations}
Substituting the individual terms into Eq.~\ref{eqnA:condition_tan_2phi_init} and performing some simplifications, 
we arrive at
\begin{align}
\label{eqnA:tan2phi_condition_general_final}
\tan(2\varphi) = \frac{2 \langle v_0 x_\tau \rangle}{\langle x_0 x_\tau \rangle + \langle v_0 v_\tau \rangle}.
\end{align}

As we saw before, the functional form of pairwise correlation terms depends on the damping regime
(see Eq.~\ref{eqnA:Sigma_st_s0} and Eq.~\ref{eqnA:pairwise_correlations}).
Substituting the expressions of $\langle v_0 x_\tau \rangle$, $\langle x_0 x_\tau \rangle$ and $\langle v_0 v_\tau \rangle$
for different regimes into 
Eq.~\ref{eqnA:tan2phi_condition_general_final}, we find
\begin{align}
\label{eqnA:tan2phi_condition_cases}
\tan(2\varphi) = 
\begin{cases}
\frac{1}{\kappa} \tanh(\kappa \tau), &{\rm if}\, \, \eta > 2\\[1.2ex]
\frac{1}{\omega} \tan(\omega \tau), &{\rm if}\, \, \eta < 2\\[1.2ex]
\tau, &{\rm if}\,\, \eta = 2
\end{cases}
\end{align}
As a reminder, $\kappa = \sqrt{\eta^2/4 - 1}$ is defined for the overdamped case ($\eta > 2$),
and $\omega = \sqrt{1 - \eta^2/4}$ is defined for the underdamped case ($\eta < 2$).

To understand how the encoding angles $\varphi_1$ and $\varphi_2$ both satisfying Eq.~\ref{eqnA:tan2phi_condition_cases}
are assigned, we look at the conditional covariance ${\rm cov}(x_0, v_0 | {\bf s}_\tau)$, which is the off-diagonal 
term of $\boldsymbol{\Sigma}_{{\bf s}_0 | {\bf s}_\tau}$ (Eq.~\ref{eqnA:Sigma_s0_given_stau_pairwise}) 
and has the following general form:
\begin{align}
{\rm cov}(x_0, v_0 | {\bf s}_\tau) = - \langle v_0 x_\tau \rangle \big( \langle x_0 x_\tau \rangle - \langle v_0 v_\tau \rangle \big).
\end{align}
Substituting the pairwise correlation terms (Eq.~\ref{eqnA:pairwise_correlations}), we obtain the conditional covariance 
at different regimes as
\begin{align}
{\rm cov}(x_0, v_0 | {\bf s}_\tau) = 
\begin{cases}
-\frac{\eta}{\kappa^2} \sinh^2(\kappa \tau) \, e^{-\eta \tau/2},  &{\rm if}\,\, \eta > 2 \\[1.2ex]
-\frac{\eta}{\omega^2} \sin^2(\omega \tau) \, e^{-\eta \tau/2},  &{\rm if}\,\, \eta < 2 \\[1.2ex]
-2\tau^2 e^{-\tau}, &{\rm if}\,\, \eta = 2
\end{cases}
\end{align}
Notably, ${\rm cov}(x_0, v_0 | {\bf s}_\tau) \le 0$ in all regimes irrespective of the prediction interval.
This implies that 
the $P({\bf s}_0 | {\bf s}_\tau)$ ellipse on the ${\bf s}_0$-plane will generally 
be ``left-tilted'' and hence, 
the angle $\varphi_1$ of its minor axis will be in the first quadrant, i.e. $\varphi_1 \in (0, \pi/2)$.
The only exception is the special case in the underdamped regime, namely $\omega \tau \in \pi n$,
where this correlation vanishes and $P({\bf s}_0 | {\bf s}_\tau)$ ellipse turns into a circle.

In light of this property, if we consider the function $\arctan(\cdot)$ with outputs in the first and second quadrants, i.e. 
between $0$ and $\pi$,
the angle $\varphi_1$ of the first principal encoding direction follows from 
the condition in Eq.~\ref{eqnA:tan2phi_condition_cases} as
\begin{align}
\label{eqnA:varphi1_cases}
\varphi_1 = 
\begin{cases}
\frac{1}{2}\arctan \left( \frac{\tanh(\kappa \tau)}{\kappa} \right), &{\rm if} \, \, \eta > 2 \\[1.2ex]
\frac{1}{2}\arctan \left( \frac{\tan(\omega \tau)}{\omega} \right), &{\rm if} \, \, \eta < 2 \\[1.2ex]
\frac{1}{2}\arctan \tau, &{\rm if} \, \, \eta = 2
\end{cases}
\end{align}
Since the output of $\arctan(\cdot)$ is in the first or the second quadrant, half of that output will be guaranteed to fall
in the first quadrant, satisfying the requirement $\varphi_1 \in (0, \pi/2)$.
This requirement means that the first encoding component 
$z_1 =  x_0 \cos \varphi_1  + v_0 \sin \varphi_1$
linearly combines $x_0$ and $v_0$ with weights of the same sign.

The second encoding component, however, having as its angle
\begin{align}
\varphi_2 = \varphi_1 + \pi/2,
\end{align}
combines $x_0$ and $v_0$ with weights of opposite signs, since $\varphi_2$ now falls in an even quadrant.\\

Having derived the principal encoding directions, we now turn to the dimensions $\lambda_1$ and $\lambda_2$
of the $P({\bf s}_0 | {\bf s}_\tau)$ ellipse, which set the bound on the predictive information available 
to the two principal encoding components. 
From Eq.~\ref{eqnA:w_Sigma_w_varphi_interm}, we can write
\begin{align}
\lambda^2 &= \frac{1}{2}
\left( 
\sigma^2_{x_0 | {\bf s}_\tau} + \sigma^2_{v_0 | {\bf s}_\tau}
\right)
+ \frac{\cos(2\varphi)}{2} \left( 
\sigma^2_{x_0 | {\bf s}_\tau} - \sigma^2_{v_0 | {\bf s}_\tau}
\right)
\Bigg( 
1 + \tan(2\varphi) 
\underbracket[0.140ex]{
\frac{2 {\rm cov}(x_0, v_0 | {\bf s}_\tau) }{
\sigma^2_{x_0 | {\bf s}_\tau} - \sigma^2_{v_0 | {\bf s}_\tau}}}_{=\tan(2\varphi)}
\Bigg).
\end{align}
Recognizing the presence of $\tan(2\varphi)$ on the right-hand side (see Eq.~\ref{eqnA:condition_tan_2phi_init})
and further noting the general identity $\cos (2\varphi) = \pm 1/\sqrt{1 + \tan^2 (2\varphi)}$
(with different signs used for $\varphi \in \{ \varphi_1, \varphi_2 \}$), we arrive at 
\begin{align}
\lambda^2 = 
\frac{1}{2}
\left( 
\sigma^2_{x_0 | {\bf s}_\tau} + \sigma^2_{v_0 | {\bf s}_\tau}
\right)
\pm \frac{1}{2}
\left| \sigma^2_{x_0 | {\bf s}_\tau} - \sigma^2_{v_0 | {\bf s}_\tau} \right|
\sqrt{1 + \tan^2(2\varphi)}.
\end{align}
Next, we substitute expressions for $\sigma^2_{x_0 | {\bf s}_\tau}$ and $\sigma^2_{v_0 | {\bf s}_\tau}$ 
written in terms of pairwise correlations (Eq.~\ref{eqnA:Sigma_s0_given_stau_pairwise}c), obtaining
\begin{align}
\label{eqnA:lambda2_via_correlation}
\lambda^2 =1 - \frac{\langle x_0 x_\tau \rangle^2 + \langle v_0 v_\tau \rangle^2 + 2 \langle v_0 x_\tau \rangle^2}{2}
\pm \frac{1}{2} \left| \langle x_0 x_\tau \rangle^2 - \langle v_0 v_\tau \rangle^2 \right|
\sqrt{1 + \tan^2(2\varphi)}.
\end{align}

As discussed earlier, the general expressions for the pairwise correlations depend on the damping regime
(see Eq.~\ref{eqnA:Sigma_st_s0} and Eq.~\ref{eqnA:pairwise_correlations}). After substituting these expressions 
for the different damping regimes and performing algebraic simplifications, we arrive at 
the final solution for $\lambda^2$:
\begin{align}
\label{eqnA:lambda_sq_final_expressions}
\lambda^2 = 
\begin{cases}
1 -  \left( 1 + \frac{\eta^2}{2\kappa^2} \sinh^2(\kappa \tau)
\pm \frac{\eta}{\kappa} \sinh(\kappa \tau) \sqrt{\cosh^2(\kappa \tau) + \frac{1}{\kappa^2}\sinh^2(\kappa \tau)}
\, \right) e^{-\eta \tau}, &{\rm if}\, \, \eta > 2 \\[3.0ex]
1 -  \left( 1 + \frac{\eta^2}{2 \omega^2} \sin^2 (\omega \tau)  
\pm \frac{\eta}{\omega} |\sin(\omega \tau)| \sqrt{\cos^2(\omega \tau) + \frac{1}{\omega^2} \sin^2(\omega \tau)}
\, \right) e^{-\eta \tau}, &{\rm if}\, \, \eta < 2 \\[3.0ex]
1 -  \left( 1 + 2 \tau^2 \pm 2 \tau \sqrt{ 1 + \tau^2} \, \right) e^{-2 \tau}, &{\rm if}\, \, \eta = 2
\end{cases}
\end{align}
Note that the shorter semi-axis length $\lambda_1$ corresponding to the first encoding component is obtained 
through the ``+'' solution of Eq.~\ref{eqnA:lambda_sq_final_expressions}, 
while the ``-'' solution gives the longer semi-axis length $\lambda_2$ 
corresponding to the second component.
Since the marginal variances of $x$ and $v$ are unity ($\sigma^2_x = \sigma^2_v = 1$), 
and hence the signal amplitude $r$ appearing in $I_{\rm max}(z_i; {\bf s}_\tau) = \log (r/\lambda_i)$
is $1$,
the semi-axis lengths $\lambda_1$ and $\lambda_2$
directly determine the maximum predictive information available to the $z_1$ and $z_2$ components,
respectively, via
\begin{subequations}
\begin{align}
I_{\rm max}(z_1; {\bf s}_\tau) &= - \log \lambda_1, \\
I_{\rm max}(z_2; {\bf s}_\tau) &= - \log \lambda_2.
\end{align}
\end{subequations}
In the limit of very short forecast intervals $\tau$, the semi-axis lengths can be approximated as 
\begin{subequations}
\begin{align}
\lambda_1 &\approx \sqrt{\frac{\eta}{6}} \, \tau^{3/2}, \\\
\lambda_2 &\approx \sqrt{2\eta} \, \tau^{1/2}.
\end{align}
\end{subequations}
Intuitively, the $\propto \tau^{1/2}$ scaling of $\lambda_2$ captures the diffusive fluctuations
of the signal derivative $v$ at short time scales, 
while $\propto \tau^{3/2}$ (the integral of $\propto \tau^{1/2}$) captures
the scaling of signal fluctuations.
The corresponding maximum predictive information values then become approximately  
$I_{\rm max}(z_1; {\bf s}_\tau) \approx -\frac{3}{2} \log \tau$
and 
$I_{\rm max}(z_2; {\bf s}_\tau) \approx -\frac{1}{2} \log \tau$,
where we ignored the contributions from constant prefactors, as they are dominated by 
the much larger terms proportional to $-\log \tau$.
We can thus see that at very short forecast intervals the $z_1$ component contains three times as much 
predictive information as the $z_2$ component, irrespective of the damping regime.

In contrast, the behavior at long forecast intervals is highly dependent on the damping regime. 
In what follows, we explore this behavior case by case.\\

\phantom{\, \\}

\noindent \underline{Underdamped regime ($\eta < 2$)}\\

In Sec. VI of the main text, we already studied the behavior of the optimal encoding angles 
and relative predictive capacities of the two encoding components as a function of the forecast interval (Fig.~13).
Here, we provide the derivation of an analytical result that we stated in that same section (Eq.~47) concerning 
the maximum value of the relative predictive information available to the $z_1$ component in the limit of 
long forecast intervals.

This normalized predictive information can be generally written as
\begin{subequations}
\begin{align}
\tilde{I}_{\rm max}(z_1; {\bf s}_\tau) &= \frac{I_{\rm max}(z_1; {\bf s}_\tau)}{
I_{\rm max}(z_1; {\bf s}_\tau) + I_{\rm max}(z_2; {\bf s}_\tau)} \\
&= \frac{\log \lambda_1}{\log \lambda_1 + \log \lambda_2} \\
&= \frac{\log \lambda_1^2}{\log \lambda_1^2 + \log \lambda_2^2}.
\end{align}
\end{subequations}
Now, from Eq.~\ref{eqnA:lambda_sq_final_expressions}, we know that $\lambda_i^2$ is of the form
$\lambda_i^2 = 1 - u_i e^{-\eta \tau}$, where $u_i$ is a finite quantity. In the limit $\tau \gg \eta^{-1}$,
we can approximate the logarithmic terms as $\log \lambda_i^2 \approx - u_i e^{-\eta \tau}$.
The normalized predictive information can then be approximated as 
$\tilde{I}_{\rm max}(z_1; {\bf s}_\tau) = u_1/(u_1 + u_2)$. Substituting the expressions for $u_i$, we obtain
\begin{subequations}
\label{eqnA:I1_rel_max_intermediate}
\begin{align}
\tilde{I}_{\rm max}(z_1; {\bf s}_\tau)
&\approx \frac{1 + \frac{\eta^2}{2 \omega^2} \sin^2 (\omega \tau)  
+ \frac{\eta}{\omega} |\sin(\omega \tau)| \sqrt{\cos^2(\omega \tau) + \frac{1}{\omega^2} \sin^2(\omega \tau)}}{
2 \left( 1 + \frac{\eta^2}{2 \omega^2} \sin^2 (\omega \tau)   \right)} \\
&= \frac{1}{2} + \frac{\eta}{2\omega} \frac{
 |\sin(\omega \tau)| \sqrt{\cos^2(\omega \tau) + \frac{1}{\omega^2} \sin^2(\omega \tau)}
}{1 + \frac{\eta^2}{2 \omega^2} \sin^2 (\omega \tau)  }.
\end{align}
\end{subequations}
Recalling the identity $\omega^2 = 1 - \eta^2/4$, one can show that the above expression increases monotonically
with $|\sin(\omega \tau)|$ and hence achieves its maximum value when $|\sin(\omega \tau)| = 1$, which happens 
when $\omega \tau = \pi/2 + k \pi$, $k \in \mathcal{N}$ 
(note that $k$ must be large enough for the condition $\tau_k \gg \eta^{-1}$ to hold).
Substituting $|\sin(\omega \tau)| = 1$ into Eq.~\ref{eqnA:I1_rel_max_intermediate} 
and using the identity $\omega^2 = 1 - \eta^2/4$, we find
\begin{subequations}
\begin{align}
\tilde{I}_{\rm max}(z_1; {\bf s}_{\tau_k}) &\approx \frac{1}{2} + \frac{\frac{\eta}{2\omega^2}
}{1 + \frac{\eta^2}{2\omega^2}} \\
&= \frac{1}{2} + \frac{\eta/2}{1 + (\eta/2)^2},
\end{align}
\end{subequations}
which is the result reported in the main text.\\

\noindent \underline{Overdamped regime ($\eta > 2$)}\\

\begin{figure}[!ht]
\centering
\includegraphics[width=1.00\textwidth]{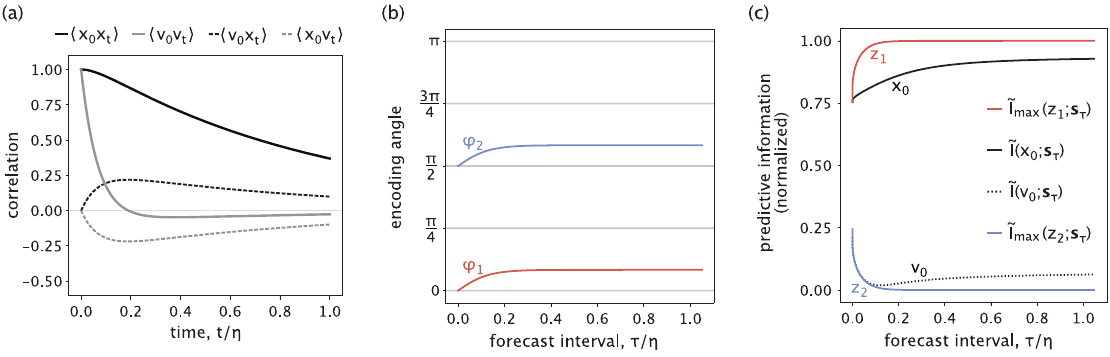}
\caption{
Features of the signal statistics and the optimal encoding strategy for example overdamped signal dynamics
($\eta = 4$).
(a) Auto- and cross-correlation functions of the signal and its derivative.
(b) Optimal encoding angles $\varphi_1$ and $\varphi_2 = \varphi_1 + \pi/2$ for the components $z_1$ and $z_2$,
respectively, shown as a function of the forecast interval $\tau$. 
The angle $\varphi$ increases monotonically with $\tau$, with its upper bound
given by $\frac{1}{2} \arctan \left( 1/ \sqrt{\eta^2/4 - 1} \right)$. 
This upper bound reaches its maximum value of $\pi/4$ in the critical damping limit ($\eta \rightarrow 2$).
(c) Normalized values of predictive information plotted against the forecast interval for different encoding strategies.
All strategies are evaluated in the zero-noise limit.
Normalization is performed by the maximum achievable predictive information, namely $I({\bf s}_0; {\bf s}_\tau)$,
as determined by the signal statistics.
The upper (red) and lower (blue) curves represent the normalized predictive information 
available to the $z_1$ and $z_2$ components, respectively, in the limit of infinite encoding capacity.
Their sum equals one, namely $\tilde{I}_{\rm max}(z_1; {\bf s}_\tau) + \tilde{I}_{\rm max}(z_2; {\bf s}_\tau) = 1$.
The two curves in between correspond to purely $x_0$-based (solid black line) and 
purely $v_0$-based (dotted black line) strategies.
$\tilde{I}(x_0; {\bf s}_\tau)$ saturates at $\frac{1}{2} (1 + \sqrt{1 - 4/\eta^2})$, while 
$\tilde{I}(v_0; {\bf s}_\tau)$ saturates at $\frac{1}{2} (1 - \sqrt{1 - 4/\eta^2})$.
The corresponding features for the underdamped regime are shown in Fig. 12 and Fig. 13 of the main text.
}
\label{figA:overdamped}
\end{figure}

Unlike in the underdamped regime where the $z_1$ component periodically shifts from being 
predominantly $x_0$-based to predominantly $v_0$-based as the forecast interval increases, 
in the overdamped regime the most predictive $z_1$ component always assigns greater weight 
to the current signal $x_0$ than to its derivative $v_0$.
This behavior can be understood intuitively by noting that, in the overdamped regime,
the signal derivative $v$ decorrelates more rapidly compared to the signal $x$,
and hence, contains less predictive information (Fig.~\ref{figA:overdamped}a).
This is also apparent from the functional form of the angle $\varphi_1$ (Eq.~\ref{eqnA:varphi1_cases}),
rewritten below for convenience:
\begin{align}
\varphi_1 = \frac{1}{2} \arctan \left( \frac{\tanh(\kappa \tau)}{\kappa} \right).
\end{align}
Since the argument of the arctan function is positive for any choice of $\tau$,
its output is always less than $\pi/2$, implying that the angle $\varphi_1$ is less than $\pi/4$.
Thus, the weight $\hat{w}_{1,1} = \cos \varphi_1$
assigned to the current signal $x_0$ is always greater than the weight $\hat{w}_{1,2} = \sin \varphi_1$
assigned to its derivative $v_0$.
Fig.~\ref{figA:overdamped}b shows the angles $\varphi_1$  and $\varphi_2 = \varphi_1 + \pi/2$ plotted 
as functions of the forecast interval. 
The monotonically increasing $\varphi_1$ suggests that the importance of the current derivative 
in the $z_1$ component increases with the forecast interval.
This happens because, although the autocorrelation of $v$ decays rapidly,
its cross-correlation with the future signal value, $\langle v_0 x_t \rangle$, 
decays at a rate similar to that of the signal's
autocorrelation $\langle x_0 x_t \rangle$ (Fig.~\ref{figA:overdamped}a), making it beneficial to incorporate the current 
derivative $v_0$ into the $z_1$ component even at long forecast intervals.
In the limit $\tau \rightarrow \infty$, we have $\tanh(\kappa \tau) \rightarrow 1$, 
and hence the large-$\tau$ limit of the angle $\varphi_1$ is 
\begin{align}
\label{eqnA:varphi1_overdamped}
\varphi_1(\tau \rightarrow \infty) = \frac{1}{2} \arctan \left( \kappa^{-1} \right),
\end{align}
where, as a reminder, $\kappa = \sqrt{\eta^2/4 - 1}$.
Note that as $\eta$ approaches $2$, which corresponds to the critically damped regime, 
the large-$\tau$ limit of $\varphi_1$ converges to $\pi/4$.

The behavior of the relative predictive capacities of the two components in response to varying forecast intervals 
is also drastically different between the underdamped and overdamped regimes. 
In the underdamped regime, the maximum predictive capacities of the $z_1$ and $z_2$ components 
periodically become equal (Fig.~13b in the main text).
In contrast, in the overdamped regime, the relative predictive capacity of the second component 
decreases monotonically with $\tau$, eventually becoming zero in the limit $\tau \rightarrow \infty$.
This can be seen in Fig.~\ref{figA:overdamped}c.

Fig.~\ref{figA:overdamped}c also shows that the relative predictive capacity 
$\tilde{I}(x_0; {\bf s}_\tau) = I(x_0; {\bf s}_\tau) / I({\bf s}_0; {\bf s}_\tau)$
of the $x_0$-based strategy increases monotonically with the forecast interval $\tau$, saturating at a value less
than $1$. To obtain this value, we first note that 
\begin{subequations}
\label{eqnA:Ix0stau}
\begin{align}
I(x_0; {\bf s}_\tau) &= \log \frac{\sigma_{x_0}}{\sigma_{x_0 | {\bf s}_\tau}} \\
&= - \log \sigma_{x_0 | {\bf s}_\tau},
\end{align}
\end{subequations}
where we used the fact that $\sigma_{x_0} = 1$ (Eq.~\ref{eqnA:S_s0_identity}).
Next, we express the bound on predictive information as
\begin{subequations}
\label{eqnA:Is0stau_lambdas}
\begin{align}
I({\bf s}_0; {\bf s}_\tau) &= H({\bf s}_0) - H({\bf s}_0 | {\bf s}_\tau) \\
&= \log \left( \frac{1}{\lambda_1 \lambda_2} \right) \\
&= -\log \lambda_1 - \log \lambda_2,
\end{align}
\end{subequations}
where $\lambda_1$ and $\lambda_2$ are the semi-axis lengths of the $P({\bf s}_0 | {\bf s}_\tau)$ ellipse.

Substituting Eq.~\ref{eqnA:Ix0stau}b and Eq.~\ref{eqnA:Is0stau_lambdas}c
into the definition $\tilde{I}(x_0; {\bf s}_\tau) = I(x_0; {\bf s}_\tau) / I({\bf s}_0; {\bf s}_\tau)$
and squaring the arguments of log functions for convenience, we obtain
\begin{align}
\label{eqnA:Ix0stau_norm_intermediate}
\tilde{I}(x_0; {\bf s}_\tau) 
&= \frac{\log \sigma^2_{x_0 | {\bf s}_\tau}}{\log \lambda_1^2 + \log \lambda_2^2}.
\end{align}
Next, we apply the Schur complement formula and use the identity 
$\boldsymbol{\Sigma}_{{\bf s}_\tau} = {\bf I}$ (Eq. \ref{eqnA:S_st_identity})
to write
\begin{subequations}
\label{eqnA:sigma2_x0_given_s_tau}
\begin{align}
\sigma^2_{x_0 | {\bf s}_\tau} &= \sigma^2_{x_0} - 
\begin{bmatrix}
\langle x_0 x_\tau \rangle & \langle x_0 v_\tau \rangle
\end{bmatrix}
\boldsymbol{\Sigma}_{{\bf s}_\tau}^{-1}
\begin{bmatrix}
\langle x_0 x_\tau \rangle \\
\langle x_0 v_\tau \rangle
\end{bmatrix} \\
&= 1 - \langle x_0 x_\tau \rangle^2 - \langle x_0 v_\tau \rangle^2.
\end{align}
\end{subequations}
Now, both $\sigma^2_{x_0 | {\bf s}_\tau}$ (Eq.~\ref{eqnA:sigma2_x0_given_s_tau}) 
and $\lambda_i^2$ (Eq.~\ref{eqnA:lambda2_via_correlation})
are of the form $1 - u$, where $u$ consists of different auto- and cross-correlation terms.
In the limit of large $\tau$, these correlation terms are very small, allowing the use of the approximation
$\log (1 - u) \approx -u$. Applying this approximation
to the terms of Eq.~\ref{eqnA:Ix0stau_norm_intermediate} yields
\begin{align}
\tilde{I}(x_0; {\bf s}_\tau) 
\approx \frac{\langle x_0 x_\tau \rangle^2 + \langle x_0 v_\tau \rangle^2}{
\langle x_0 x_\tau \rangle^2 + \langle v_0 v_\tau \rangle^2 + 2 \langle v_0 x_\tau \rangle^2
}
\end{align}
when $\tau \gg \eta$. Substituting the different correlation terms from Eq.~\ref{eqnA:pairwise_correlations}
(see also Eq.~\ref{eqnA:Sigma_st_s0})
and performing algebraic simplifications, we arrive at
\begin{align}
\label{eqnA:I_x0_rel_overdamped_long_tau}
\tilde{I}(x_0; {\bf s}_\tau) \approx \frac{1}{2} \left( 1 + \sqrt{1 - \frac{4}{\eta^2}}\right).
\end{align}
This result indeed shows that the purely $x_0$-based strategy cannot retrieve the full predictive information 
at long forecast intervals, and that incorporating the current signal derivative $v_0$ into the $z_1$ 
component is necessary.

We also see from Eq.~\ref{eqnA:I_x0_rel_overdamped_long_tau} 
that $\tilde{I}(x_0; {\bf s}_\tau)$ approaches $1/2$ in the critically damped limit
($\eta \rightarrow 2$). With similar steps, one can show that 
$\tilde{I}(v_0; {\bf s}_\tau)$ too approaches $1/2$ when $\eta \rightarrow 2$.
This means that, in the critically damped regime,
the current derivative $v_0$ is as important as $x_0$ for predicting the distant future,
which is why it is assigned an equal weight in the $z_1$ encoding component
($\varphi_1 \rightarrow \pi/4$, 
follows from Eq.~\ref{eqnA:varphi1_overdamped} when $\eta \rightarrow 2$ or $\kappa \rightarrow 2$).

\end{document}